\documentclass[a4paper,11pt]{article}
\usepackage{jcappub}
\usepackage{amssymb}
\usepackage{amsmath,color}
\usepackage{epsfig}
\usepackage{graphicx,epsf,epsfig}
\usepackage{bbm}
\usepackage{subfigure}
\usepackage{multirow}
\usepackage{setspace}
\usepackage{verbatim}
\usepackage{epstopdf}
\usepackage{mathrsfs,dsfont,bm}
\usepackage{hyperref}
\usepackage{multirow}

\title{\boldmath The Gravitational-Wave Physics II: Progress}

\author[13]{Ligong Bian,}
\emailAdd{lgbycl@cqu.edu.cn}

\author[1,5,8]{Rong-Gen Cai,}
\emailAdd{cairg@itp.ac.cn}

\author[2]{Shuo Cao,}
\emailAdd{caoshuo@bnu.edu.cn}

\author[2]{Zhoujian Cao,}
\emailAdd{zjcao@bnu.edu.cn}

\author[2]{He Gao,}
\emailAdd{gaohe@bnu.edu.cn}

\author[1,5,8]{Zong-Kuan Guo,}
\emailAdd{guozk@itp.ac.cn}

\author[4,3]{Kejia Lee,}
\emailAdd{kjlee@pku.edu.cn}

\author[3]{Di Li,}
\emailAdd{dili@nao.cas.cn}

\author[8,5]{Jing Liu,}
\emailAdd{liujing@ucas.ac.cn}

\author[3,6]{Youjun Lu,}
\emailAdd{luyj@nao.cas.cn}

\author[1]{Shi Pi,}
\emailAdd{shi.pi@itp.ac.cn}

\author[7,6,3]{Jian-Min Wang,}
\emailAdd{wangjm@mail.ihep.ac.cn}

\author[1]{Shao-Jiang Wang,}
\emailAdd{schwang@itp.ac.cn}

\author[9]{Yan Wang,}
\emailAdd{ywang12@hust.edu.cn}

\author[10]{Tao Yang,}
\emailAdd{tao.yang@apctp.org}

\author[1,5]{Xing-Yu Yang,}
\emailAdd{yangxingyu@itp.ac.cn}

\author[3]{Shenghua Yu,}
\emailAdd{shenghuayu@bao.ac.cn}

\author[11,12]{Xin Zhang}
\emailAdd{zhangxin@mail.neu.edu.cn}

\affiliation[1]{CAS Key Laboratory of Theoretical Physics, Institute of Theoretical Physics, Chinese Academy of Sciences, Beijing 100190, China}
\affiliation[2]{Department of Astronomy, Beijing Normal University, Beijing 100875, China}
\affiliation[3]{National Astronomical Observatories, Chinese Academy of Sciences, 20A Datun Road, Beijing 100101, China}
\affiliation[4]{Kavli Institute for Astronomy and Astrophysics, Peking University, Beijing 100080, China}
\affiliation[5]{School of Physical Sciences, University of Chinese Academy of Sciences, No.19A Yuquan Road, Beijing 100049, China}
\affiliation[6]{School of Astronomy and Space Sciences, University of Chinese Academy of Sciences, 19A Yuquan Road, Beijing 100049, China}
\affiliation[7]{Key Laboratory for Particle Astrophysics, Institute of High Energy Physics, Chinese Academy of Sciences, 19B Yuquan Road, Beijing, 100049, China}
\affiliation[8]{School of Fundamental Physics and Mathematical Sciences, Hangzhou Institute for Advanced Study (HIAS), University of Chinese Academy of Sciences, Hangzhou 310024, China}
\affiliation[9]{MOE Key Laboratory of Fundamental Physical Quantities Measurements, Hubei Key Laboratory of Gravitation and Quantum Physics, National precise gravity measurement facility (PGMF), Department of Astronomy and School of Physics, Huazhong University of Science and Technology, Wuhan 430074, China}
\affiliation[10]{Asia Pacific Center for Theoretical Physics, Pohang 37673, Korea}
\affiliation[11]{Department of Physics, College of Sciences, Northeastern University, Shenyang 110819, China}
\affiliation[12]{MOE Key Laboratory of Data Analytics and Optimization for Smart Industry, Northeastern University, Shenyang 110819, China}
\affiliation[13]{Department of Physics, Chongqing University, Chongqing 401331, China}

\abstract{It has been a half-decade since the first direct detection of gravitational waves, which signifies the coming of the era of the gravitational-wave astronomy and gravitational-wave cosmology. The increasing number of the detected gravitational-wave events has revealed the promising capability of constraining various aspects of cosmology, astronomy, and gravity. Due to the limited space in this review article, we will briefly summarize the recent progress over the past five years, but with a special focus on some of our own work for the Key Project ``Physics associated with the gravitational waves'' supported by the National Natural Science Foundation of China. In particular, (1) we have presented the mechanism of the gravitational-wave production during some physical processes of the early Universe, such as inflation, preheating and phase transition, and the cosmological implications of gravitational-wave measurements; (2) we have put constraints on the neutron star maximum mass according to GW170817 observations; (3) we have developed a numerical relativity algorithm based on the finite element method and a waveform model for the binary black hole coalescence along an eccentric orbit.}

\begin{document}
\maketitle
\flushbottom

\section{Introduction}

Over the past half-decade, the direct detections of the gravitational wave (GW) events from the mergers of binary black holes (BBHs) \cite{Abbott:2016blz} and binary neutron stars (BNSs) \cite{TheLIGOScientific:2017qsa} have signified the coming of the era of the GW astronomy and GW cosmology, and ever since then, the increasing number of the detected GW events from Laser Interferometer Gravitational Wave Observatory (LIGO) first observing run (O1), O2 and O3 data \cite{LIGOScientific:2018mvr,Abbott:2020niy} has manifested its great potential to probe the unknown realms of the cosmology, astrophysics, and gravity. Due to the limited space of this paper, it is hardly feasible to give a thorough review for the numerous progress in this field, and we will only make a brief overview for the important progress over the past five years with a special focus on some of our own work. In the near future, the coming data from the Pulsar Timing Array (PTA), the Square Kilometre Array (SKA) and the Five-hundred-meter Aperture Spherical radio Telescope (FAST \cite{nan11,li18ieee}) will reveal the new era of radio astronomy. In the far future around the 2030s, the third generation of ground-based GW detectors [like the Einstein Telescope (ET) \cite{Punturo:2010zz} and the Cosmic Explorer (CE) \cite{Evans:2016mbw}] and space-borne GW detectors from the ongoing programs [like the Laser Interferometer Space Antenna (LISA) \cite{Armano:2016bkm,Audley:2017drz}, Taiji \cite{Hu:2017mde,Guo:2018npi,Taiji-1} \footnote{See \cite{Gong:2011zzd,Gong:2014mca} for earlier proposal of the Advanced Laser Interferometer Antenna (ALIA) mission and \cite{Gui:2021gjy} for its relation to the Taiji mission.} and TianQin \cite{Luo:2015ght,Luo:2020bls,Mei:2020lrl}] and planning proposals [like the DECi-hertz Interferometer Gravitational wave Observatory (DECIGO) \cite{Kawamura:2006up,Kawamura:2011zz} and the Big Bang Observer (BBO) \cite{Crowder:2005nr}] might be able to answer some important questions on the primordial Universe, multimessenger astronomy, and gravity tests.

\begin{figure}
\centering
\includegraphics[width=0.7\textwidth]{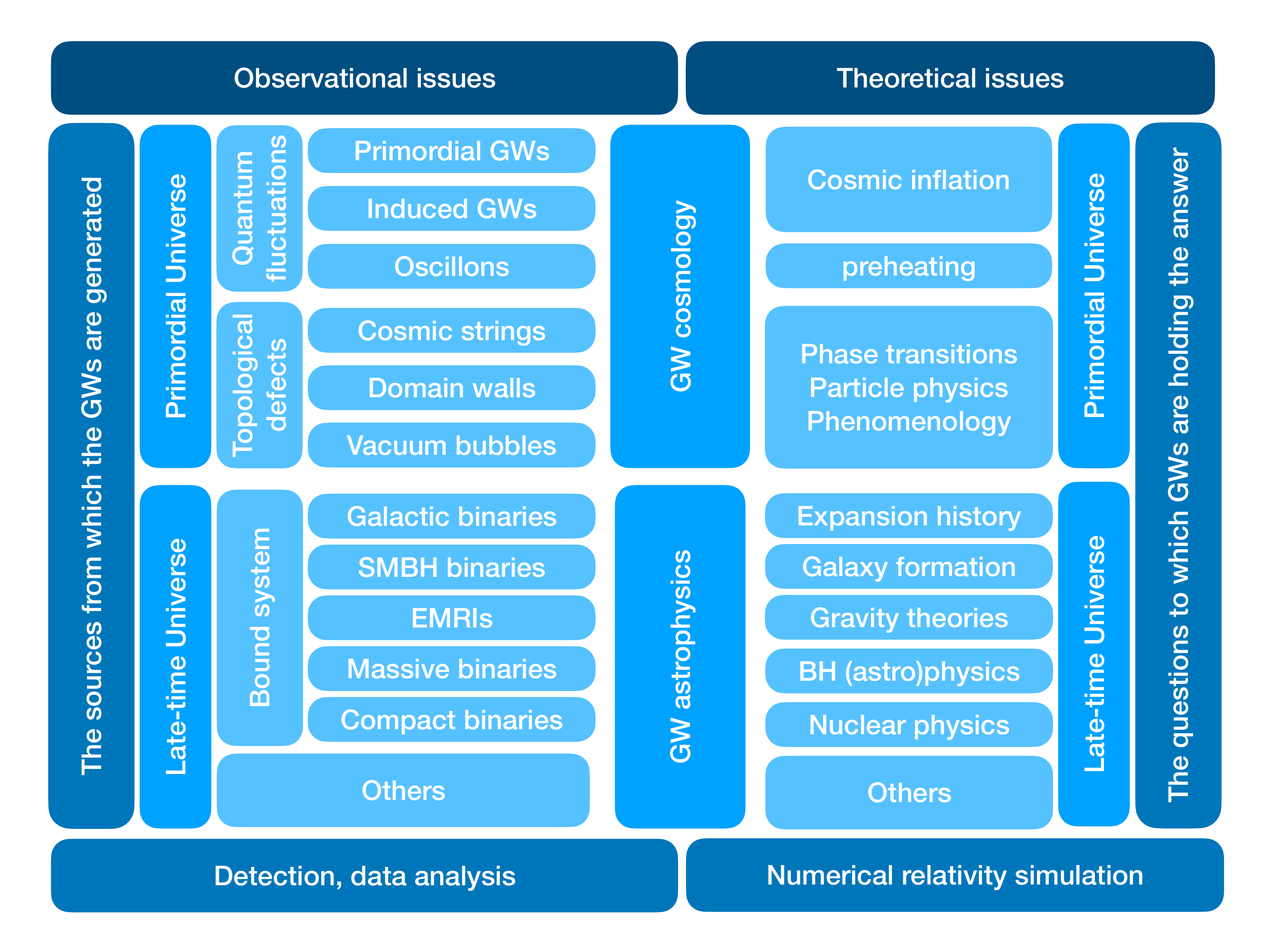}\\
\caption{An overview for GW studies of observational and theoretical issues, the former contains the GW sources, detection and data analysis, while the latter contains the theoretical problems (GW cosmology/astrophysics) and numerical relativity simulation. }\label{fig:class}
\end{figure}

The GW studies could be classified by either the sources from which the GWs are generated or the questions to which the GWs are holding the answer. In the former catalog, the GWs from the primordial Universe could be generated by the quantum fluctuations at large/small scales (primordial GWs, induced GWs, GWs from oscillons) and various topological defects (cosmic strings, domain walls, vacuum bubbles),  while the GWs from the late Universe could be generated by the mergers of galactic binaries, supermassive binary black holes (SMBBHs), extreme-mass-ratio-inspirals (EMRIs), massive binaries, and compact binaries [solar-mass BHs, neutron stars (NSs), white dwarfs (WDs), pulsars, primordial black holes (PBHs), etc.]. In the latter catalog, GWs from the primordial Universe could be used to explore the fundamental physics of cosmic inflation, preheating, and cosmic phase transitions, while the GWs from the late Universe could be used to constrain the black hole (astro)physics,  modified gravity theories, galaxy formation theories, late-time expansion history, nuclear physics, etc. The former catalog could be regarded as one end to the observational issues, which also include other issues like detection and data analysis. The latter catalog could be regarded as the other end to the theoretical issues, which also include other issues like numerical relativity simulations. This end-to-end picture is depicted in Fig. \ref{fig:class}, which will be partially addressed in this short review as a sequential status report to our previous review \cite{Cai:2017cbj}.

For the recent progress on the GW observations, we make a brief review on the first BH-BH event \cite{Abbott:2016blz}, the first NS-NS event \cite{TheLIGOScientific:2017qsa}, the first two BH-NS events \cite{LIGOScientific:2021qlt}, and the recent NANOGrav 12.5-yr results \cite{Arzoumanian:2020vkk}. The first BH-BH event (also known as the first direct detection of a GW event) was detected on 2015/09/14 (hereafter GW150914 \cite{Abbott:2016blz}) by LIGO from a transient GW signal matching the waveform of the binary stellar-mass black hole merger predicted by the numerical simulation of general relativity. Following up the detection of the GW150914 event, the test for general relativity \cite{LIGOScientific:2016lio} and the implications for astrophysics \cite{LIGOScientific:2016vpg} and the stochastic GW background from BBH \cite{LIGOScientific:2016fpe} are extensively discussed. The first NS-NS event (also known as the first multi-messenger GW event) was detected on 2017/08/17 (hereafter GW170817 \cite{TheLIGOScientific:2017qsa}) by LIGO-Virgo from an unprecedented joint GW and electromagnetic wave observation matching the waveform of BNS inspiral and its transient counterparts \cite{LIGOScientific:2017ync,LIGOScientific:2017zic}. Following up the detection of the GW170817 event, the Hubble constant \cite{LIGOScientific:2017adf}, the neutron star radii, and the neutron star equation of state \cite{LIGOScientific:2018cki} were first measured by GW. The test for general relativity \cite{LIGOScientific:2018dkp} and the implication for stochastic GW background \cite{LIGOScientific:2017zlf} are also discussed. The first two BH-NS events were detected on 2020/01/05 and 2020/01/15 (hereafter GW200105 and GW200115 \cite{LIGOScientific:2021qlt}) by LIGO-Virgo, providing the first convincing observational evidence for the existence of neutron star-black hole (NSBH) systems. Although the spin measurement for GW200105 is insufficient to tell the isolated binary evolution from the dynamical interaction, the spin direction of BH in GW200115 seems to be opposite to the direction of the binary orbit. The recent search for an isotropic stochastic GW background in North American Nanohertz Observatory (NANOGrav) 12.5-yr data shows strong evidence for a common-spectrum stochastic process with power-law fitting over the red-noise processes in each pulsar \cite{NANOGrav:2020bcs}. The implications for inflation, cosmic strings, phase transitions, and PBHs are discussed extensively in the literature, some of which are reviewed in the context below.

The outline of this review is as follows: In section \ref{sec:GWcosmology}, we review  the GW backgrounds from some topological defeats like vacuum bubbles and domain walls in \ref{subsec:GWFOPT}, scalar-induced GWs in \ref{subsec:IGW}, and GWs during preheating era in \ref{subsec:GWpreheating}, as well as GW multimessenger from strong lensing time delay in \ref{subsec:SLTD} and standard siren in \ref{subsec:StandardSiren}. In section \ref{sec:GWastrophysics}, we review various aspects of GW astrophysics, for example, GWs from NS-WD binaries in \ref{subsec:NSWD}, NS binaries in \ref{subsec:NSmass}, SMBBHs in \ref{subsec:SMBBHs}, and stellar compact binaries in \ref{subsec:SCB} as well as PTA/SKA astronomy in \ref{subsec:PTA1}, \ref{subsec:PTA2}, and \ref{subsec:PTA3}. In section \ref{sec:NR}, we review the numerical relativity and gravitational waveform template with related work from the finite element method in \ref{subsec:FiniteElement}, gravitational waveform for eccentric compact binaries in \ref{subsec:ECB}, and GW memory model for compact binaries in \ref{subsec:GWmemory}.

\section{Fundamental physics and GW cosmology}\label{sec:GWcosmology}

\subsection{Cosmic first-order phase transition}\label{subsec:GWFOPT}

The standard model (SM) of particle physics is known to be incomplete, either from the baryon asymmetry or the dark matter (DM), and our world is meant to be symmetry-broken, either with the electroweak symmetry or even the supersymmetry to be probed. Since the local searches for the new physics beyond the SM (BSM) have so far returned null results, either from particle acceleration colliders or DM detections, we therefore looked into the global Universe back to the early time, either from the electromagnetic waves or the GWs. The observations of electromagnetic waves (for example, the cosmic microwave background radiations and the large scale structures surveys) have established the boundaries for the new physics (for example, the upper bound on the inflationary scale, the lower bound on the reheating temperature, the upper bound on the new light degrees of freedom, the amount of DM fraction), the detailed structures of the new physics could only be depicted by the GWs that are transparent for the early Universe before the recombination. One of the GW backgrounds comes from the cosmic first-order phase transitions (see \cite{Mazumdar:2018dfl} for a comprehensive review and \cite{Hindmarsh:2020hop} for a pedagogical lecture on the cosmic first-order phase transitions), which was proposed for detection \cite{Binetruy:2012ze,AmaroSeoane:2012je} in the European LISA (eLISA) and the European New Gravitational Wave Observatory (NGO) missions \cite{AmaroSeoane:2012km} and summarized by the eLISA working group in \cite{Caprini:2015zlo} and updated recently in \cite{Caprini:2019egz} by the rejoined LISA working group \cite{Audley:2017drz}. See also \cite{Cai:2017cbj,Weir:2017wfa} for a brief review. In this short review, we will not devote ourselves to the historical developments and particle-physics model buildings but only to overview the important progress in the recent five years within the following four catalogs and highlight three of our work \cite{Cai:2020djd,Di:2020nny,Liu:2020mru} for future perspective.

\subsubsection*{Phase transition dynamics on bubble nucleations and percolations}

The cosmic first-order phase transition proceeds with the stochastic nucleations \cite{Coleman:1977py,Callan:1977pt,Linde:1980tt,Linde:1981zj} of true-vacuum bubbles in the false-vacuum background with an exponentially varying probability \cite{Cutting:2018tjt} linear \cite{Linde:1981zj,Leitao:2015fmj} or quadratic \cite{Megevand:2016lpr,Jinno:2017ixd}  in time elapse around the percolation time \cite{Leitao:2012tx,Leitao:2015fmj,Ellis:2018mja,Ellis:2020awk,Wang:2020jrd}, the later of which usually leads to a very strong first-order phase transition in a supercooled Universe that would have been slow \cite{Kobakhidze:2017mru,Cai:2017tmh} but strongly constrained by a necessary condition \cite{Guth:1982pn,Turner:1992tz,Ellis:2018mja} for the phase transition to be completed properly. The constraints on such a temporarily short duration of false vacuum domination of supercooled slow phase transition might be relaxed if the PBHs production channels are considered \cite{Baker:2021nyl,Kawana:2021tde,Liu:2021svg}. Note that the gravitational effect \cite{Coleman:1980aw} might plays a role for the phase transitions realized in the usual particle physics models \cite{Lee:2021nwg}, and other phase transition models from inflationary era \cite{Baccigalupi:1997re,Chialva:2010jt,Jiang:2015qor,An:2020fff}, matter-dominated eras \cite{Barenboim:2016mjm,Ellis:2020nnr} and non-standard cosmology \cite{Artymowski:2016tme} are also studied in addition to the usual phase transition models during radiation-dominated era. In particular, for the phase transition completed during inflation, a unique GW signal could be produced with an oscillatory feature at large wave numbers \cite{An:2020fff}. 

\subsubsection*{Microscopic/Macroscopic dynamics on bubble expansion}

The macroscopic dynamics of the bubble expansion in the thermal fluid could be captured essentially by the hydrodynamics with the bubble wall velocity as an input free parameter to solve for the fluid velocity profile, which in turn gives rise to the energy budget \cite{Espinosa:2010hh,Ellis:2019oqb,Ellis:2020nnr,Cai:2020djd}  of the total released vacuum energy into the kinetic energies of the expanding bubble wall \cite{Ellis:2019oqb,Ellis:2020nnr,Cai:2020djd} and thermal fluid motions \cite{Espinosa:2010hh,Ellis:2019oqb,Ellis:2020nnr} as well as the thermal energy dissipation. Realistic description for the bubble expansion requires going beyond the flat spacetime background \cite{Guo:2020grp} (see \cite{Cai:2018teh} for earlier trial) and the simple bag model of equation-of-state \cite{Leitao:2014pda,Giese:2020rtr,Giese:2020znk,Wang:2020nzm}.

The microscopic dynamics of the bubble expansion is governed by the Boltzmann equation \cite{Turok:1992jp,Dine:1992wr,Liu:1992tn,Moore:1995ua,Moore:1995si,Moore:2000wx,Konstandin:2014zta} with an out-of-equilibrium term characterizing the wall-plasma interactions against the driving force from the released vacuum energy, from which the bubble wall velocity could be obtained for some simple BSM models \cite{John:2000zq,Cline:2000nw,Carena:2000id,Carena:2002ss,Konstandin:2005cd,Cirigliano:2006dg,Kozaczuk:2015owa,Dorsch:2018pat,Laurent:2020gpg,Friedlander:2020tnq,Wang:2020zlf} with the help of the planar wall approximation and flow ansatz for the particle distribution function. See also \cite{Bea:2021zsu,Bigazzi:2021fmq} for recent attempts to infer the bubble wall velocity from the holographic point of view. However, a phenomenological approach \cite{Ignatius:1993qn,Moore:1995si,KurkiSuonio:1996rk,Megevand:2009ut,Megevand:2009gh,Espinosa:2010hh,Huber:2011aa,Megevand:2012rt,Megevand:2013hwa,Huber:2013kj} is usually adopted in general to conveniently parameterize the out-of-equilibrium term so as to reproduce the scaling behavior \cite{Bodeker:2009qy,Megevand:2013hwa,Bodeker:2017cim,Mancha:2020fzw,Hoeche:2020rsg,Vanvlasselaer:2020niz} of the friction term in terms of the Lorentz factor of the bubble wall velocity. Therefore, an effective picture \cite{Ellis:2019oqb,Ellis:2020nnr,Cai:2020djd} is obtained for the bubble expansion as we will highlight shortly below:

$\bullet$ The numerical simulations for the first-order phase transition aim at solving the combined equations of motion for the scalar field and thermal plasma,
\begin{align}
\square\phi-\frac{\partial V_\mathrm{eff}}{\partial\phi}&=\delta f;\\
\partial_\mu T_\mathrm{p}^{\mu\nu}+\partial^{\nu}\phi\frac{\partial V_T}{\partial\phi}&=-\partial^\nu\phi\cdot\delta f,
\end{align}
respectively, where the out-of-equilibrium term $\delta f$ is defined by
\begin{align}
\delta f=\sum\limits_{i=\mathrm{B,F}}g_i\frac{\mathrm{d}m_i^2}{\mathrm{d}\phi}\int\frac{\mathrm{d}^3\vec{k}}{(2\pi)^3}\frac{\delta f_i}{2E_i(\vec{k})}
\end{align}
in terms of the deviation of the particle distribution function from equilibrium, $f_i=f_i^\mathrm{eq}+\delta f_i$. Usually it is hard to compute $\delta f_i$ directly from a concrete particle physics model, hence a phenomenological parameterization is adopted in a model-independent manner by $\delta f=\eta_Tu^\mu\partial_\mu\phi$, where $\eta_T$ is some dimensionless function of the scalar $\phi$, temperature $T$ and fluid velocity $u^\mu$. This parameterization is specifically chosen so that the effective equation-of-motion for the position $R(t)$ of the bubble wall,
\begin{align}
\sigma\gamma^3\ddot{R}+\frac{2\sigma\gamma}{R}=\Delta p_\mathrm{dr}-\Delta p_\mathrm{fr},
\end{align}
from integrating the scalar equation-of-motion in the vicinity of bubble wall could reproduce the scaling behavior of the friction term $\Delta p_\mathrm{fr}\propto\gamma$ with respect to the Lorentz factor $\gamma=1/\sqrt{1-\dot{R}^2}   $ that roughly matches the microscopic estimations $\Delta p_\mathrm{fr}=P_{1\to1}+P_{1\to2}$ consisting of the particle transmission/reflection $P_{1\to1}\approx\Delta p_\mathrm{LO}$ at the leading-order \cite{Bodeker:2009qy} and transition splitting $P_{1\to2}\approx\gamma\Delta p_\mathrm{NLO}$ at the next-to-leading-order \cite{Bodeker:2017cim}. Recent estimation to all orders reveals a friction of form $\Delta p_\mathrm{fr}\approx\Delta p_\mathrm{LO}+\gamma^2\Delta p_{N\mathrm{LO}}$ \cite{Hoeche:2020rsg} or $\Delta p_\mathrm{fr}=(\gamma^2-1)T\Delta s$ from thermodynamic considerations \cite{Mancha:2020fzw}, which could be reproduced from a different parameterization \cite{Cai:2020djd},
\begin{align}
\delta f=-\tilde{\eta}_T(u^\mu\partial_\mu\phi)^2,
\end{align}
with function $\tilde{\eta}_T$ of mass dimension 3. Numerical simulations using this parameterization should be feasible in future. To see immediately the impact from a general form of friction term $\Delta p_\mathrm{fr}=\Delta p_\mathrm{LO}+h(\gamma)\Delta p_{N\mathrm{LO}}$ on the energy budget of the phase transition, note that the effective equation-of-motion (EOM) of the bubble wall should be modified as \cite{Cai:2020djd}
\begin{align}\label{eq:ExpansionEOM}
\left(\sigma+\frac{R}{3}\frac{\mathrm{d}\Delta p_\mathrm{fr}}{\mathrm{d}\gamma}\right)\gamma^3\ddot{R}+\frac{2\sigma\gamma}{R}=\Delta p_\mathrm{dr}-\Delta p_\mathrm{fr}
\end{align}
to respect the conservation law of the total energy $E=4\pi R^2\sigma\gamma-\frac43\pi R^3(\Delta p_\mathrm{dr}-\Delta p_\mathrm{fr})$, which could be solved directly as \cite{Cai:2020djd}
\begin{align}
\frac{h(\gamma)-h(1)}{h(\gamma_\mathrm{eq})-h(1)}+\frac{3\gamma}{2R}=1+\frac{1}{2R^3}
\end{align}
with $R$ already normalized with respect to the initial bubble size $R_0$ and $\gamma_\mathrm{eq}$ defined by $h(\gamma_\mathrm{eq})=(\Delta p_\mathrm{dr}-\Delta p_\mathrm{LO})/\Delta p_{N\mathrm{LO}}$. Then the efficiency factor for the bubble collisions could be calculated directly from \cite{Cai:2020djd}
\begin{align}
\kappa_\mathrm{col}=\left(1-\frac{\alpha_{\infty}}{\alpha}\right)\int_1^{R_\mathrm{col}}\left(\frac{\mathrm{d}R}{R_\mathrm{col}}\right)\left[1-\frac{h(\gamma(R))}{h(\gamma_\mathrm{eq})}\right]
\end{align}
where $R_\mathrm{col}$ is the bubble radius at collisions, $\alpha=\Delta p_\mathrm{dr}/\rho_\mathrm{rad}=(\Delta V_\mathrm{eff}(\phi_+)-\Delta V_\mathrm{eff}(\phi_-))/\rho_\mathrm{rad}$ is the strength factor of the total released vacuum energy with respect to the background radiation, $\gamma_\mathrm{eq}$ is the asymptotic Lorentz factor that balances the driving pressure with the friction force $\Delta p_\mathrm{dr}=\Delta p_\mathrm{fr}$,  and $\alpha_\infty=\Delta p_\mathrm{LO}/\rho_\mathrm{rad}$. The new term in \eqref{eq:ExpansionEOM} defines a transition radius, 
\begin{align}
R_\sigma=\frac32\frac{h(\gamma_\mathrm{eq})-h(1)}{h'(\gamma(R_\sigma))}R_0,
\end{align}
when the bubble wall slows down its acceleration and starts to approach a terminal velocity. Although our modified EOM \eqref{eq:ExpansionEOM} reproduces previous estimations \cite{Ellis:2019oqb,Ellis:2020nnr} in the limit of a large collision radius $R_\mathrm{col}\gg R_\sigma$, the difference in the efficiency factor $\kappa_\mathrm{col}$ could be announced if the bubbles collide at a radius $R_\mathrm{col}<100 R_\sigma$, which is important for the strong first-order phase transition where the bubbles collide with each other when they are still rapidly accelerating ever before having approached the terminal velocity.  Numerical simulations should be carried out for checking this effective picture.

\subsubsection*{Numerical simulations and analytic auxiliary modeling on bubble collisions}

Numerical simulations on bubble collisions in vacuum background \cite{Kosowsky:1991ua,Kosowsky:1992rz,Kosowsky:1992vn}  and thermal fluid \cite{Kamionkowski:1993fg,Huber:2008hg} have taken advantage of the envelope approximation, which was abandoned later in \cite{Hindmarsh:2013xza,Hindmarsh:2015qta,Hindmarsh:2017gnf,Cutting:2019zws} with the discovery of sound waves \cite{Hogan:1986qda} as the dominated contribution \cite{Weir:2016tov,Cutting:2018tjt} over the bubble collisions \cite{Witten:1984rs} to the stochastic background of GWs from the cosmic first-order phase transition if the bubble wall velocity could terminate at some constant velocity as opposed to the runaway expansion. See also \cite{Cutting:2020nla} for checking the thin-wall approximation and \cite{Lewicki:2019gmv,Lewicki:2020jiv} for a new oscillating feature and inclusion of a complex scalar field. The last gravitational-wave source comes from the magneto-hydrodynamic (MHD) turbulence \cite{Witten:1984rs,Kamionkowski:1993fg}, whose numerical simulation only made possible until very recent time \cite{Niksa:2018ofa,Pol:2019yex,Brandenburg:2021tmp,Brandenburg:2021bvg}. See also \cite{Kahniashvili:2005qi,Kahniashvili:2008er,Kahniashvili:2008pf,Kahniashvili:2008pe,Caprini:2009pr,Kahniashvili:2009qi,Kisslinger:2015hua,Brandenburg:2017neh,Kahniashvili:2020jgm,Ellis:2020uid} for the helical MHD turbulence.

On the other hand, the extraction of the energy-density spectrum from the numerical simulation results requires some analytic auxiliary models to generate some well-motivated parameterization formulas. Earlier analytic calculations \cite{Caprini:2007xq,Caprini:2009fx} on bubble wall collisions have been renewed in \cite{Jinno:2016vai} with the aid of thin-wall and envelope approximations, the latter of which was further abandoned in \cite{Jinno:2017fby}. The analytic treatments from the sound shell model \cite{Hindmarsh:2016lnk,Hindmarsh:2019phv}\cite{Guo:2020grp}, bulk flow model \cite{Jinno:2017fby,Konstandin:2017sat} and hybrid model \cite{Jinno:2019jhi,Jinno:2019bxw,Jinno:2020eqg} have depicted the lifetime evolution of the sound waves before the formation of the MHD turbulence \cite{Kosowsky:2001xp,Dolgov:2002ra,Nicolis:2003tg,Caprini:2006jb,Gogoberidze:2007an,Caprini:2009yp}. See also \cite{Di:2020nny} for a numerical simulation with inclusion of a gauge field for the generation of primordial magnetic field as we will highlight shortly below: 

$\bullet$ In literatures, it was proposed that magnetic field (MF) that may seed the observed cosmic scale MF~\cite{Vachaspati:2001nb,Durrer:2013pga,Grasso:2000wj,Subramanian:2015lua}. To verify if the MF and GW can be generated together during the first-order phase transition, Ref. \cite{Di:2020nny} performed the lattice simulation of the magnetic field and GW production from bubble collisions and oscillations stages during the phase transition by considering the evolution of gauge fields and Higgs field on three-dimensional lattice. It was shown that the Higgs gradient effect not only dominates the GW production, its effects on the magnetic field production are also significant as seen from Fig.2 in Ref. \cite{Di:2020nny}. The study further show that the observations of cosmic MF and GW are complementary to probe new physics admitting first-order phase transition, after taking into account the affects of MHD on the generated MF. For the GW calculation, Ref. \cite{Di:2020nny} adopted the straightforward procedure from Ref. \cite{GarciaBellido:2007af} rather than the envelope approximation. The EOM of tensor perturbations $h_{ij}$ reads
\begin{equation}
\ddot{h}_{ij} - \nabla^2 h_{ij} = 16 \pi G T^{\mathrm{TT}}_{ij}\;,
\end{equation}
where the superscript $\mathrm{TT}$ denotes the transverse trace-less projection, and the energy-momentum tensor (right hand side) is given by
\begin{equation}
	\begin{split}
T_{\mu\nu}=\partial_\mu \Phi^\dagger \partial_\nu \Phi&-g_{\mu\nu}\frac{1}{2}\rm{Re}[(\partial_i  \Phi^\dagger \partial^i \Phi )]\;.\label{ttt}
\end{split}
\end{equation}
Here $\Phi$ is the Higgs field, and the subdominant gauge field contributions are neglected. Nevertheless, the contribution from the magnetic fields still affects the evolution of the Higgs field through the combined EOMs
\begin{align}
\partial_0^2\Phi&=D_iD_i\Phi-\frac{dV(\Phi)}{d\Phi},\\
\partial_0^2B_i&=-\partial_jB_{ij}+g'\,\mathrm{Im}[\Phi^\dagger D_i\Phi],\\
\partial_0^2W_i^a&=-\partial_kW_{ik}^a-g\,\epsilon^{abc}W_k^bW_{ik}^c+g\,\mathrm{Im}[\Phi^\dagger \sigma^a D_i\Phi],
\end{align}
with the solutions subjected to Gauss constraints,
\begin{align}
\partial_0\partial_jB_j-g'\,\mathrm{Im}[\Phi^\dagger\partial_0\Phi]&=0,\\
\partial_0\partial_j W_j^a+g\,\epsilon^{abc}W_j^b\partial_0W_j^c-g\,\mathrm{Im}[\Phi^\dagger\sigma^a\partial_0\Phi]&=0,
\end{align}
where the temporal gauge $W_0^a=B_0=0$ is used.
The energy spectrum of GW from the bubble collision and oscillation stage can be extracted from
\begin{align}
\Omega_{\mathrm{GW}}&=\dfrac{1}{\rho_{c}}\frac{d\rho_\text{GW}(k)}{d\ln k}\nonumber\\
&=\frac{k^3}{32\pi G L^3\rho_c}\int d\Omega\Lambda_{ij,lm}(\hat{k})\dot{u}_{ij}(t,k)\dot{u}_{lm}(t,k)\;,
\end{align}
where $h_{ij}(t,k)=\Lambda_{ij,lm}(\hat{k})u_{ij}(t,\mathbf{k})$, with $\Lambda_{ij,lm}=P_{il}(\hat{k})P_{jm}(\hat{k})-\frac{1}{2}P_{ij}(\hat{k})P_{lm}(\hat{k})$ and the spatial projection operator $P_{ij}=\delta_{ij}-\hat{k}_i\hat{k}_j$($\hat{k}_i=k_i/k$). Here $G$ is the Newtonian gravitational constant, $\rho_c$ is the critical density, and $L$ is lattice spacing in the simulation. The GW production are shown to be affected by bubble wall thickness and bubble wall velocity. 

\subsubsection*{Particle physics phenomenology on the phase transition models}
There are an immense amount of particle physics models with cosmic first-order phase transitions, most of which have been summarized in \cite{Caprini:2015zlo,Caprini:2019egz} for their phenomenological detection. We will not review here a completed list of these BSM models with higher operators extensions, scalar extensions, supersymmetric extensions, warp extra dimensions, composite Higgs, and dark/hidden sector, but only to mention some phenomenological considerations on model selection.

Since the phenomenological predictions could be largely affected by some theoretical uncertainties from the renormalization group running \cite{Cai:2017tmh}, renormalization scale dependence \cite{Croon:2020cgk,Gould:2021oba}, the gauge dependence and infrared divergences \cite{Ekstedt:2020abj} as well as careful treatments on phase transition dynamics and macroscopic thermal parameters \cite{Guo:2021qcq}, a non-perturbative approach to phase transition description using dimensional reduction has been developed for some BSM models \cite{Losada:1996ju,Cline:1996mga,Laine:1996ms,Bodeker:1996pc,Laine:1998qk,Laine:2012jy,Carena:2012np,Brauner:2016fla,Andersen:2017ika,Niemi:2018asa,Gorda:2018hvi,Kainulainen:2019kyp,Gould:2019qek,Schicho:2021gca}.  On the other hand, the observational perspectives on gravitational-waves signals extraction have been tested for some particle physics models \cite{Alanne:2019bsm,Schmitz:2020syl,Schmitz:2020rag} with LISA sensitivity curves. Recently, the model selections and comparisons have been confronted with the real data from the  NANOGrav 12.5-yr data \cite{Nakai:2020oit,Addazi:2020zcj,Ratzinger:2020koh,Bian:2021lmz,Neronov:2020qrl,Barman:2020jrf,Chiang:2020aui} and LIGO \cite{Romero:2021kby,Huang:2021rrk}. To better extract BSM physics from the phase transition models, more theoretical understandings are needed for the possible features of anisotropies \cite{Geller:2018mwu} and the non-Gaussianity \cite{Kumar:2021ffi} beyond the simple power-law shapes of the energy density spectrum. See also \cite{Liu:2020mru} for GW anisotropies from domain walls as we will highlight shortly below:

$\bullet$  Along with the spontaneous symmetry breaking during phase transitions, topological defects will finally form, as an important consequence of phase transitions \cite{Kibble:1976sj}. For example, cosmic strings/domain walls are one/two-dimensional topological defects that can form in case of the continuous/discrete symmetry is spontaneously broken \cite{Vilenkin:1984ib}. Since the energy density is highly concentrated, the motion of topological defects driven by their tension results in continuous GW productions. Different from the transitory GW production from bubbles and sound waves, GWs are continuously produced as long as the topological defects do not annihilate. The characteristic GW energy spectrums of cosmic strings and domain walls are reviewed in Refs. \cite{Auclair:2019wcv,Saikawa:2017hiv}, and through GWs one can detect or give more strict constraints on their tension \cite{Sanidas:2012ee,Abbott:2017mem}. Moreover, GWs from topological defects can successfully explain the common-spectrum process observed by NANOGrav \cite{Ellis:2020ena,Blasi:2020mfx,Bian:2021lmz} (Note that an inflationary interpretation of the NANOGrav signal is not excluded for a sufficiently low reheating scale \cite{Vagnozzi:2020gtf}.).

\begin{figure}
\centering
\includegraphics[width=0.49\textwidth]{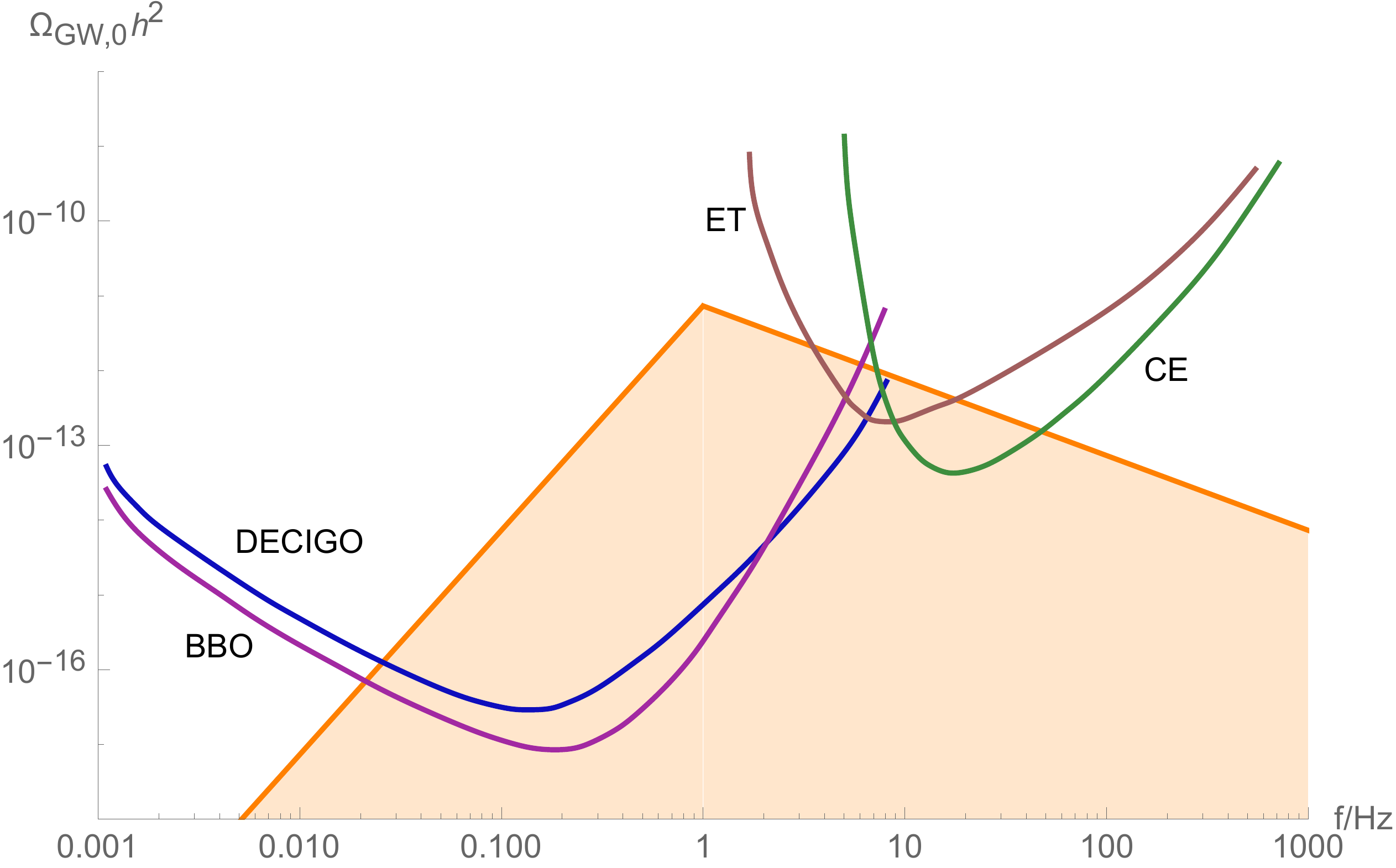}
\includegraphics[width=0.49\textwidth]{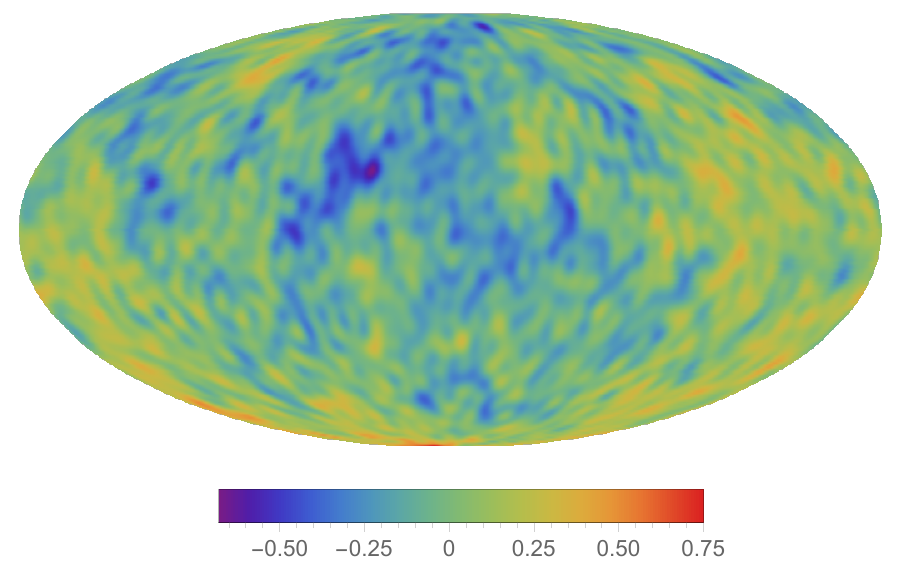}\\
\caption{In the left panel, the orange line presents an example of the present energy spectrum of GWs from domain walls, $\Omega_{\mathrm{GW},0}h^{2}$, using the approximation method in Ref.~\cite{Saikawa:2017hiv}, where the tension and annihilation temperature of domain walls are $(5\times 10^{10}\mathrm{GeV})^{3}$ and $10^{7}$GeV. This SGWB can be observed by DECIGO~\cite{Kawamura:2011zz}, BBO~\cite{Yagi:2011wg}, ET~\cite{Punturo:2010zz} and CE~\cite{Reitze:2019iox}. The right panel shows a random realization of the SGWB using the first 50 $l$-modes, where the angular power spectrum is $l(l+1)C_{l}=0.085$, predicted by models with $\phi_{i}\sim H_{\mathrm{inf}}$. Copied from Ref. \cite{Liu:2020mru} with permission. }
\label{fig:aniso}
\end{figure}

Instead of the energy spectrum of GWs from topological defects, we also find the anisotropies of those stochastic GW backgrounds carry unique information about inflation. The anisotropies in stochastic GW backgrounds could be generated by the sources \cite{Bethke:2013aba,Bethke:2013vca,Geller:2018mwu,Jenkins:2018lvb} and the processes during propagation \cite{Contaldi:2016koz,Bartolo:2019oiq,Bartolo:2019yeu,DallArmi:2020dar}, which in most cases are still challenging to observe \cite{Allen:1996gp,Seto:2004np,Hotinli:2019tpc,Chu:2020qiw,Mentasti:2020yyd,Alonso:2020rar}. 
We explore the SGWB produced from unstable cosmic domain walls, which annihilates before dominating the Universe \cite{Vilenkin:1981zs,Gelmini:1988sf}.
The discrete symmetry is spontaneously broken before inflation, and the formation of DWs is realized because $\phi$ could cross the potential barrier due of quantum fluctuations. The corresponding probability that $\phi$ crosses the barrier reads \cite{Espinosa:2015qea}
\begin{equation}
	\label{eq:prob}
	P\left( t\right) 
	=\dfrac{1}{2}\operatorname{erfc}\left(\frac{\sqrt{2} \pi \phi_{i}}{H_{\mathrm{inf}} \sqrt{N(t)}}\right),
\end{equation}
where $\phi_{i}$ is the initial value of the inflaton, and we assume $\phi_{i}>0$ without loosing generality. $H_{\mathrm{inf}}$ is the Hubble parameter during inflation and $N(t)\equiv H_{\mathrm{inf}} (t-t_{i})$ where $t_{i}$ is the horizon-crossing time of the CMB scale. After inflation the Hubble parameter $H$ decreases with time, and once $H$ becomes smaller than the effective mass of $\phi$, then $\phi$ settles down in different vacua and DWs form. Large scale perturbations of $\phi$ remain constant at superhorizon scales, and then result in perturbations of the energy density of DWs, so that anisotropies in SGWBs arise. The angular spectrum of GWs is defined by
\begin{equation}
	\label{eq:plat}
	l(l+1)C_{l}=\frac{\pi}{2}\langle\delta\Omega_{\mathrm{GW}}^{2}\rangle,
\end{equation}
where $\delta\Omega_{\mathrm{GW}}(\mathbf{x})\equiv(\Omega_{\mathrm{GW}}(\mathbf{x})-\overline{\Omega_{\mathrm{GW}}})/\overline{\Omega_{\mathrm{GW}}}$ denotes fluctuations of the GW energy spectrum. The GW energy spectrum, is proportional to quantum fluctuations of $\phi$,
\begin{equation}
	\label{eq:deltaGW}
	\delta\Omega_{\mathrm{GW}}(\mathbf{x})
	=c_{1}\delta\phi(\mathbf{x}),
\end{equation}
where $c_{1}$ can be obtained from the derivative of Eq.~\ref{eq:prob} with respect to $\phi$. During inflation quantum fluctuations of $\phi$ leads to $\langle\delta\phi^{2}\rangle=H_{\mathrm{inf}}^{2}/(4\pi^{2})$, then the angular power spectrum could be obtained as \cite{Liu:2020mru}
\begin{equation}
	\label{eq:cl}
	l(l+1) C_{l}\approx\left\{
	\begin{aligned}
		&\frac{\pi}{N_{\mathrm{peak}}}\alpha_{\mathrm{peak}}^{2},&\alpha_{\mathrm{peak}}\gg 1,\\
		&\dfrac{1}{N_{\mathrm{peak}}},&\alpha_{\mathrm{peak}}\ll 1, \\
	\end{aligned}
	\right.
\end{equation}
where $\alpha_{\mathrm{peak}}\equiv\frac{\sqrt{2} \pi \phi_{i}}{H_{\mathrm{inf}}\sqrt{N_{\mathrm{peak}}}}$,  $N_{\mathrm{peak}}=\ln(f_{\mathrm{peak}}/H_{0})$ and $f_{\mathrm{peak}}$ is the peak frequency of GWs from domain walls. Since the $e$-folding number of inflation is expected to be $50$-$60$, the angular power spectrum is at least $10^{-2}$ at the CMB scales, see Fig.~\ref{fig:aniso} as an example of random realization of a SGWB with $l(l+1)C_{l}=0.085$, which could be used as a  probe of the inflationary energy scale. Since primordial GWs are too weak to be detected for low-scale inflation, observing such anisotropies in stochastic GW backgrounds provides a novel method to detect the inflationary energy scale even it is orders of magnitude lower than the grand unified theory scale~\cite{Liu:2020mru}.

\subsection{Induced gravitational waves}\label{subsec:IGW}

The perturbations of the metric in general relativity (GR) can be decomposed into scalar and tensor types. In an appropriate gauge, the spatial part of the flat FLRW metric can be written as $dl^2=a^2((1+2\mathcal{R})\delta_{ij}+h_{ij})$, where $\mathcal{R}$ is the scalar-type curvature perturbation and $h_{ij}$ is the tensor perturbation~\footnote{We will not enter the topic of gauge choice in this subsection. We recommend the interested readers to read Refs. \cite{Hwang:2017oxa,Wang:2019zhj,Gong:2019mui,Tomikawa:2019tvi,DeLuca:2019ufz,Inomata:2019yww,Yuan:2019fwv,Nakamura:2019zbe,Lu:2020diy,Ali:2020sfw,Giovannini:2020qta,Giovannini:2020soq,Chang:2020tji,Chang:2020iji,Chang:2020mky,Domenech:2020xin} for details.}. At the linear order, there are no interactions between these different types of perturbations, which evolve independently. By the recent observation of CMB \cite{Aghanim:2018eyx}, the curvature perturbation is measured to be $\mathcal{R}\sim10^{-5}$ to very high accuracy, while there is still no evidence for the detection of the tensor-type primordial perturbation, which puts an upper bound of $\Omega_\text{GW}\lesssim10^{-16}$ to the energy density spectrum of the primordial GWs for frequencies higher than $10^{-15}~\text{Hz}$. Another form of GW is the secondary wave induced by the scalar-scalar-tensor type interaction present in the non-linear order of the perturbed action, which are called induced GWs or secondary GWs \cite{Matarrese:1992rp,Matarrese:1993zf,Matarrese:1997ay,Noh:2004bc,Carbone:2004iv,Nakamura:2004rm,Ananda:2006af,Osano:2006ew}. If the curvature perturbations are as small as their values at the CMB scales, the induced GWs are much smaller than the first-order primordial GWs ($\Omega_\text{GW}^\text{(ind)}\sim10^{-24}$) and are completely negligible. However, in some inflation models, the curvature perturbation gets enhanced at scales much smaller than 1 Mpc, which can induce detectable GWs as well as substantial PBHs.
For instance, in Ref. \cite{Pi:2017gih} we considered an non-minimally coupled scalar field with a flat concave potential in Starobinsky $R^2$-gravity. The scalaron potential dominates in the early stage of inflation which makes the predictions of tensor-to-scalar ratio $r$ and spectral tilt $n_s$ on large scales favorable by Planck constraint, while on small scales the field roll down to the valley of the scalaron and starts to slowly roll in the other direction, which induces a huge enhancement of the power spectrum of the curvature perturbation that can generate abundant PBHs.  
Another example is to realize the parametric resonance in a single-field inflationary model with a small periodic structure upon the potential, of which the equation of the curvature perturbation has the form of the Mathieu equation, and the power spectrum of the curvature perturbation is enhanced thereby in the instability tongues of the parameter space \cite{Cai:2019bmk}.

The PBHs are generated by the gravitational collapse of the high-$\sigma$ peaks of the curvature perturbation at its horizon reentry before recombination \cite{Zeldovich:1967lct,Hawking:1971ei,Carr:1974nx,Meszaros:1974tb,Carr:1975qj,Khlopov:1985jw}. For recent discussions on the PBH formation, see for instance \cite{Young:2014ana,Germani:2018jgr,Wu:2020ilx,DeLuca:2019qsy,Ezquiaga:2019ftu,He:2019cdb,Musco:2020jjb,Taoso:2021uvl,Riccardi:2021rlf}. The masses of PBHs are roughly the horizon mass when the peak wavelength of the curvature perturbation reenters the Hubble horizon, which depends on the concrete inflation models that generates them \cite{Yokoyama:1998pt,Garcia-Bellido:2016dkw,Cheng:2016qzb,Garcia-Bellido:2017mdw,Cheng:2018yyr,Dalianis:2018frf,Tada:2019amh,Xu:2019bdp,Mishra:2019pzq,Bhaumik:2019tvl,Liu:2020oqe,Atal:2019erb,Fu:2020lob,Vennin:2020kng,Ragavendra:2020sop,Gao:2021dfi,GarciaBellido:1996qt,Kawasaki:1997ju,Frampton:2010sw,Giovannini:2010tk,Clesse:2015wea,Inomata:2017okj,Gong:2017qlj,Inomata:2017vxo,Espinosa:2017sgp,Kawasaki:2019hvt,Palma:2020ejf,Fumagalli:2020adf,Braglia:2020eai,Anguelova:2020nzl,Romano:2020gtn,Gundhi:2020zvb,Gundhi:2020kzm,Kannike:2017bxn,Pi:2017gih,Gao:2018pvq,Cheong:2019vzl,Cheong:2020rao,Fu:2019ttf,Dalianis:2019vit,Lin:2020goi,Fu:2019vqc,Aldabergenov:2020bpt,Aldabergenov:2020yok,Yi:2020cut,Gao:2020tsa,Dalianis:2020cla,Kawasaki:2012wr,Kohri:2012yw,Ando:2017veq,Ando:2018nge,Chen:2019zza,Cai:2018tuh,Chen:2020uhe,Cai:2019jah,Cai:2019bmk,Kusenko:2020pcg,Cotner:2016cvr,Cotner:2017tir,Cotner:2018vug,Cotner:2019ykd,Gao:2021vxb,Solbi:2021wbo,Kawasaki:2021ycf,Ng:2021hll}. The peak wavelength in turn, determines the frequency of the induced GWs, which is connected to the PBH mass by $M_\text{PBH}/M_\odot=(f/10^{-8}\text{Hz})^{-2}$. The cross check of the energy density spectrum of the induced GWs and the PBH abundance under such a mass-frequency relation is very important, especially considering the possibility that the PBHs can circumvent the measurement of baryonic matter on the CMB thus might contribute a substantial amount or all of the fraction of cold DM (CDM).

Due to Hawking radiation, light PBHs with $M_\text{PBH}\lesssim10^{16}~\text{g}$ have already evaporated completely during the current cosmic age, which leaves strong constraints on big bang nucleosynthesis and intergalactic $\gamma$-rays.
Besides, the current observational constraints do not exclude the existence of a substantial amount of PBHs in several interesting ``mass  windows'' \cite{Green:2004wb,Frampton:2009nx,Carr:2009jm,Carr:2016hva,Carr:2016drx,Poulter:2019ooo,Wang:2019kaf,Tisserand:2006zx,Graham:2015apa,Koushiappas:2017chw,Authors:2019qbw,DeLuca:2020qqa,Serpico:2020ehh,Mena:2019nhm,Murgia:2019duy,Cai:2020fnq}, which can yield fruitful phenomena. For instance, the LIGO O3a data set of LIGO/Virgo implies that there might be two populations of black holes \cite{Abbott:2020gyp}, which can be explained by the combination of the astrophysical black holes and PBHs of $\sim20$ solar mass \cite{Wong:2020yig,Hutsi:2020sol,DeLuca:2021wjr}.
PBHs might be the supermassive or stupendously large BHs which seed the galaxy or even structure formation \cite{Bean:2002kx,Kawasaki:2012kn,Nakama:2017xvq,Carr:2018rid,Nakama:2019htb,Carr:2020erq,Atal:2020yic}.  The planetary-mass PBHs could be the lensing objects of the microlensing events observed by the Optical Gravitational Lensing Experiment (OGLE) \cite{2017Natur.548..183M,Niikura:2019kqi,Domenech:2020ers,Bhattacharya:2020lhc}, or even the Planet 9 \cite{Scholtz:2019csj}. The formation of solar-mass PBHs is greatly enhanced due to the softening of the equation-of-state parameter during the QCD phase transition \cite{Byrnes:2018clq,Carr:2019kxo}, and can provide the hotspots for baryogenesis \cite{Carr:2019hud,Garcia-Bellido:2019vlf}. The PBH abundance can not exceed that of CDM, and is further constrained by some additional observations according to the PBH masses \cite{Carr:2016drx,Sasaki:2018dmp,Carr:2020gox,Green:2020jor,Carr:2020xqk}. 
The PBH mass function, $f_\text{PBH}$, is defined as the PBH energy density normalized by the DM density. According to the observational constraints, especially the observations on the microlensing events in the halo of our galaxy \cite{Niikura:2017zjd}, the only window that affords $f_\text{PBH}\approx1$, i.e. PBHs can serve as all the DM, is $10^{16}$ g $<M_\text{PBH}<10^{22}$ g, the so-called asteroid-mass PBHs. Because of the finite-size effect and wave effect, it is impossible to observe the lensing events by visible lights when the asteroid-mass PBHs are the lensing objects \cite{Niikura:2017zjd,Katz:2018zrn,Montero-Camacho:2019jte,Sugiyama:2019dgt,Laha:2019ssq,DeRocco:2019fjq,Dasgupta:2019cae,Ray:2021mxu}. Therefore the indirect constraints from the induced GWs associated with the asteroid-mass PBHs, which accidentally lie in the millihertz band of the space-based interferometers, becomes almost the only tool to detect the PBH-as-DM scenario.


As the PBH formation depends on the high-$\sigma$ peaks of the probability distribution function (PDF) of the curvature perturbation, it crucially depends on the shape of the PDF. A typical deviation of the usually presumed Gaussian PDF is the quadratic local non-Gaussianity, which can be written as $\mathcal{R}=\mathcal{R}_g+F_\text{NL}\left(\mathcal{R}_g^2-\left\langle\mathcal{R}_g^2\right\rangle\right)$ \cite{Luo:1992er,Verde:1999ij,Verde:2000vr,Komatsu:2001rj,Bartolo:2004if,Boubekeur:2005fj,Byrnes:2007tm}. It is shown that for positive local non-Gaussianity ($F_\text{NL}>0$), the PBH abundance will be greatly enhanced, while for negative value it will be suppressed \cite{Young:2013oia}. In Ref. \cite{Cai:2018dig}, we discussed the PBH abundance with quadratic local non-Gaussianity, and calculated the GWs induced by such non-Gaussian scalar perturbations. We found that after taking into account the local non-Gaussianity with positive nonlinear parameter $F_\text{NL}$, the induced GWs are also enhanced, but not as much as that of the PBHs.

\begin{figure}
\centering
\includegraphics[width=0.6\textwidth]{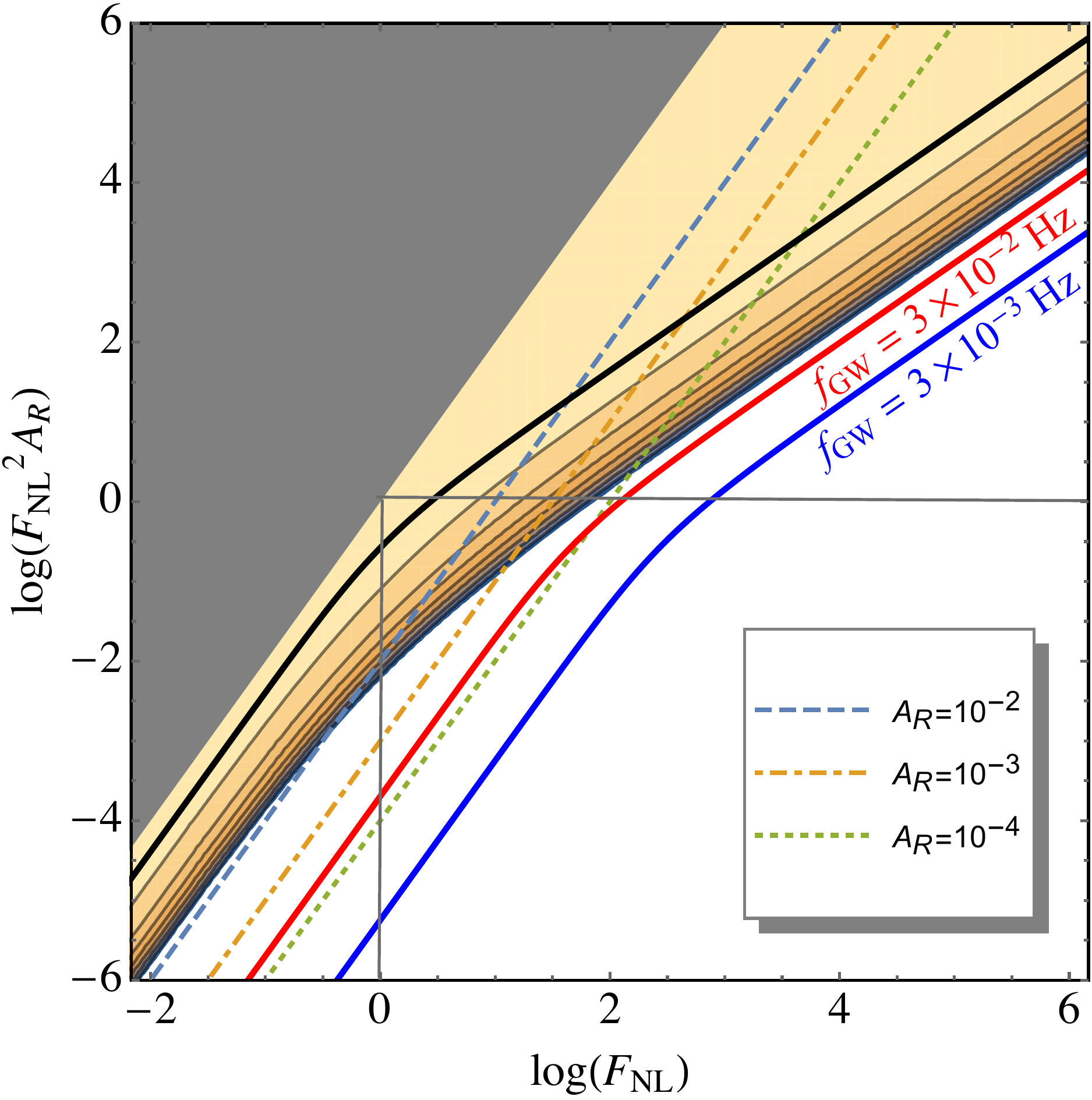}
\caption{The PBH abundance as a function of $F_\text{NL}$ and $F_\text{NL}^2\mathcal{A}_\mathcal{R}$, where $\mathcal{A}_\mathcal{R}$ is the amplitude of the power spectrum of the curvature perturbation in the Newtonian gauge. The border between the colored and white regions corresponds to $f_\text{PBH}=1$, i.e. PBHs are all the DM. The dashed lines are for $\mathcal{A}_\mathcal{R}=10^{-2}$, $10^{-3}$, and $10^{-4}$ from left to right, while the shaded area is unphysical since $\mathcal{A}_\mathcal{R}>1$. The thick black curve is the absolute constraint that the GW energy density be smaller than the current density of radiation, while the red and blue curves are the sensitivity bound of LISA at $f_\text{GW}=3\times10^{-2}~\text{Hz}$ and $3\times10^{-3}~\text{Hz}$, respectively; they correspond to PBH masses $M_\text{PBH}=10^{20}~\text g$ and $10^{22}~\text g$. Copied from Ref. \cite{Cai:2018dig} with permission.}
\label{f:IGW1}
\end{figure}

In the asteroid-mass window, if PBHs can be all the DM ($f_\text{PBH}=1$), the required amplitude of the power spectrum of the curvature perturbation should be $\mathcal{O}(10^{-2})$. If there is positive non-Gaussianity, the PBH formation will be greatly enhanced and in turn, the curvature perturbation we require will be smaller than the Gaussian case as the PBH abundance is fixed. When considering the energy density spectrum $\Omega_\text{GW}$ of the induced GWs at millihertz, we found that when $F_\text{NL}>0$ with a fixed $f_\text{PBH}$, the suppression of $\Omega_\text{GW}$ from the suppression of the curvature perturbation can not be fully compensated by the enhancement from the non-Gaussian part, which means that increasing the non-linear parameter $F_\text{NL}$ will suppress the induced GW, if the PBH abundance is fixed. See Figure \ref{f:IGW1} for details.

\begin{figure}
\centering
\includegraphics[width=0.6\textwidth]{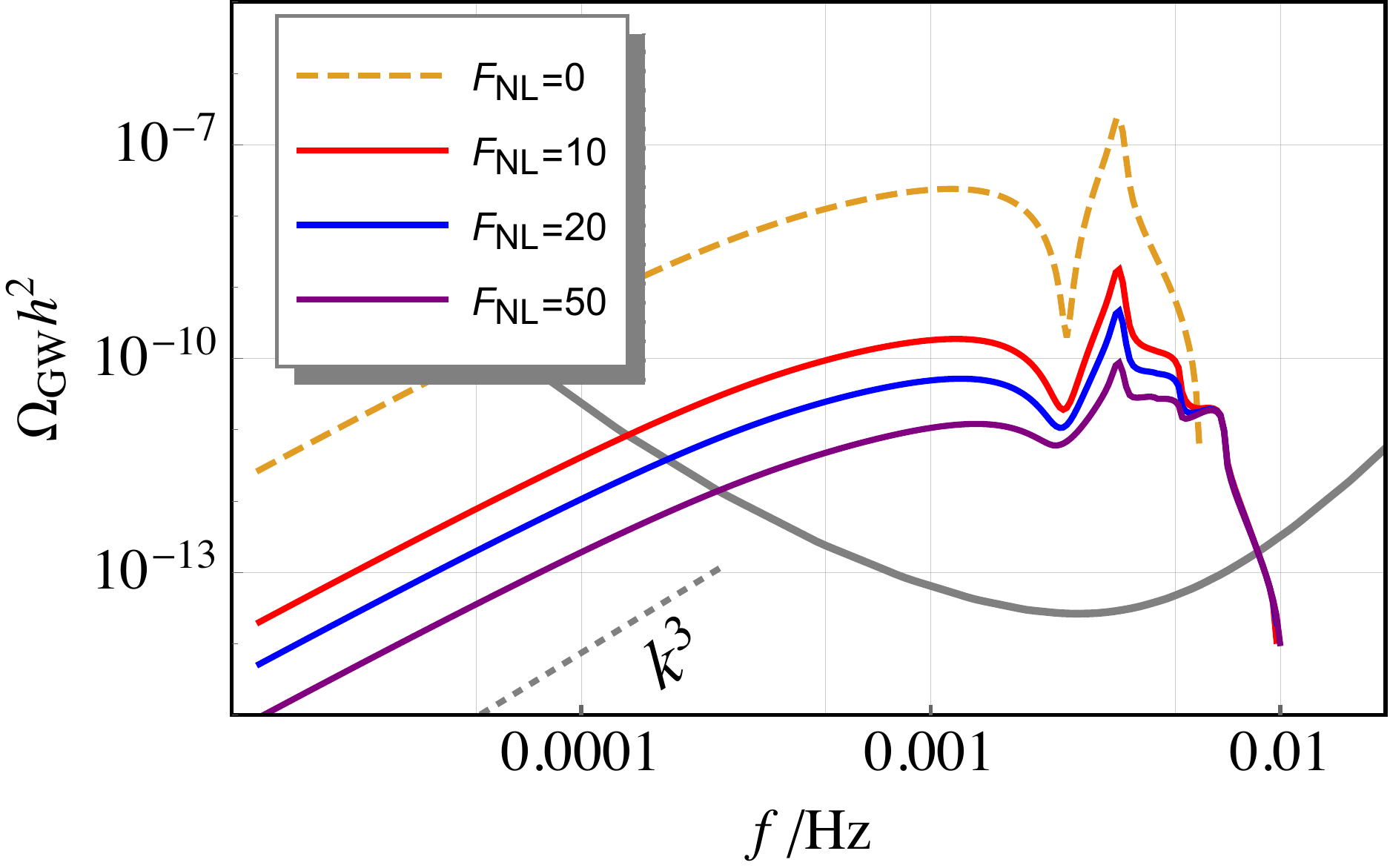}\\
\caption{Typical energy density spectrum of the GWs induced by a non-Gaussian curvature perturbation at second order with $F_\text{NL}>0$. The width of peak is fixed at $10^{-4}~\text{Hz}$. The abundance of the PBHs is fixed to be $f_\text{PBH}=1$ for $M_\text{PBH}=10^{22}~\text g$. We draw the induced GW energy density spectrum $\Omega_\text{GW}h^2$ for $F_\text{NL}=0$ (orange dashed), $10$ (red), $20$ (blue), and $50$ (purple). The gray curve is the sensitivity bound of LISA from Ref. \cite{Thrane:2013oya}. A reference line of the $k^3$ slope is also drawn for comparison. Copied from Ref. \cite{Cai:2018dig} with permission. }
\label{f:IGW2}
\end{figure}

An important implication of our result shown above is that the energy density spectrum of the induced GWs is bounded from below when $F_\text{NL}$ is very large, and this lower bound is still higher than the sensitivity curve of the space interferometers like LISA \cite{Bender1998LISALI,AmaroSeoane:2012je,AmaroSeoane:2012km,Armano:2016bkm,Audley:2017drz},  Taiji \cite{Hu:2017mde,Guo:2018npi,Taiji-1} and TianQin \cite{Luo:2015ght,Luo:2020bls,Mei:2020lrl}. Therefore, our result actually reaches an important conclusion, that if DM consists mainly of PBHs, which is only possible for the asteroid-mass window, the corresponding millihertz induced GWs must be detectable by the space interferometers, regardless of the quadratic local non-Gaussianities \cite{Cai:2018dig}. On the contrary, the non-detection of such induced GWs in the space interferometers will close the only window of PBH-as-DM scenario \cite{Kalaja:2019uju,Sato-Polito:2019hws,Gow:2020bzo,Unal:2020mts}. Based on our conclusion, the indirect detection of the asteroid-mass PBH abundance by the millihertz induced GW becomes a very important scientific goal for the space-based interferometers \cite{Bartolo:2018rku,Caldwell:2019vru,Barausse:2020rsu,Taijiwhitepaper} as shown in Figure \ref{f:IGW2}.

To identify the physical origin of the stochastic GWs, the study of their spectral shapes are very important. In Ref. \cite{Cai:2019cdl} we studied the infrared behavior of the stochastic GWs, and found that on scales much larger than any scales of the GW source, the stochastic GWs behave like a white noise with an infrared scaling of $\Omega_\text{GW}\sim f^3$, if the tensor mode reenter the Hubble horizon in the radiation dominated era. This implies the infrared scaling of any stochastic GW spectrum can be used to probe the thermal history of the universe, as a deviation from $\Omega_\text{GW}\sim f^3$ implies a deviation from $w=1/3$ \cite{Domenech:2019quo,Domenech:2020kqm}. On the other hand, the induced GW has some characteristic spectral shape near its peak, which depends crucially on the width of the power spectrum of the curvature perturbation that induces it \cite{Pi:2020otn}. The infrared scaling reduces to $\Omega_\text{GW}\sim f^2$ if the curvature perturbation power spectrum is a $\delta$-function peak \cite{Kohri:2018awv}. For a narrow peak, there is a breaking frequency $f_b$ 
below which the power goes from 2 to 3. $f_b$ moves towards the peak frequency $f_p$ when the width increases, and disappears when the width is of order 1. For a broad peak in the power spectrum of the curvature perturbation with a lognormal shape, we derived an analytical formula for $\Omega_\text{GW}$ which fits the numerical integral well. This result can be used to speed up the signal searching of the stochastic GWs in the future.

\begin{figure}
\centering
\includegraphics[width=0.32\textwidth]{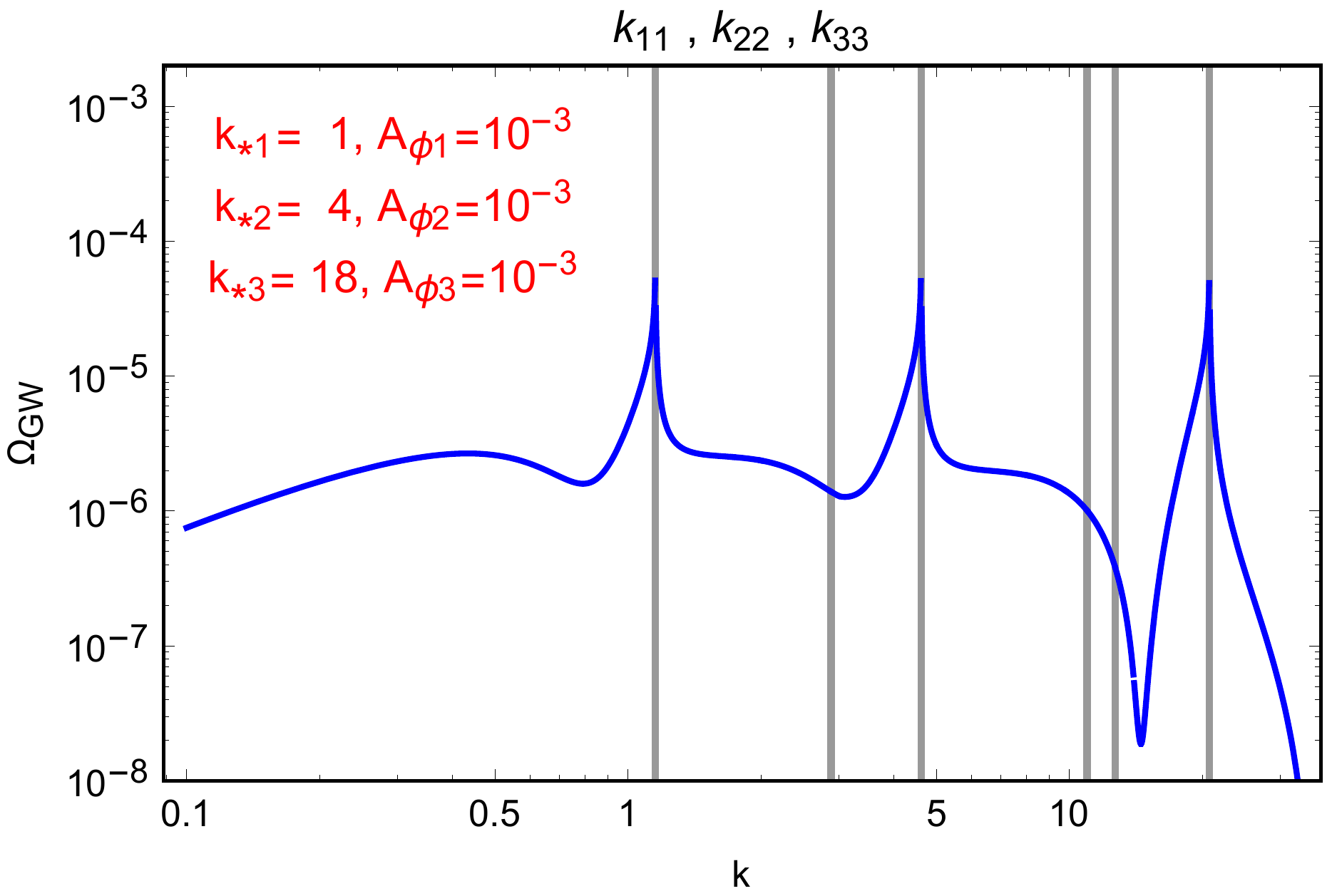}
\includegraphics[width=0.32\textwidth]{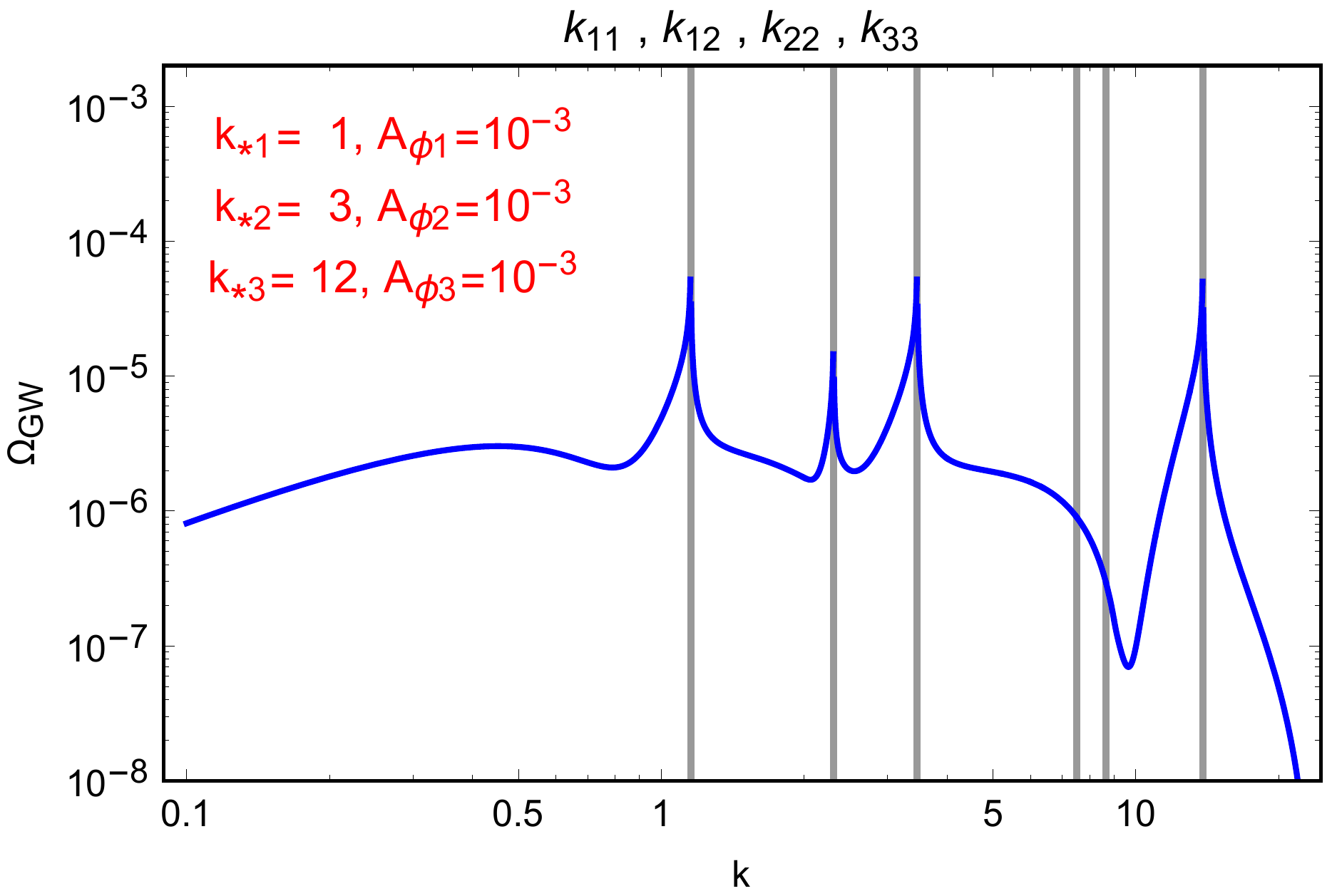}
\includegraphics[width=0.32\textwidth]{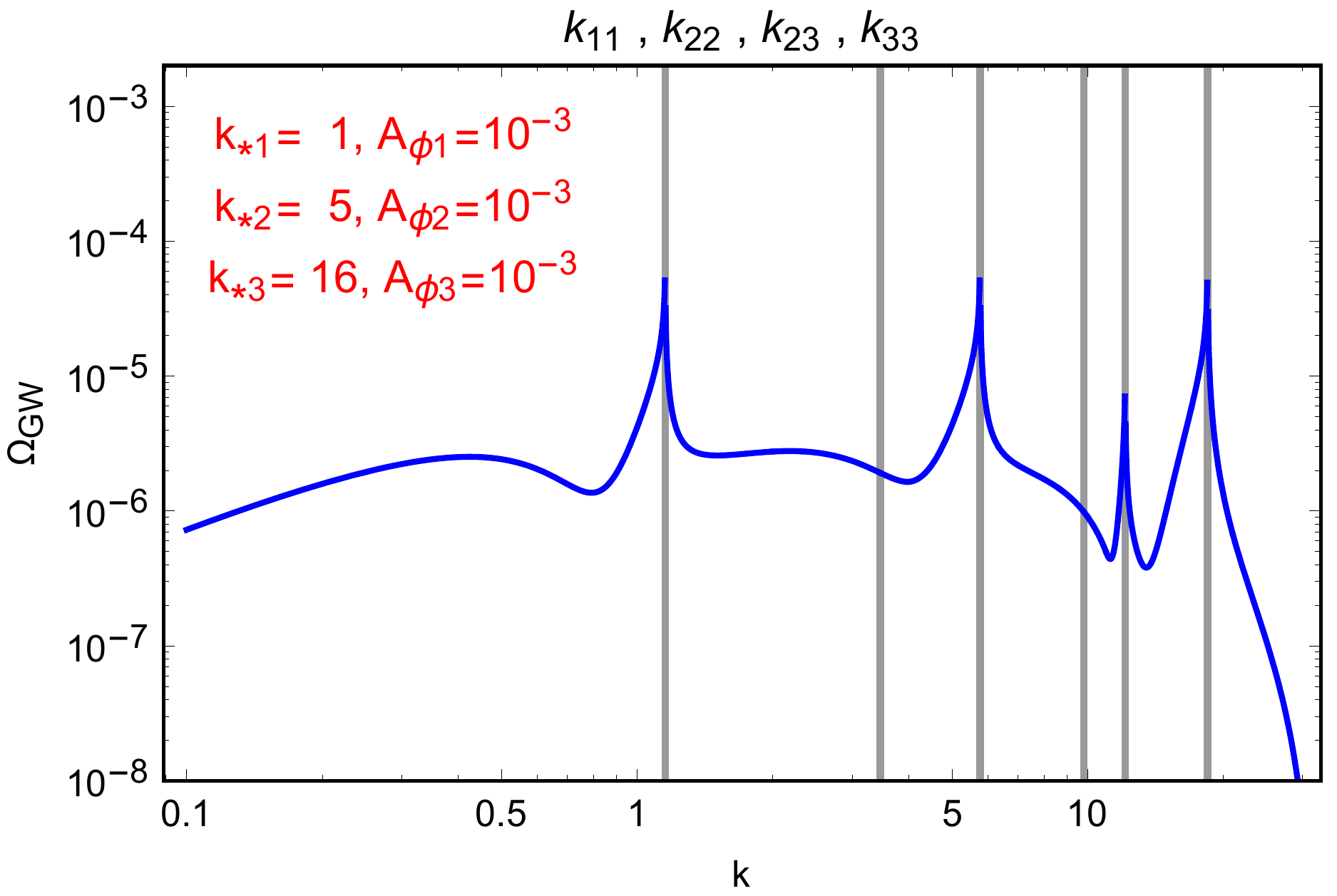}\\
\includegraphics[width=0.32\textwidth]{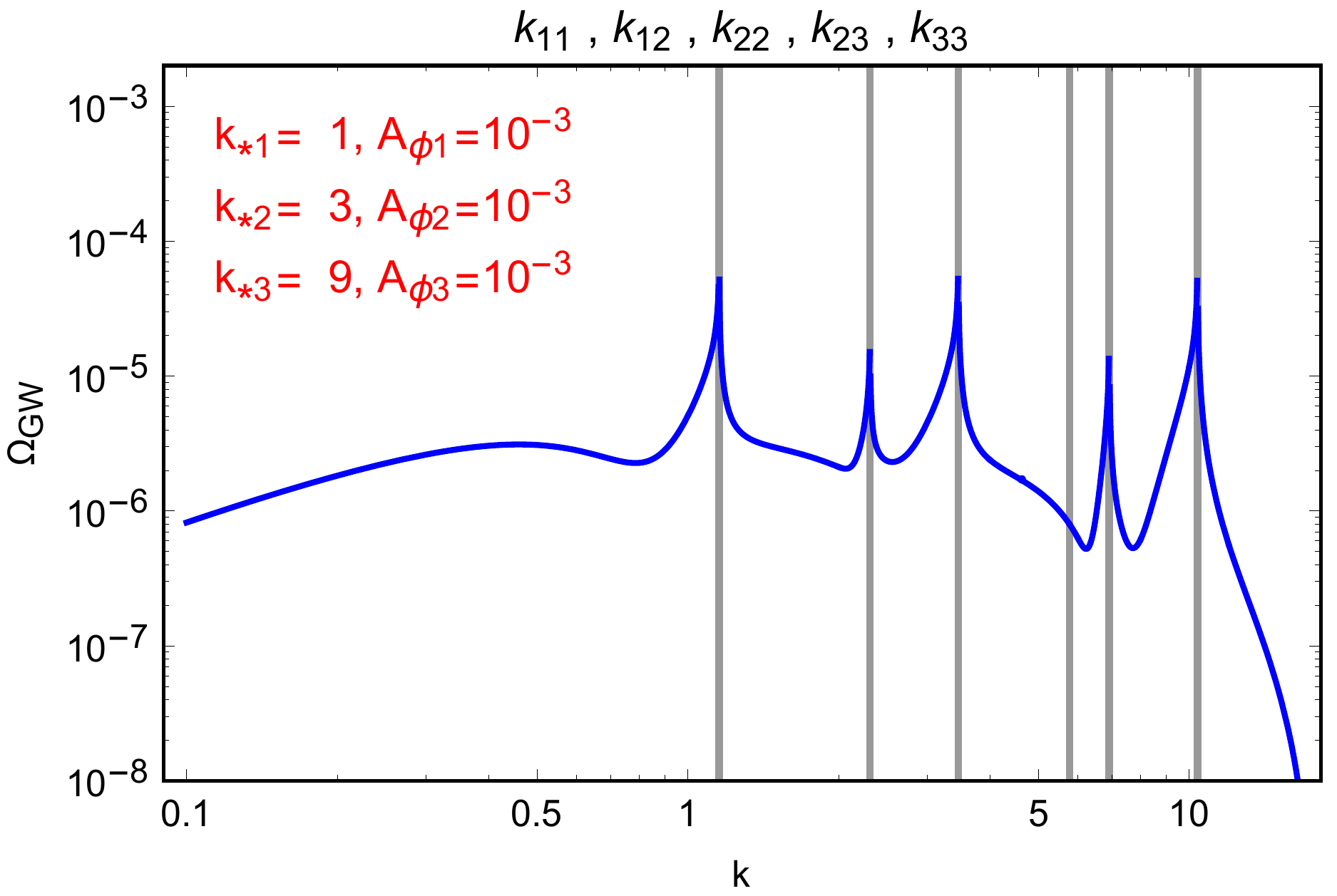}
\includegraphics[width=0.32\textwidth]{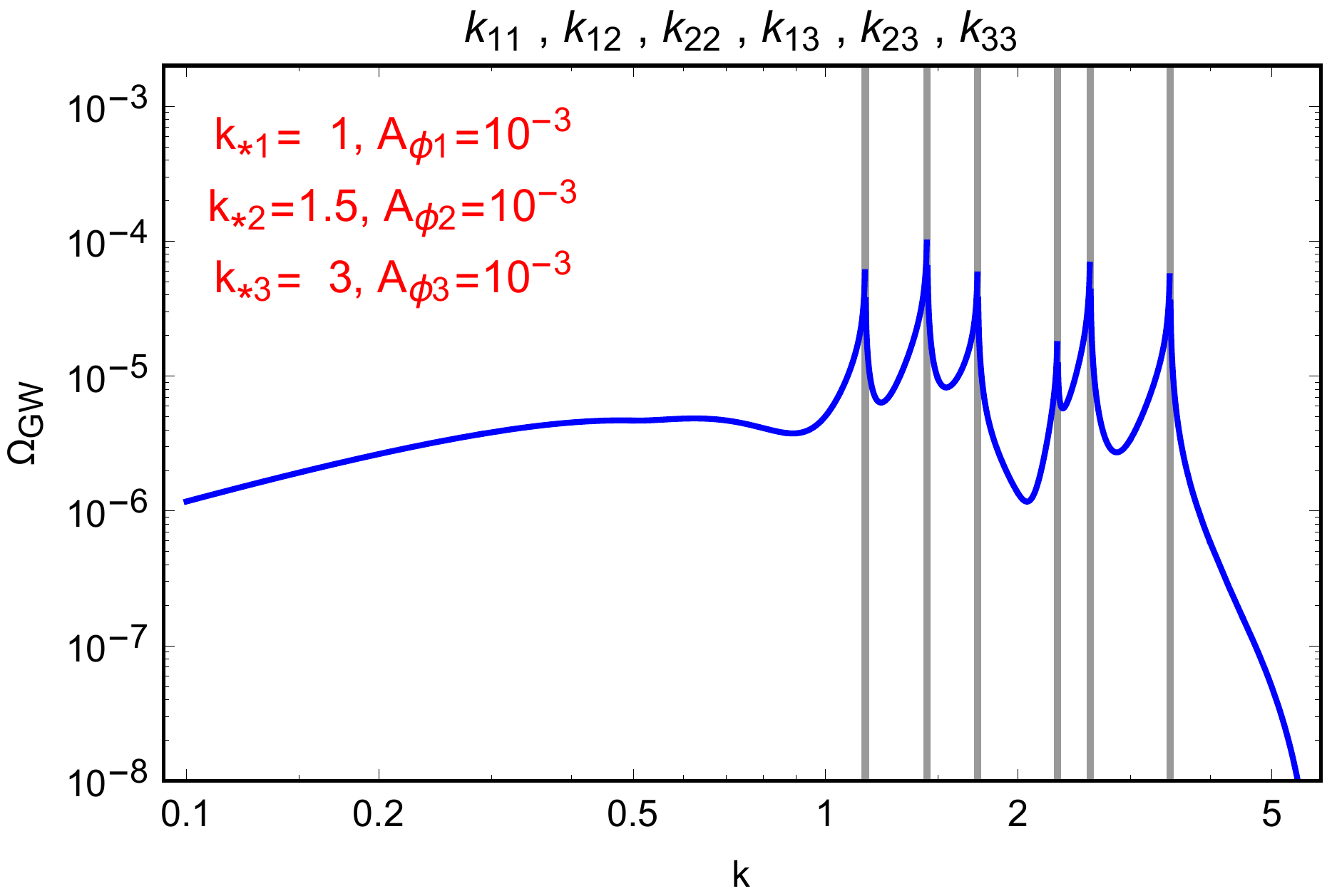}
\includegraphics[width=0.32\textwidth]{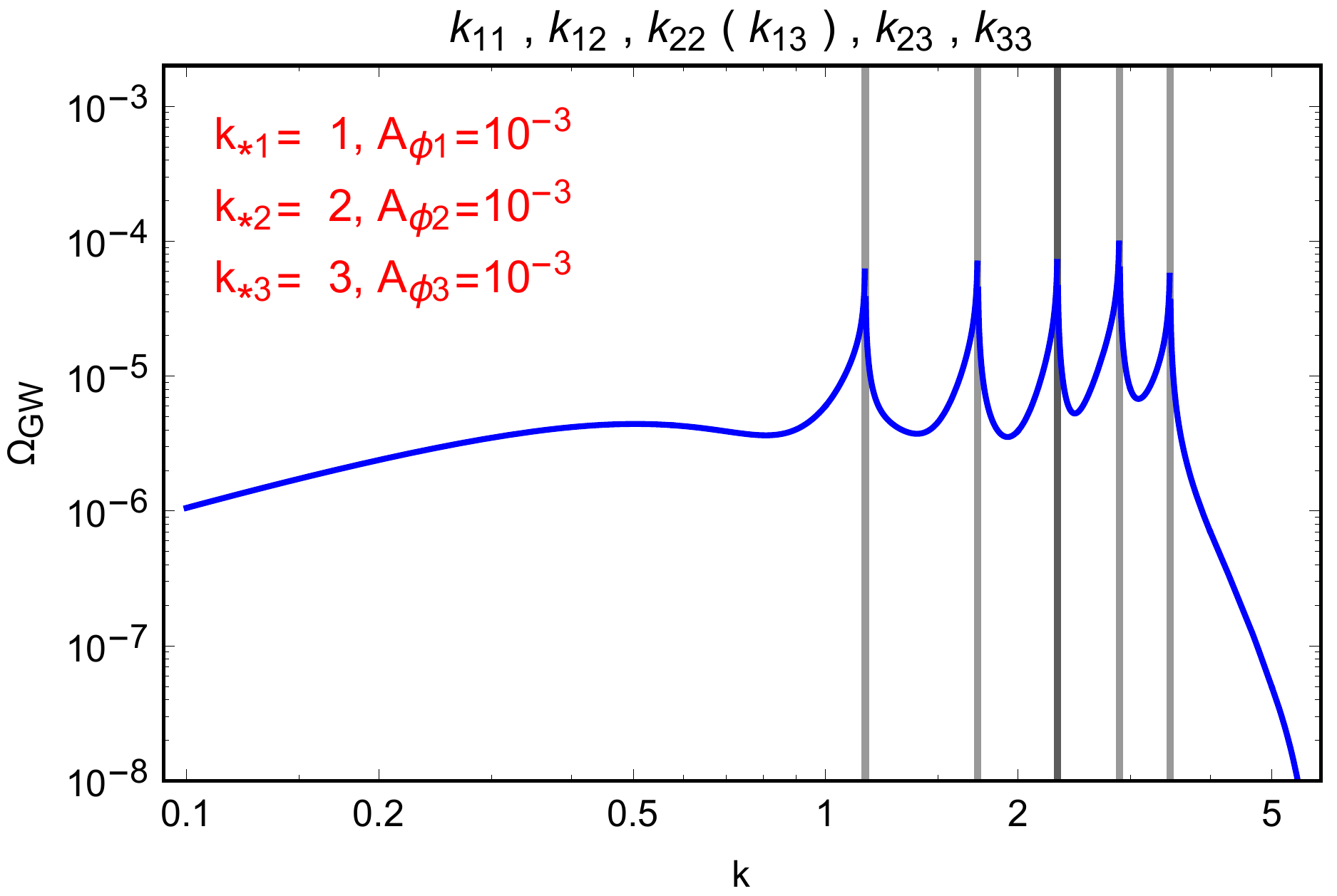}\\
\caption{The energy density spectrum of induced GWs from curvature perturbations with triple $\delta$-peaks at $k_{*1}<k_{*2}<k_{*3}$. The gray lines denote the positions of those would-be peaks at $k_{ij}$ with $i,j=1,2,3$. Copied from Ref. \cite{Cai:2019amo} with permission. }
\label{fig:3delta}
\end{figure}

\begin{figure}
\centering
\includegraphics[width=0.7\textwidth]{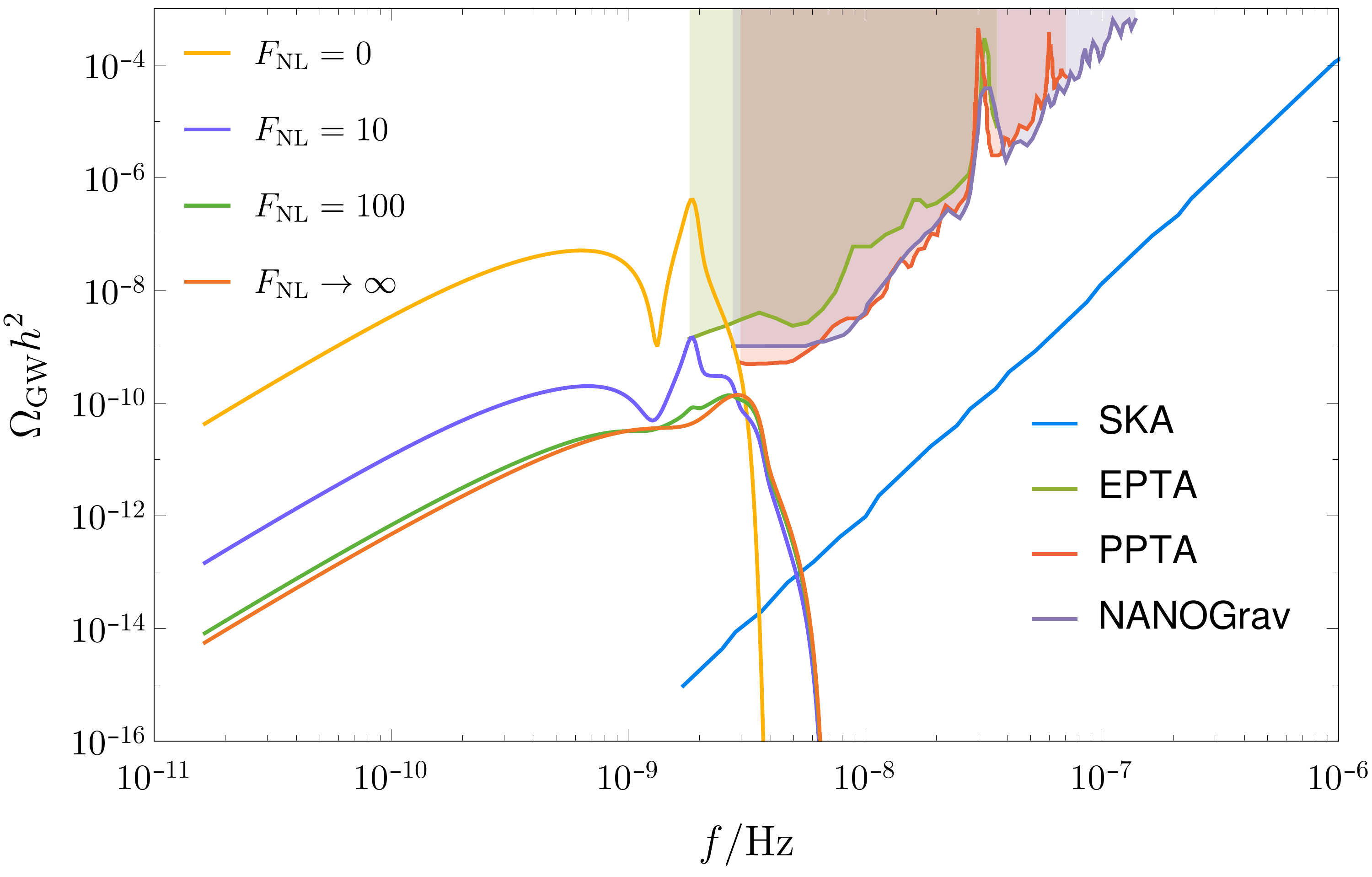}\\
\caption{The GW spectrum with $F_\text{NL}=0,~10,~100$ and $F_\text{NL}\rightarrow \infty$ fit from LIGO detections with respect to the sensitivities of current/future PTA projects. The current constraints (shaded) are given by EPTA \cite{Lentati:2015qwp}, PPTA \cite{Shannon:2015ect}, NANOGrav\cite{Arzoumanian:2018saf}, and the future sensitivity curve of SKA is depicted following \cite{Moore:2014lga}. The recent result from NANOGrav 12.5-yr data is not shown \cite{Arzoumanian:2020vkk}. Copied from Ref. \cite{Cai:2019elf} with permission. }
\label{fig:IGWPTA}
\end{figure}

The combination of a series of spectral peaks in the power spectrum of the curvature perturbation may bring a more distinctive feature in the energy density spectrum of the induced GWs: the resonance peaks (see Figure~\ref{fig:3delta}). In some inflationary models, the spectrum of curvature perturbations has multiple sharp peaks \cite{Achucarro:2010da,Chen:2011zf,Gao:2012uq,Huang:2016quc,Domenech:2018bnf,Fan:2019udt}. Such peaks usually indicate the excitation of extra degree(s) of freedom during inflation whose effective mass is larger than the Hubble parameter. When the spectrum of the curvature perturbations has only one narrow peak, one can find that there is also a narrow peak in the spectrum of GWs as shown in Figure~\ref{fig:IGWPTA}. One may naively guess that in the multiple-peak case, the numbers of peaks in the spectrum of induced GWs and curvature perturbation are equal. Unfortunately, it is not always true, and there may be more peaks in $\Omega_\text{GW}$. In Ref. \cite{Cai:2019amo}, a multiple-peak structure in the energy density spectrum of induced GWs is analytically identified, which exhibits at most $C_{n+1}^{2}$ and at least $n$ peaks at wave-vectors $k_{ij}\equiv (k_{*i}+k_{*j})/\sqrt{3}$ due to resonant amplification and momentum conservation, when there are $n$ narrow peaks located at $k_{*i}$ in the power spectrum of the curvature perturbation. An example of 3 $\delta$-peaks is shown in Figure~\ref{fig:3delta}, where at least 3 peaks and at most $C^2_4=6$ peaks are apparently observed, depending on the positions of the peak wavenumbers $k_{*i}$. It is straightforward to apply our result to the models with an oscillatory modulation in the power spectrum of the curvature perturbation, which is a typical feature in the multi-field inflation models with a curved field space proposed recently in Refs. \cite{Palma:2020ejf,Fumagalli:2020adf} to enhance the power spectrum and generate PBHs. The resonance peaks in the energy density spectrum of induced GWs studied in our paper \cite{Cai:2019amo} are verified in such models in Refs. \cite{Fumagalli:2020nvq,Braglia:2020taf,Fumagalli:2021cel,Dalianis:2021iig}.

It is possible that the binary black holes observed by LIGO/Virgo are of the primordial origin \cite{Bird:2016dcv,Sasaki:2016jop}. If these black holes are primordial, the enhanced curvature perturbation on small scales can induce nanohertz GWs, which is sensitive for PTAs. The origin of the black holes observed by LIGO/VIRGO is the key issue to tackle this problem. In Ref. \cite{Cai:2019elf}, it is shown that by assuming all the LIGO O1 and O2 events are from primordial origins, the power spectrum of the curvature perturbation can be reconstructed by the merger rate derived from the event rate. The induced GWs however, would be in seemingly mild tension with current constraints from PTA, if the curvature perturbation is Gaussian. Introducing local non-Gaussianity of the curvature perturbation with a non-linear parameter $f_{\mathrm{NL}} \gtrsim \mathcal{O}(10)$ can relieve the tension. Nevertheless, even the non-Gaussianity is very large, such induced GWs must be detectable by the SKA in a decade or less, as is shown in Figure~\ref{fig:IGWPTA}. Recent analysis on the LIGO GWTC-2 catalog implies that only $3\%$ to $55\%$ of the black holes can be primordial, with $f_\text{PBH}\sim10^{-3}$ at $18M_\odot$ \cite{Hutsi:2020sol,Franciolini:2021tla}. As we anticipated, the induced GWs associate with such PBHs are inconsistent with the renewed nanohertz GW result from NANOGrav 12.5-yr data \cite{Arzoumanian:2020vkk}, unless there is a positive local non-Gaussianity of $f_\text{NL}\gtrsim2.1$ \cite{Inomata:2020xad}. Of course, as the abundance of the PBHs depends crucially on the spectral shape, statistics, formation process, etc., it is also possible for the NANOGrav signal to be consistent with the LIGO/VIRGO events without invoking non-Gaussianity, when a modified Gaussian window function and a broader width is chosen \cite{Kohri:2020qqd}.

\subsection{GWs produced during preheating}\label{subsec:GWpreheating}

Preheating is a violent process next to inflation, and the amplified energy density perturbations during preheating produce considerable GWs. To set the initial conditions of the hot Big-Bang Universe, the vacuum energy transfers into radiation and reheats the universe after inflation, which is referred to as reheating \cite{Albrecht:1982mp, Shtanov:1994ce}. Many inflationary models predict the existence of preheating process at the beginning of reheating. In the preheating scenario \cite{Kofman:1994rk}, the inflaton field begins to oscillate around the minimum of its potential after inflation and ultimately decays into elementary particles in the standard model of particle physics. Ref. \cite{Kofman:1997yn} thoroughly investigates the case the inflaton $\phi$ is coupled to a scalar field $\chi$ by the coupling $\frac{1}{2}g^{2}\phi^{2}\chi^{2}$ (see also \cite{Huang:2019lgd} for an intermediate decay via vector particles). Ref. \cite{Dufaux:2008dn} considers tachyonic preheating after hybrid inflation. Refs. \cite{Adshead:2018doq,Cuissa:2018oiw,Adshead:2019lbr} consider the case of the coupling between the gauge fields and the inflaton. The parametric resonance induced by non-minimal coupling is studied in Refs. \cite{Fu:2017ero,GarciaBellido:2008ab, Jin:2020tmm}. During preheating, the Fourier modes of a scalar matter field coupled to the inflaton field grow exponentially via parametric resonance \cite{Zhu:2018smk}. The modes are quickly pumped up to a large amplitude. Such highly pumped modes correspond to large, time-dependent density inhomogeneities in configuration space, which can source significant GWs \cite{Khlebnikov:1997di}.

Preheating is the first process after inflation, happening at the energy scale much higher than the colliders could reach \cite{Starobinsky:1980te,Mukhanov:1981xt,Linde:1983gd}, so this period of history is very unclear and model-dependent. The uncertain equation of state of the Universe during preheating affects the model prediction of the $e$-folding numbers of inflation, the amplitude of the power spectrum of scalar perturbations, and the scalar spectral index, so that the CMB constraints on inflationary models also depends on preheating \cite{Lozanov:2016hid,Lozanov:2017hjm}. The complex dynamics of preheating also results in other interesting consequences, for example, the production of topological defects \cite{Tkachev:1998dc,Dufaux:2010cf}, primordial magnetic fields \cite{DiazGil:2007dy,DiazGil:2008tf} and PBHs \cite{Bassett:2000ha,Green:2000he,Kou:2019bbc,Auclair:2020csm,Nazari:2020fmk}. Then, from GWs generated during the preheating we now have a new opportunity to explore new physics and the history of the very early Universe. 

In case of the resonance strength being strong enough, the present peak frequency of such a GW signal is proportional to the energy scale of inflation while the present peak amplitude is independent of the energy scale of inflation \cite{Easther:2006gt,Easther:2006vd,Cai:2021gju}. In the single-field model, in general the peak frequency of GWs is so high that it is hard to be detected by interferometers. In hybrid inflation, since the energy scale ranges from the grand unification theory (GUT) scale to the TeV scale, GWs produced during preheating for low-scale inflationary models is expected to be detected by future ground-based or even space-based interferometers \cite{GarciaBellido:2007dg,GarciaBellido:2007af}.

When the scalar potential satisfies the ``opening up'' condition \cite{Amin:2010jq}, oscillons, localized nontopological quasisolitons, can be generated during preheating, which lead to a stochastic GW background. This case is referred to as oscillon preheating and studied in symmetric smooth potentials \cite{Zhou:2013tsa} as well as asymmetric potentials \cite{Antusch:2016con,Antusch:2017flz,Antusch:2017vga,Amin:2018xfe}.

\begin{figure}
\centering
\includegraphics[width=0.7\textwidth]{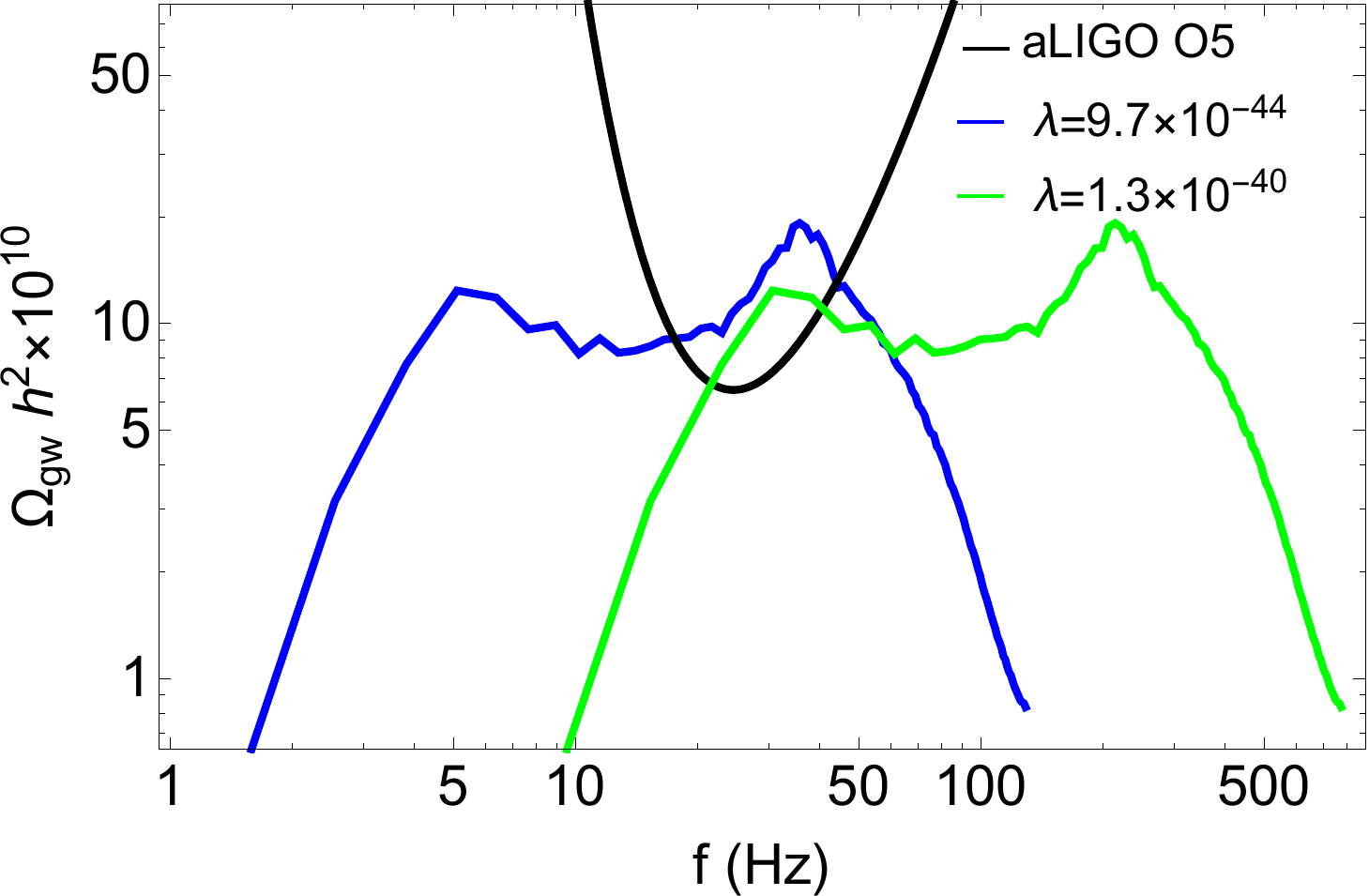}\\
\caption{Energy spectra of GWs today, predicted by the linear potential with $\lambda=9.7 \times 10^{-44}$ (blue) and $\lambda=1.3 \times 10^{-40}$ (green). The black curve is the expected sensitivity curve of the fifth observing run (O5) of the aLIGO-Virgo detector network. Copied from Ref. \cite{Liu:2017hua} with permission. }
\label{fig:aLIGO}
\end{figure}

We find oscillons naturally arise in such a cuspy potential,
\begin{align}
V(\phi) = \lambda M_{\rm pl}^{4-p}|\phi|^p
\label{eq:cuspy}
\end{align}
with $p=1,2/3,2/5$ and $M_{\rm pl}=(8\pi G)^{-1/2}$, the nonsmooth oscillations can trigger the amplification of fluctuations of the inflaton field itself at the moment when $\phi(t)=0$, so that oscillons copiously form during oscillations of the inflaton field, which sources a significant stochastic GW background \cite{Liu:2017hua}. Interestingly, these cuspy potentials lead to a characteristic energy spectrum of GWs with double peaks (see Fig.~\ref{fig:aLIGO}), which can be distinguished from other potentials by measuring the shape of the energy spectrum of GWs. Actually, the cuspy potentials can be mimicked by the following more general form of potential in the asymptotically smooth limit,
\begin{align}
V(\phi) = \frac{m^2M^2}{p}\left[\left(1+\frac{\phi^2}{M^2}\right)^{p/2}-1\right].
\end{align}
When $\phi/M$ is large, the potentials can be approximated by Eq.~\eqref{eq:cuspy}, while when $\phi/M$ is small, the potential becomes smooth near the minimum. We find that the smoothness of the potentials near the point $\phi(t)=0$ suppresses the energy spectrum of GWs \cite{Liu:2018rrt}. That is, cuspy potential yields stronger GW signals due to nonsmooth oscillations. In hybrid inflation, the energy scale of inflation ranges from the GUT scale to the electroweak scale. Consequently, oscillon formation generates a stochastic GW background with a typical frequency today of the order of $10^{-3}-10^{9}$ Hz. Future ground-based and space-based interferometers provide the possibility to search for the GW signals with the double-peak energy spectrum.

\subsection{Strongly lensed GW multimessenger}\label{subsec:SLTD}

Like the strong gravitational lensing of optical signals \cite{2019ApJ...886...94L,2020MNRAS.496..708L,2021MNRAS.503.1319G,2021MNRAS.503.2179Q}, GW lensing attracted many interests in the last decade. The lensing of GW is very crucial for cosmology, fundamental physics, and astrophysics (see some example work in \cite{Sereno:2011ty,Ding:2015uha,Collett:2016dey,Fan:2016swi,Liao:2017ioi,Yang:2018bdf,Hannuksela:2020xor,Yu:2020agu,Cao:2020sky,Xu:2021bfn}). Though currently, we have not yet found any sufficient evidence for the GW lensing signal from the released LIGO data \cite{Hannuksela:2019kle,Abbott:2021iab}, the development of the methodology of applying GW lensing to cosmology still deserves many studies. In this section, we focus on the GW strong lensing and introduce two of our work in this aspect.

The GR has been tested very precisely on solar system scales \cite{Bertotti:2003rm,Shapiro:2004zz}. However, the long-range nature of gravity on the extra-galactic scale is still loosely constrained and poorly understood. The parameterized post-Newtonian (PPN) framework \cite{Thorne:1970wv} provides us with a systematic way to quantify the deviation from GR. The traditional strong gravitational lensing of quasars provides us with a unique opportunity to probe modifications to GR over a range of redshifts and on/above kiloparsec scales with the PPN parameterization \cite{Bolton:2006yz,Smith:2009fn,Schwab:2009nz,Cao:2017nnq,Collett:2018gpf}. Recently, the strong lensing of GW has attracted many interests, and the strongly lensed multimessenger shows significant improvements to the traditional electromagnetic (EM) experiments on cosmology, especially for the measurement of $H_0$ \cite{Liao:2017ioi}.

Ref. \cite{Yang:2018bdf} proposed a new multimessenger approach using data from both GW, and the corresponding EM counterpart to constrain the modified gravity (MG) theory from the scale-dependent phenomenological parameter $\gamma_{\rm PPN}$. The author calculated the time-delay predictions by choosing various values of the phenomenological parameters for MG and then compare them with that from GR,
\begin{equation}
\Delta t_{i,j(\rm MG)}=\frac{1+z_l}{c}\frac{D_l D_s}{D_{ls}}\Delta \phi_{i,j (\rm MG)}\,,
\label{eq:dt}
\end{equation}
here the time delay $\Delta t_{i,j}$ is measured from the strongly lensed GWs and the Fermat potential difference $\Delta \phi_{i,j}$ is reconstructed from the lensed EM domain. $D_X$ is the angular diameter distance. This strategy takes the most advantage of the information from both the GW and EM domains.  For the third generation ground-based GW observatory Einstein Telescope, with only one typical event, and assuming that the dominated error from the stellar velocity dispersions is 5\%, one can probe an 18\% MG effect on a scale of 10 Kpc (68\% confidence level). If assuming GR and a singular isothermal sphere mass model, our approach can distinguish an 8\% MG effect. This work showed that the strongly lensed GW multimessenger plays an important role in revealing the nature of gravity on the galactic and extra-galactic scales.

Based on the prediction of GR, the absorption and dispersion of GWs could be neglected in a perfect-fluid Universe \cite{Ehlers1996}. Such a theoretical point of view has been widely applied in some recent work, i.e., the test of Etherington distance duality relation and the opacity of the Universe at higher redshifts, with the combination of GW and EM signals \cite{Qi:2019spg,Qi:2019wwb}. Up to now the hypothesis of transparent GWs in the EM domain remains experimentally untested, since the current GW detections have not yielded positive results \cite{Abbott:2016blz,TheLIGOScientific:2017qsa}. On the other hand, a large number of independent measurements of GWs indicate that DM, which constitutes a dominant component of virialized objects, could gravitationally interact with itself and with normal matter in galaxies and galaxy clusters. Many ideas have been proposed to explore the possibilities of DM self-interaction (SI) generating the cosmic accelerated expansion \cite{Zimdahl:2000zm,Cao:2010pi} and the non-zero cosmological shear viscosity \cite{Hawking:1966qi,PhysRevD.96.084033,Mir_n_Granese_2021}, if DM can be treated as non-ideal fluids with a viscosity term of $\eta$. In the framework of such methodology, an efficient graviton-matter conversion would be achieved with a simple relation
\begin{align}
\beta \equiv 16\pi G\eta
\end{align}
between the GW damping rate ($\beta$) and DM viscosity ($\eta$). Recently a method of measuring the viscosity of DM in cosmological context has been suggested \cite{Lu:2018smr} and implemented on the current eleven GW events released by LIGO and Virgo Collaborations (on the assumption that DM in the Universe is treated as perfect fluids). When the GW damping rate is taken into account, the viscosity-free luminosity distance ($D_L$) inferred from the standard siren GW signal will be modified to
\begin{align}\label{DLeff}
D_{L,eff}(z, \beta) = D_L(z) e^{\beta D(z)/2},
\end{align}
where $D_{L,eff}$ and $D$ represent the so-called effective luminosity distance and comoving distance, respectively. However, such strategy is hard to realize from the observational point of view, considering the failure of precise measurement of viscosity-free distances and redshift determination for inspiraling and merging binary black holes (BH). This motivates the need to probe the viscosity of DM with other plausible mechanisms.

\begin{figure}
\centering
\includegraphics[width=\textwidth]{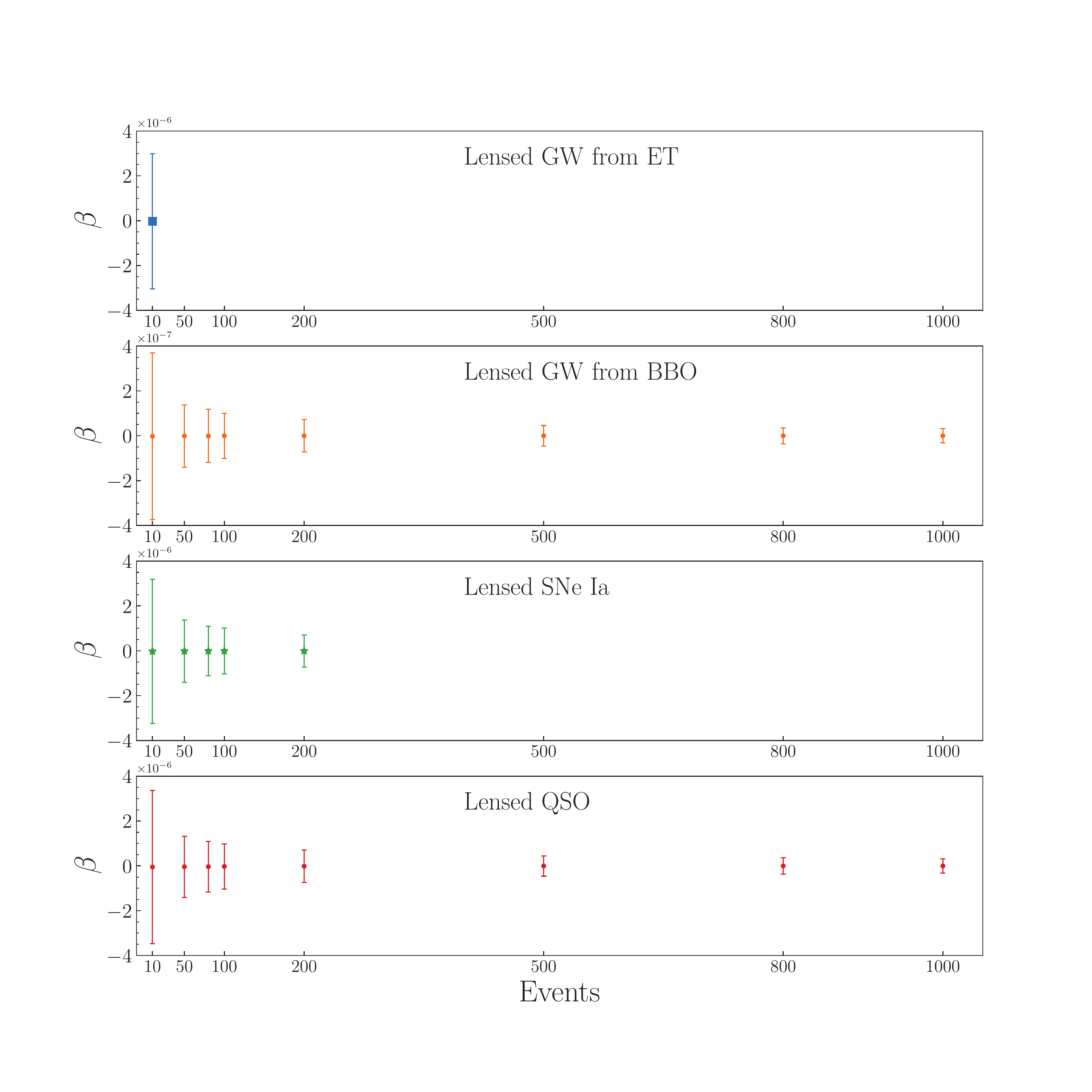}\\
\caption{Constraints on the viscosity of DM with different number of strongly lensed transients: the results of the GW damping rate based on the studies of \cite{Cao:2020sky}. Copied from Ref. \cite{Cao:2020sky} with permission. }
\label{figDM}
\end{figure}

In order to draw firm and robust conclusions about the non-gravitational behavior of DM, \cite{Cao:2020sky} proposed a new strategy to measure the viscosity of DM with strongly lensed GWs produced by the inspiralling NS.  GWs damped by viscosity from a neutron star merger would reach the observer, with the generation of electromagnetic radiation from a short and intense burst of $\gamma$ rays. One should note that besides electromagnetic counterparts, the identification of host galaxies for the majority of GW events could also contribute to unique redshift determination in the EM window. Our idea relies on the time-delay distance in a specific GW-galaxy strong lensing system (with the background GW source at redshift $z_s$ and the lensing galaxy at redshift $z_l$)
\begin{equation}\label{Ddelay}
D_{\mathrm{\Delta t}}(z_l, z_s) \equiv\frac{D_{A}(z_l)
D_{A}(z_s)}{D_{A}(z_l, z_s)},
\end{equation}
which could be measured precisely and accurately for time-variable GW sources, due to the well-reconstructed Fermat potential difference ($\Delta\phi_{i,j}$) between multiple GW signals
\begin{equation}
\Delta t_{i,j} = \frac{D_{\mathrm{\Delta
t}}(1+z_{\mathrm{l}})}{c}\Delta \phi_{i,j}. \label{relation}
\end{equation}
More importantly, besides the configuration of multiple signals that could be used to derive cosmological distances \cite{2012JCAP...03..016C,2012ApJ...755...31C,2015ApJ...806..185C}, the time difference ($\Delta t_{i,j}$) in the arrival times of two signals (at angular coordinates $\boldsymbol{\theta}_i$ and $\boldsymbol{\theta}_j$on the sky) can also be measured with unprecedented accuracy \cite{Liao:2017ioi}, which are not affected by GW damping effect and DM viscosity. Such advantage of strongly lensed GW signals has been widely discussed concerning fundamental physics\cite{Fan:2016swi}, cosmology \cite{Cao:2019kgn}, and dark matter \cite{Liao:2018ofi}. In this analysis, the measured time-delay distance [Eq.~(\ref{Ddelay})] from each strongly lensed GW (unaffected by viscous DM damping) is calibrated with the effective luminosity distances of unlensed GW signals (affected by viscous DM damping) [Eq.~(\ref{DLeff})].

Can the time delays be measured with sufficient precision to yield the viscosity of DM ? To determine the answer, we evaluate the performance of the third-generation ground-based GW detectors, like the Einstein Telescope (ET) \cite{Taylor:2012db} and space-based detectors, like the Big Bang Observer (BBO) \cite{Cutler:2009qv} and DECihertz Interferometer Gravitational wave Observatory
(DECIGO) \cite{Kawamura:2018esd}. The simulations process of different GW samples follows the procedure presented in \cite{Cao:2019kgn,Collett:2015roa}, based on the redshift distributions of unlensed and lensed double compact objects (DCO) from the conservative SFR function \cite{Dominik:2013tma,Ding:2015uha,Pi_rkowska_Kurpas_2021}. For each specific GW event, the uncertainties of different observables (time delays, lens modeling, the line of sight contamination, etc.) will be propagated to the uncertainty of different distances ($D_{\mathrm{\Delta t}}, D_{L,eff}$). For only 10 strongly lensed GWs from ET, the GW damping rate can be constrained to the precision of $\Delta \beta=10^{-6}\,{\rm Mpc}^{-1}$. Such analysis will be significantly improved to $\Delta \beta=10^{-8}\,{\rm Mpc}^{-1}$ with 1000 strongly lensed GW events detected by the BBO. More interestingly, such stringent measurements of GW damping in a viscous Universe provide another perspective to the scatter cross-section of self-interacting DM ($\sigma_{\chi}/m_{\chi}$):
\begin{align}
\frac{\sigma_{\chi}}{m_{\chi}}=\frac{6.3\pi G\left \langle v \right
\rangle}{\beta}.
\end{align}
Note that \cite{Kaplinghat:2015aga} recently reported a unified solution to small-scale structure from dwarfs to clusters, in which DM particles interact with each other following a hydrodynamic description and Maxwellian distribution \cite{Tulin:2017ara}. Compared with the current operating GW detectors (LIGO and Virgo network), the third generation ground-based ET would yield more precise measurements of DM SI cross-section, especially for DM in galaxy clusters ($\Delta(\sigma_{\chi}/m_{\chi}) \sim 10^{-3}$ cm$^2$/g). Within the reach of the second generation space-based BBO (with more detected lensed GW events at much higher redshifts ($z\sim 5$)), both of galaxy-scale and cluster-scale DM SI are expected to be detected at high confidence levels: $\Delta(\sigma_{\chi}/m_{\chi}) \sim 10^{-6}$ cm$^2$/g for dark matter in dwarf galaxies and low-surface-brightness galaxies, and $\Delta(\sigma_{\chi}/m_{\chi}) \sim 10^{-5}$ cm$^2$/g in galaxy clusters. As can be clearly seen from the comparison between strongly lensed transient sources in GW and EM domain [supernovae (SNe) Ia or quasars] [Fig.~\ref{figDM}], such recipe for measuring the viscosity of DM at different scales could helpfully alleviate the strong conflict between the collisionless CDM paradigm and N-body simulations of the small-scale structures of the Universe (known as the cusp-vs-core problem, the missing satellite problem, and the too-big-to-fail problem) \cite{Spergel:1999mh}.

\subsection{Standard siren cosmology}\label{subsec:StandardSiren}

Using GWs to measure the Hubble constant had been proposed by Schutz \cite{Schutz:1986gp} in 1986. This feature of GWs benefits from the fact that one can infer the luminosity distance directly from the GW waveform without the external calibration, which is unavoidable in the utilization of SNe Ia as standard candles. As GW detections can be thought of as aural rather than optical, a more appropriate terminology for a GW standard candle is a ``standard siren'' \cite{Hughes:2002yy}. The standard sirens rely on a very clear underlying physics, i.e., GR. The radiation emitted during the inspiral phase is well described using the post-Newtonian expansion of GR \cite{Blanchet:2013haa}. On the contrary, SNe Ia standard candles are poorly understood in physics and systematics \cite{Drell:1999dx,Benetti:2004bk}.

However, one disadvantage of the standard sirens is the missing redshift information which is completely entangled with the mass and frequency parameter in the waveform. For using standard sirens to investigate cosmology, one needs the redshift information from an independent strategy. The most promising way is from the EM counterparts associated with the GWs. We call this type of standard sirens ``bright sirens''. For instance, it has long been argued that BNSs and NS-BH mergers are likely to be accompanied by a gamma-ray burst (GRB) \cite{Eichler:1989ve,Fox:2005kv,Nakar:2005bs,Berger:2006ik}. The applications of using short GRBs as the EM counterparts of GW standard sirens on cosmology were investigated in details in \cite{Dalal:2006qt,Nissanke:2009kt,Sathyaprakash:2009xt,Zhao:2010sz,Cai:2016sby}. The GRB counterpart to the GW source can not only provide a precise sky localization, which is useful for determining the redshift to the source galaxy, but also significantly improve the GW determination of luminosity distance by breaking the degeneracies between distance, position, and orientation angles. Thus it can measure the expansion history of our Universe back to redshift up to around 2 \cite{Sathyaprakash:2009xt,Zhao:2010sz}.

The first joint observations of GW from a BNS GW170817 with its EM counterpart GRB 170817A \cite{TheLIGOScientific:2017qsa,GBM:2017lvd,Monitor:2017mdv} mark a significant breakthrough for multimessenger astronomy. By identifying the host galaxy NGC 4993, the first measurement of the Hubble constant from standard sirens has been reported \cite{Abbott:2017xzu}. Recently, \cite{Graham:2020gwr} reported the first plausible optical EM counterpart to a (candidate) BBH merger GW190521, detected by the Zwicky Transient Facility (ZTF). The corresponding measurements of the Hubble constant and other cosmological parameters have also been investigated \cite{Chen:2020gek,Mukherjee:2020kki}.  Though at the present stage they cannot resolve the Hubble tension due to the large uncertainty, the future second-generation GW detector network LIGO-Hanford+advanced LIGO-Livingston+advanced Virgo+ KAGRA+LIGO-India (HLVKI) could provide a much tighter constraint \cite{Chen:2017rfc}.

In addition to BNS, the massive black hole binaries (MBHBs) are assumed to produce observable EM emissions at the merger from the production of an optical accretion-powered luminosity flare, and also the radio flares and jets. The EM emissions are based on results from general-relativistic simulations of merging MBHBs in an external magnetic field \cite{Palenzuela:2010nf}. Several studies found that MBHBs could emit radiation in different bands of the EM spectrum both at merger and during long-lasting (ranging from weeks to months) afterglows \cite{Dotti:2011um,Giacomazzo:2012iv}. Moreover, pre-merger EM observational signatures could even be spotted during their inspiral phase \cite{Kocsis:2007yu,OShaughnessy:2011nwl,Kaplan:2011mz,Haiman:2017szj}. The standard sirens of MBHB detected with future space-based detector LISA on cosmology have been investigated \cite{Holz:2005df,Klein:2015hvg,Tamanini:2016zlh,Tamanini:2016uin,Caprini:2016qxs,Cai:2017yww}. Compared to the BNS standard sirens with the ground-based detector, the MBHB standard sirens with a space-based detector can approach redshift up to 6--7, thus measuring the expansion history of Universe back to a much earlier time \cite{Tamanini:2016zlh}.

The bright sirens not only could measure the expansion history of our Universe, thus constraining such as the Hubble constant and the dark-energy equation of state,  but also test gravity theories. For instance, the GW170817 with its EM counterparts has put a very tight constraint of the speed on GW, $(c_T-c)/c<\mathcal{O}(10^{-15})$ \cite{Monitor:2017mdv}. It has a very strong implication on dark energy model and modified gravity theory \cite{Creminelli:2017sry,Ezquiaga:2017ekz,Amendola:2017orw,Langlois:2017dyl,Baker:2017hug,Kase:2018aps,Copeland:2018yuh} or even some dark matter model \cite{Cai:2017buj}. In addition to the speed of GW, the bright sirens can constrain the modified gravity theory from the propagation of GW in the cosmic distance. Several studies have shown the potential of the BNS and MBHB standard sirens on testing GR through GW propagation from the ground/space-based GW detectors \cite{Saltas:2014dha,Nishizawa:2017nef,Arai:2017hxj,Belgacem:2017ihm,Amendola:2017ovw,Belgacem:2018lbp,Lagos:2019kds,Nishizawa:2019rra,Belgacem:2019pkk,Belgacem:2019tbw,Belgacem:2019zzu}.

With advanced LIGO and advanced Virgo reaching their target sensitivity, and other detectors such as Kamioka Gravitational wave detector (KAGRA) and LIGO-India joining the search in the near future, the second-generation ground-based detector network HLVKI would be in operation. On a longer timescale, around the 2030s the third-generation ground-based detectors, such as ET \cite{Punturo:2010zz} and CE \cite{Evans:2016mbw}, and the space interferometer LISA will be ready for operation. During the same period, Chinese space-based GW detectors Taiji \cite{Hu:2017mde,Guo:2018npi,Taiji-1} [proposed by the Chinese Academy of Sciences (CAS)] and TianQin \cite{Luo:2015ght,Luo:2020bls,Mei:2020lrl} [proposed by the Sun Yat-Sen University (SYSU)] will be launched. The synergic operation of these GW networks would contribute greatly on cosmology and the test of GR \cite{Chen:2017rfc,Belgacem:2019tbw,Ruan:2019tje,Ruan:2020smc,Wang:2020dkc,Orlando:2020oko,Wang:2021srv,Yang:2021qge}.

\subsubsection*{Bright sirens}

In this subsection, we wish to highlight some work of using GWs with EM counterparts as bright sirens to probe the evolution of the universe and test GR.

$\bullet$ For forecasting the ability of third-generation ET on constraining the cosmological parameters such as Hubble constant, matter density and the dynamics of dark energy, \cite{Cai:2016sby} simulated a series of GW standard sirens of BNS and NS-BH with short GRB as the EM counterparts. Using the inspiral phase of waveform which is computed in the post-Newtonian formalism up to 3.5 PN, the Signal-to-Noise-Ratio (SNR) can be calculate from the matched filtering with an optimum filter in the ideal case of Gaussian noise,
\begin{equation}
\rho^2=\frac{5}{6}\frac{(G\mathcal{M}_c)^{5/3}\mathcal{F}^2}{c^3\pi^{4/3}d_L^2(z)}\int^{f_{\rm max}}_{f_{\rm min}}df\frac{f^{-7/3}}{S_n(f)} \,,
\label{eq:SNR}
\end{equation}
here $\mathcal{M}_c$ is the redshifted chirp mass $\mathcal{M}_c=(m_1m_2)^{3/5}/(m_1+m_2)^{1/5}(1+z)$. $d_L$ is the luminosity distance. $S_n(f)$ is the one-sided noise power spectral density (PSD) of the detector. The factor $\mathcal{F}$ is to characterize the detector response, $\mathcal{F}^2=\frac{(1+\cos^2\iota)^2}{4}F^2_++\cos^2\iota F^2_\times$. $F_+$ and $F_\times$ are the antenna response functions to the GW + and $\times$ polarizations.
One can easily estimate the uncertainty of the inferred luminosity distance from the Fisher information matrix. In that work, the authors considered the beam feature of short GRB which can help to break the degeneracy between inclination angle $\iota$ and luminosity distance $d_L$. But they still give a conservative estimation of $\sigma_{d_L}^{\rm inst}=2d_L/\rho$ from the instrument together with the weak lensing uncertainty $0.05zd_L$. From these distance uncertainties and the merger rate distributions of BNS and NS-BH, the authors simulated 100-1000 standard sirens from ET. Then they adopted the Markov chain Monte Carlo (MCMC) approach to constrain the cosmological parameters $H_0$ and $\Omega_m$ under the baseline $\Lambda$CDM model. From the nonparametric Gaussian (GP) process regression method, they reconstructed the equation of state of dark energy $w(z)$ in the redshift range 0--1. The results showed that with about 500-600 GW events from ET one can constrain the Hubble constant with precision comparable to Planck 2015 results. Using GP and with 1000 GW events, one can constrain $w(z)$ with an error of 0.03 in the low redshift region. This work gave an impression of what we can do with the mock future standard sirens to constrain the cosmological parameters by different data interpretation techniques.

$\bullet$ The interaction between dark energy and DM is a very crucial problem to understand the nature of the dark sector of our Universe. \cite{Cai:2017yww} performed a forecast analysis of the ability of the LISA space-based interferometer to reconstruct the dark sector interaction using MBHB standard sirens at high redshift.  Using the MBHB catalogs constructed from three different astrophysical scenarios (i.e., light-seeds popIII, heavy-seeds with time delay Q3d and without time delay Q3nod) for the evolution of massive black hole mergers based on the semi-analytic model \cite{Barausse:2012fy}, this work constructed the catalogs of MBHB standard sirens by LISA, with an electromagnetic counterpart detectable by future telescopes. Then the authors employed Gaussian process methods to reconstruct the dark sector interactions in a nonparametric way,
\begin{equation}
q=2\left(\frac{3D''^2}{D'^5}-\frac{D'''}{D'^4}\right)(1+z)^2+4\frac{D''}{D'^4}(1+z) \,,
\end{equation}
where $D$ is the normalized comoving distance $D=H_0d_L/(c(1+z))$. $q\equiv Q/H_0^3$ is the dimensionless interacting term between the matter and vacuum energy. The continuity equations for DM and the vacuum energy are $\dot{\rho}_m+3H\rho_m=-Q$ and $\dot{\rho}_v=Q$. Using MBHB standard siren alone, LISA can reconstruct the interaction well from  $z\sim1$ to $z\sim3$ (for a 5-year mission) and to $z\sim4$ or even $z\sim5$ (for a 10-year mission). When combined with the simulated Dark Energy Survey (DES) SNe Ia datasets, the low redshift below 1 can be also covered and reconstructed well. These results suggested that MBHB standard sirens from LISA  are a very promising tool to test and constrain possible deviations from the standard $\Lambda$CDM dynamics, especially at high redshift.

$\bullet$ The cosmic anisotropy with a dipole amplitude have been constrained with CMB, SNe Ia and large scale structure data sets \cite{Ade:2015hxq,Cai:2011xs,Cai:2013lja,Fernandez-Cobos:2013fda,Bengaly:2016amk}. GW standard sirens as excellent indicators of the cosmic distance are very suitable for the test of anisotropy of our Universe. \cite{Cai:2017aea} simulated a series of standard sirens of BNS and NS-BH from ET and DECIGO, and of MBHBs from LISA. These standard sirens can be used as a probe of the cosmic anisotropy with a dipole form
\begin{equation}
d_L(\hat{z})=d_L^0(z)[1+g(\hat{n}\cdot\hat{z})] \,,
\end{equation}
where they parameterized the dipole modulation simply by its amplitude $g$ and direction $\hat{n}$ given by $\hat{n}=(\cos\phi\sin\theta,~\sin\phi\sin\theta,~\cos\theta)$. $d_L^0$ is the isotropic luminosity distance calculated from the fiducial cosmology. Thus it is very straightforward to adopt MCMC to constrain the amplitude and direction of the dipole anisotropy from the mock GW data generated with different sky locations. The results of this work can be summarized as follows. For LISA, the cosmic anisotropy can be detected at $3\sigma$ confidence level (C.L.) if the dipole amplitude is larger than 0.03, 0.06, and 0.025 (for MBHB seed models Q3d, pop III, and Q3nod, respectively). In the meanwhile, the dipole direction can be constrained roughly around 20\% at $2\sigma$ C.L.. For ET with no less than 200 GW events, one can detect the cosmic anisotropy at $3\sigma$ C.L. if the dipole amplitude is larger than 0.06, and the dipole direction can be constrained within 20\% at $3\sigma$ C.L. if the dipole amplitude is about 0.1. For DECIGO with no less than 100 GW events, the cosmic anisotropy can be detected at $3\sigma$ C.L. for dipole amplitude larger than 0.03, and the dipole direction can even be constrained within 10\% at $3\sigma$ C.L. if dipole amplitude is larger than 0.07. This work manifested the potentials of using the standard sirens approach in the the detection of the cosmic anisotropy.

$\bullet$ Recent work has shown that the modified GW propagation can be used as a powerful probe of dark energy and modified gravity. In theories where GR is modified on cosmological scales, from standard sirens we do not measure the same luminosity distance as electromagnetic probes. The general models of modified gravity that pass the speed-of-gravity test still can modify the friction term of the equation of motion of GW,
\begin{equation}
h_A^{\prime\prime}+2[1-\delta(\eta)]\mathcal{H}h_A^{\prime}+k^2h_A=0 \,.
\end{equation}
Here $\delta(\eta)$ is used to denote the deviation from GR. Then one can show, from the canonical inference of the GW amplitude, the luminosity distance is actually
\begin{equation}
d_L^{\rm gw}(z)=d_L^{\rm em}(z)\exp\left\{-\int_0^z\frac{dz^{\prime}}{1+z^{\prime}}\delta(z^{\prime})\right\} \,.
\end{equation}
In \cite{Belgacem:2019zzu}, the authors investigated using the technique of Gaussian processes to reconstruct the $\delta$ function by combining the GW luminosity distance from simulated joint GW-GRB detections with the  electromagnetic luminosity distance from simulated DES data, without assuming any parameterizations. This work showed that from future HLVK and ET detectors, the $\delta$ function which denotes the deviation from GR in terms of the tensor propagation can be reconstructed very precisely with the Gaussian process.

\begin{figure}
\centering
\includegraphics[width=\textwidth]{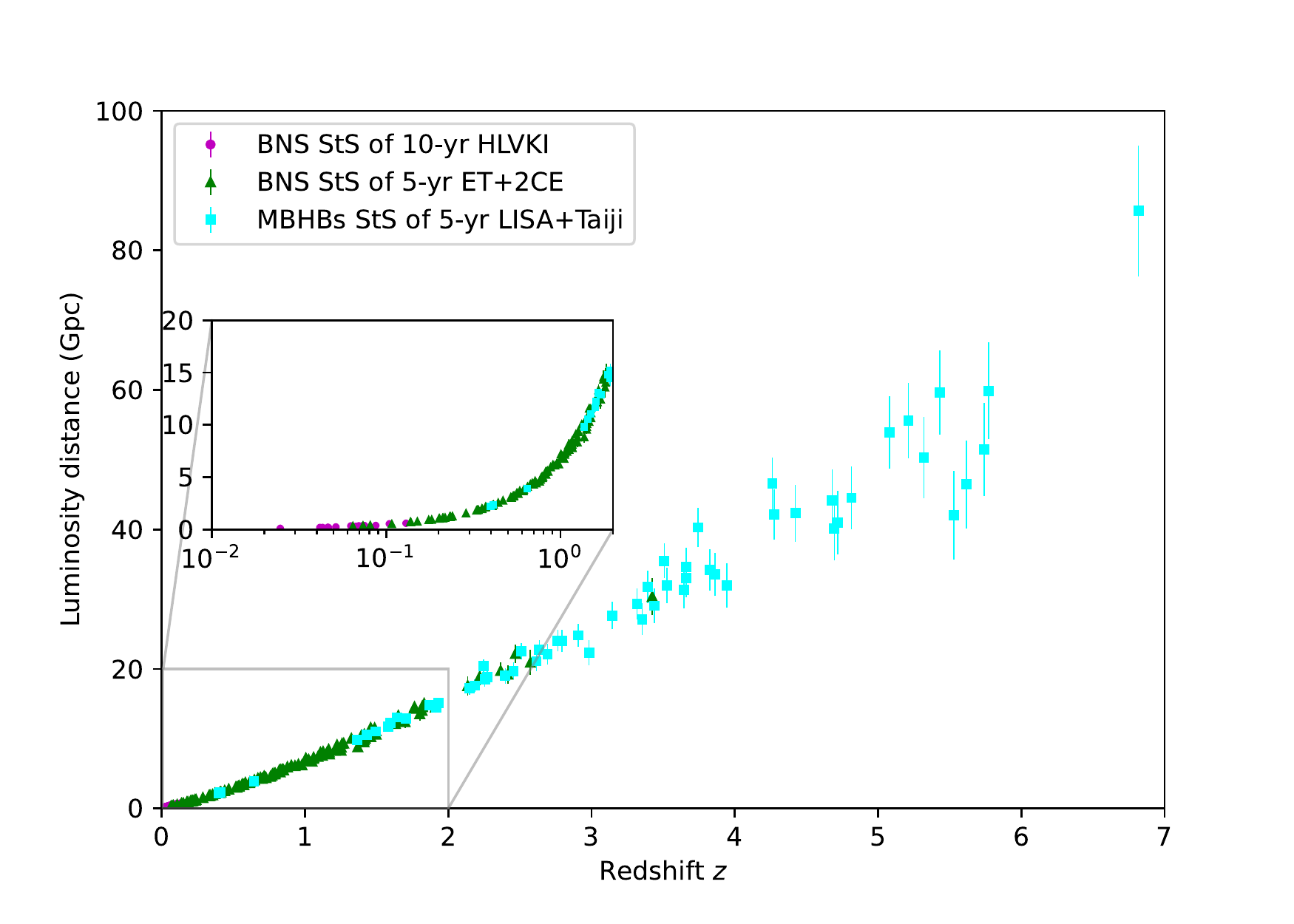}\\
\caption{The Hubble diagram of one realization of mock standard sirens from future GW detector networks. Copied from \cite{Yang:2021qge} with permission. }
\label{fig:Hubble_diagram}
\end{figure}

$\bullet$ In addition to the second-generation ground-based network HLVKI, the third-generation ground-based ET+2CE and space-based LISA-Taiji networks have also been proposed and investigated. The synergy of the networks would not only improve the SNR thus detecting more GW events, but also help to measure the GW parameters more precisely. \cite{Yang:2021qge} constructed the catalogs of standard sirens with the joint GW+EM detections for 10 years detections of HLVKI, 5 years detections of ET+2CE, and 5 years of detections of LISA-Taiji, which are estimated to be available and released in the 2030s. With a combined Hubble diagram from these ground and space-based detector networks which can explore the expansion history of our Universe from redshift 0 to 7, as shown in figure~\ref{fig:Hubble_diagram}.  The author adopted several methodologies such as MCMC, Gaussian process, and Artificial Neural Networks to investigate the potential of future bright sirens on cosmology and modified gravity theory. The results show that the combined standard siren alone can constrain the Hubble constant at the precision level of $0.34\%$, 1.76 times more tightly than the current most precise measurement from \textit{Planck}+baryon acoustic oscillation (BAO)+Pantheon. The joint standard siren with current EM experiments will improve the constraints of cosmological parameters significantly. The modified gravity theory can be constrained with $0.46\%$ error from the GW propagation. This work showed the bright sirens in the 2030s are powerful probes of cosmology and gravity theory in addition to the traditional EM experiments.

\subsubsection*{Standard sirens combined with other cosmological probes}

In the above part, several examples of using GW standard sirens to explore the cosmic evolution and test gravity theories are provided; but the focus is on the cases of solely using the standard sirens as a cosmological probe. Actually, the more important role that the GW standard sirens will play in the exploration of the universe in the future relies on the fact that once they are combined with other EM cosmological probes, the degeneracies between cosmological parameters can then be well broken. Thus, from a perspective of comprehensive analysis on cosmology, GW standard sirens would provide an extremely important cosmological probe to precisely measure cosmological parameters in the next decades. In the following, we shall focus on the discussions on the role that standard sirens would play in precisely measuring cosmological parameters in synergy with other cosmological probes.

Precisely measuring cosmological parameters has always been one of the most important tasks in cosmology because almost all the important scientific questions in cosmology are relevant to precision measurements of cosmological parameters. Currently, there are two main problems concerning the measurements of cosmological parameters: (i) tensions appear between the early and late universe observations \cite{Aghanim:2018eyx,Verde:2019ivm,Li:2020tds,Guo:2018ans,DiValentino:2021izs}. For example, for the Hubble constant $H_0$, the Planck-$\Lambda$CDM results \cite{Aghanim:2018eyx} are in significant tension with the local measurements (which prefer a higher value) \cite{Riess:2020fzl}; (ii) in the extended cosmological models (beyond $\Lambda$CDM), extra cosmological parameters (from ``new physics'') cannot be tightly constrained by current cosmological observations, since parameter degeneracies are usually serious. For example, dark-energy equation of state (EoS) $w$, total neutrino mass $\sum m_\nu$, and other extra parameters still cannot be well constrained by the current observations \cite{Zhao:2018fjj,Feng:2019mym,Feng:2019jqa,Li:2018ydj,Guo:2018uic,Guo:2019dui,Zhang:2020mox,Li:2020gtk,Zhang:2019ipd,Xu:2020uws,Zhang:2021yof}, due to the strong parameter degeneracies. Since GW standard sirens can measure absolute cosmological distances, they can be used to effectively break the parameter degeneracies existing in the constraints from the traditional EM cosmological observations \cite{Zhang:2019ylr}. Hence, it can be expected that GW standard siren observations could play an important role in precisely measuring cosmological parameters. Next, we highlight some work concerning using GW standard sirens to break cosmological parameter degeneracies.

\begin{figure}
\centering
\includegraphics[width=\textwidth]{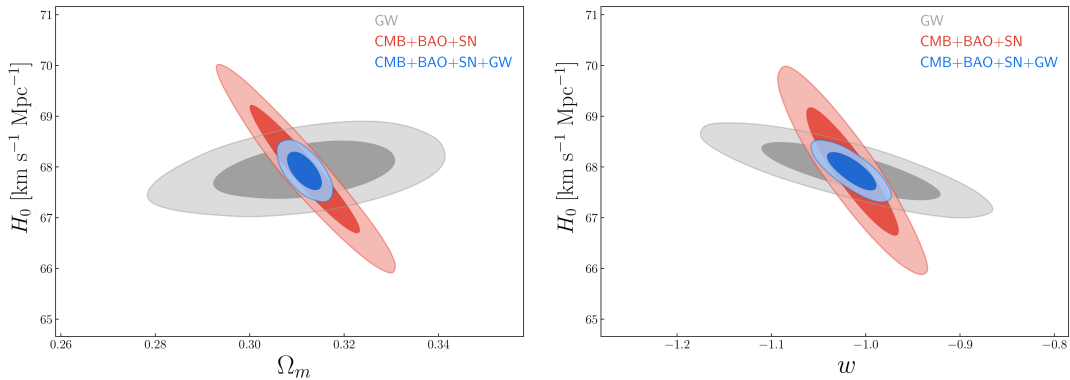}\\
\caption{Two-dimensional marginalized contours (68.3\% and 95.4\% confidence level) in the $\Omega_m$--$H_0$ and $w$--$H_0$ planes for the $w$CDM model, by using the data of GW, CMB+BAO+SN, and CMB+BAO+SN+GW. Here, GW represents 1000 standard sirens of BNS mergers simulated for CE's 10-year observation. Copied from Ref. \cite{Jin:2020hmc} with permission. } \label{fig1}
\end{figure}

$\bullet$ In Refs. \cite{Zhang:2018byx,Wang:2018lun,Du:2018tia,DiValentino:2018jbh,Zhang:2019ple,Zhang:2019loq,Li:2019ajo,Yang:2019vni,Yang:2019bpr,Jin:2020hmc}, a series of analyses have been made for the cosmological parameter estimation in various dark energy models using the combination of GW and EM observations (e.g., CMB, BAO, and SN). These analyses show that standard sirens alone cannot provide very tight constraints on the cosmological parameters except $H_0$, but the combination of GW and EM observations can effectively break the parameter degeneracies and thus greatly improve the constraint accuracies. For example, Jin \emph{et al.} \cite{Jin:2020hmc} find that the future GW standard siren data from CE can well break the cosmological parameter degeneracies generated by the EM observations, as shown in Figure \ref{fig1} (1000 simulated standard siren data are used). When adding the GW data to the data combination of CMB+BAO+SN, the constraint precisions of $\Omega_{\rm m}$ and $H_0$ could be improved from 1.59\% and 0.49\% to 0.71\% and 0.21\%, respectively, in the $\Lambda$CDM model; the constraint precision of $w$ can be improved from 3.15\% to 1.68\% in the $w$CDM model. The standard siren observation can also be combined with the fast radio burst observation to provide an independent low-redshift probe \cite{Zhao:2020ole}.

\begin{figure}
\centering
\includegraphics[width=\textwidth]{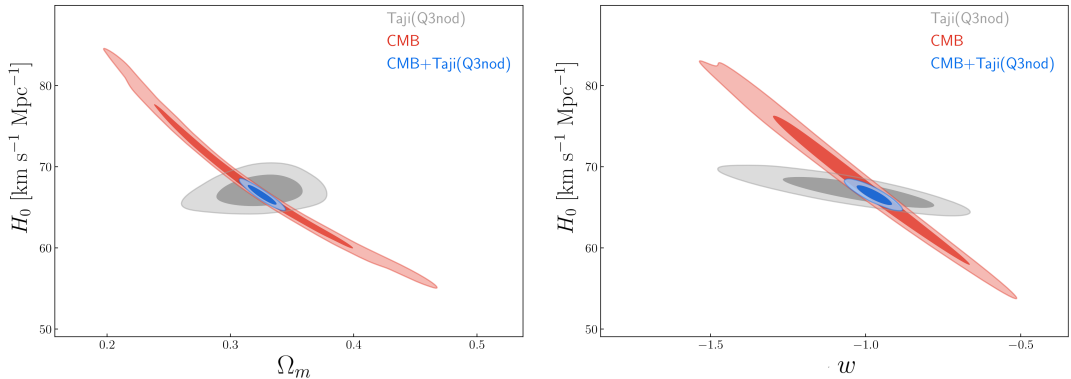}\\
\caption{Two-dimensional marginalized contours (68.3\% and 95.4\% confidence level) in the $\Omega_m$--$H_0$ and $w$--$H_0$ planes for the $w$CDM model, by using the data of Taiji, CMB, and CMB+Taiji. Here, Taiji represents the simulated standard sirens of MBHB mergers detected by the Taiji mission (5-year observation) based on the Q3nod model. Copied from Ref. \cite{Zhao:2019gyk} with permission. } 
\label{zhao}
\end{figure}

$\bullet$ {The work discussed above focus on the standard sirens of using the GWs emitted by BNSs, now we turn to the scenario with the GWs emitted by MBHBs. We have mentioned that MBHB mergers are expected to produce powerful EM radiation \cite{Blandford:1977ds,Meier:2000wk,Palenzuela:2010nf,OShaughnessy:2011nwl,Barausse:2012fy} and thus one can determine the redshifts of the sources through identifying the EM counterparts. If GW detectors can locate the positions of sources within $10\ {\rm deg^2}$, it can reach the same orders of magnitude as the fields of view of the Vera C. Rubin Observatory (formerly LSST) \cite{Abell:2009aa}, the European Extremely Large Telescope (E-ELT) \cite{Liske:2008ph}, and the SKA \cite{Bacon:2018dui}, making it possible to identify EM counterparts. In Ref. \cite{Tamanini:2016zlh}, the authors show that the number of events that EM counterparts can be observed is only a few dozen within the 5-year observation. Even so, the high-redshift MBHB standard siren data would provide an important supplement to low-redshift observations.} In Ref. \cite{Zhao:2019gyk}, Zhao \emph{et al.} take the Taiji mission as a representative to study the capability of space-based GW detection of breaking the degeneracies of cosmological parameters. 
As shown in Figure \ref{zhao}, although the Taiji data alone cannot constrain cosmological parameters well enough, it can effectively break the parameter degeneracies inherent in the CMB data, thus greatly improving the constraint accuracies of parameters. For example, for the constant dark-energy EoS $w$, the data combination of Taiji+CMB can constrain it to an accuracy of $4\%$, which is comparable to the result of about 3\%--4\% from the CMB+BAO+SN data combination. In addition, Wang \emph{et al.} \cite{Wang:2019tto} also conduct a similar analysis on TianQin and reach the same conclusion.

$\bullet$ Moreover, the space-based GW observatories, such as LISA, Taiji, and TianQin, could form a space-based GW detection network in the future \cite{Ruan:2020smc}. In Ref. \cite{Ruan:2020smc}, Ruan {\it et al.} find that the LISA-Taiji network may achieve four orders of magnitude improvement on the source localization (solid angle increased by three orders and luminosity distance increased by one order) compared to the single detector. The improvement by the LISA-Taiji network compared to the single Taiji mission in constraining cosmological parameters is discussed in Ref. \cite{Wang:2021srv}. It is shown that, even in the conservative scenario where only the inspiral phase is used to locate GW sources, the constraint precision of $H_0$ could reach 1.3\% by solely using the standard sirens from the LISA-Taiji network. Moreover, the CMB+network data could improve the constraint on $w$ by 56.7\% compared with the CMB+Taiji data, and the constraint precision of $w$ reaches about 4\%, which is comparable with the result of CMB+BAO+SN. We can conclude that the GW detection network composed of multiple space-based GW observatories will play an important role in understanding the nature of dark energy in the future.

\begin{figure}
\centering
\includegraphics[width=0.6\textwidth]{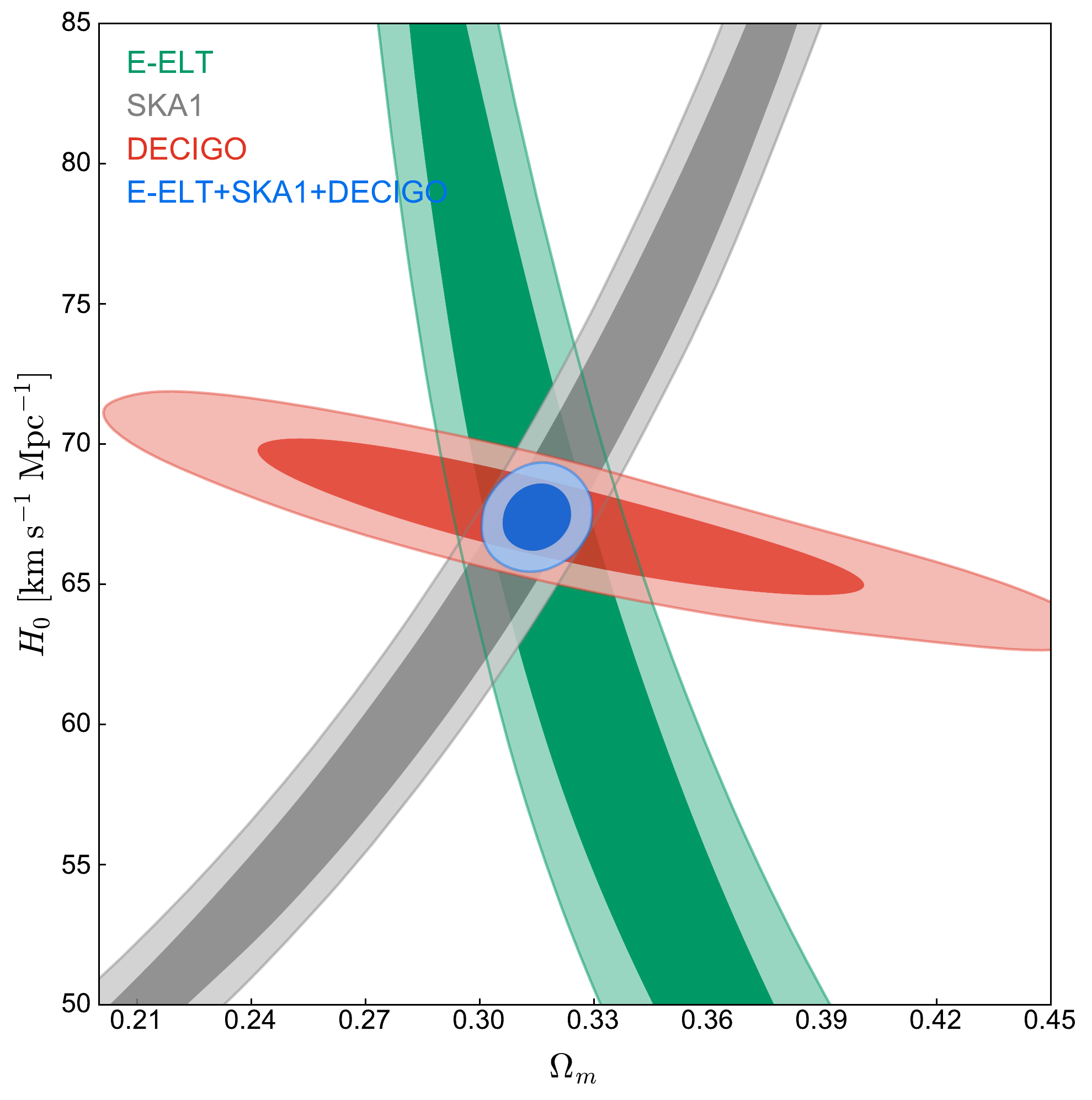}\\
\caption{Two-dimensional marginalized contours (68.3\% and 95.4\% confidence level) in the $\Omega_{\rm m}$--$H_0$ plane for the $\Lambda$CDM model from E-ELT, SKA1, DECIGO, and E-ELT+SKA1+DECIGO. Here, DECIGO represents the simulated standard sirens of BNS mergers detected by the DECihertz Interferometer Gravitational wave Observatory (1-year observation). Copied from Ref. \cite{Qi:2021iic} with permission. }
\label{fig2}
\end{figure}

$\bullet$ Actually, for the dark-energy EoS parameter, using the $d_{\rm L}-z$ relation to constrain $w(z)$ will lead to information loss due to the two integrals in the expression of distance relating to $w(z)$. In Ref. \cite{Qi:2021iic}, the authors use the multi-messenger (GW standard sirens) and multi-wavelength (optical band, ELT; radio band, SKA) observations to probe the nature of dark energy through direct measurements of the Hubble parameter $H(z)$. As shown in Figure \ref{fig2}, each of these three observations cannot constrain $H_0$ and $\Omega_{\rm m}$ well, but the combination of them can effectively break the parameter degeneracies. For the dark-energy EoS parameters in the $w$CDM and CPL models, the authors find that the joint data analysis gives better constraints than the results of CMB+BAO+SN. It is concluded that future GW observations combined with other EM observations will provide great help in revealing the nature of dark energy.

\subsubsection*{Dark sirens}

The results above are derived based on the analyses in which EM counterparts of GW sources are assumed to be directly detected. Actually, even without EM counterparts, the corresponding redshift information of GW sources can still be acquired with a galaxy catalog using Bayesian inference \cite{DelPozzo:2012zz,Chen:2017rfc,Fishbach:2018gjp,Gray:2019ksv,Soares-Santos:2019irc}. By using the BBH detections from the first and second observing runs of the Advanced LIGO--Advanced Virgo detector network, the joint standard siren--dark siren measurement on $H_0$ is improved by about 4\% over the result from GW170817 \cite{Abbott:2019yzh}. Some other GW-only standard siren analyses are also made assuming that the properties of the source population are known, such as the mass distribution of compact binaries \cite{Chernoff:1993th,Taylor:2011fs,Farr:2019twy,You:2020wju,Ezquiaga:2020tns} or the EoS of neutron star \cite{Messenger:2011gi,DelPozzo:2015bna,Wang:2020xwn}. The dark siren method can be used to examine the results from standard sirens alone and to extend the application potential of GWs in cosmology.

In \cite{Wang:2020dkc}, the authors investigated the potential of using dark sirens from the LISA-Taiji network to measure the Hubble constant. Inspired by the fact that the LISA-Taiji network can improve the localization of the GW sources significantly, which is very helpful for the identification of host galaxy and thus obtaining the redshift information of dark sirens. Considering 3 different massive black hole formation models with different black hole seedings and time delays, it is showed that within a 5-year operation time, the LISA-Taiji network is able to constrain $H_0$ to within 1\% accuracy and possibly beats the scatters down to 0.5\% or even better. 

In conclusion, the GW standard siren observation can directly measure the absolute distance, which is very important for the study of cosmology. If the EM counterparts of GW events can be identified, the $d_{\rm L}-z$ relation can be established to constrain the cosmological parameters. The multi-messenger standard sirens alone can accurately measure the Hubble constant, but they cannot provide precise measurements for other parameters, such as the EoS of dark energy. Nevertheless, according to a series of studies, we find that the combination of GW standard sirens and EM observations can effectively break the parameter degeneracies and thus greatly improve the constraint accuracies. In the next two decades, multi-band GW observations combined with the optical, near-infrared, and radio observations will usher in a new era of cosmology.

\section{GW astrophysics}\label{sec:GWastrophysics}

The detection of GWs  by ground-based GW observatories has opened a new window to probe various astrophysical GW sources. The main astrophysical GW sources include the stellar compact binaries [SCBs; double-white-dwarfs (DWDs), NS-WD binaries, BNSs, NS-BH binaries, BBHs, etc.], SMBBHs, EMRIs, and so on. GW observations can provide rich information for investigating the formation and evolution of these sources, which may not be able to obtain only by EM observations. We mainly focused on investigating the formation and evolution of these SCBs and SMBBHs across the cosmic time, the physical processes involved in their formation and evolution, the observation/detection (methods) of the EM signatures and GWs radiated from these sources, and the implications of their GW and EM observations.

We have made significant progress in various aspects of GW astrophysics in the past several years. For example, we have carefully studied the evolution of NS-WD binaries, especially on the orbital evolution of ultra-compact X-ray binaries and its GW radiation \cite{2019SSPMA..49a9501Y,Yu:2021hwl}. We have investigated in details about the EM properties of BNS and NS-BH mergers and the constraints on the merger products and equation of state (EOS) of NSs that can be obtained from both GW and EM  observations of GW170817, the first GW source with multi-wavelength EM counterparts detected \cite{Gao:2017fcu, Gao:2018tva, Ma:2017qsp, Ai:2019rre, Gao:2019vby}. We have studied the formation and evolution of stellar BBHs (sBBHs) and BNSs and its relation to the properties of the host galaxies of BBH mergers \cite{Cao:2017ztl}, and we predicted the SGWBs resulting from sBBHs and BNSs \cite{Zhao:2020iew}. We have helped to organize the Chinese PTA (CPTA) collaboration to search for nano-Hertz GW radiation and have been collecting data through FAST and other radio telescopes in China. We have performed serial work on the solar system ephemeris (SSE) by using PTA data \cite{Caballero:2018lvc, Guo:2019sbp}, which is necessary for further extract GW signals from the CPTA observations. We have investigated the detection of individual SMBBHs by current and future PTAs and its possible cosmological application to constrain cosmological parameters \cite{Feng:2020nyw, Feng:2019vnf, Yan:2019sbx}.
We have found a number of SMBBH candidates \cite{Li:2016hcm, Li:2017eqf}, investigated the detailed observational properties of some SMBBH candidates \cite{2020MNRAS.491.4023S, 2021A&A...645A..15S}, and introduced several new methods to identify BBH candidates and distinguish it from alternative interpretations \cite{ Songsheng:2019yea, 2020ApJS..247....3S, 2020A&A...643L...9W, 2021ApJ...910..101J}, investigated the cosmic evolution of SMBBHs and predicted the SGWBs resulting from inspiralling SMBBHs and SMBBH merger rate \cite{Chen:2020qlp}. 

Below we highlight some of our recent work, ranging from the evolution of NS-WD/ultra-compact binaries in the Milky Way, the detected BNS GW170817, solar system ephemeris (SSE) for PTA GW observations, GW background and SMBBH observations by future PTAs, SMBBHs searching and its PTA detection, to the estimation of SGWBs.

\subsection{Orbital evolution of NS -- Roche-Lobe filling WD binaries and their GWs}\label{subsec:NSWD}

Neutron star (NS)-white dwarf (WD) binaries may have extremely short orbital periods and radiate GWs in the frequency range of $10^{-4}-1$\,Hz. These compact binaries are therefore of interest for the GW detections of LISA \cite{Nelemans:2009hy, Nelemans:2006kv}, Taiji \cite{Ruan20}, and TianQin \cite{Wang:2019ryf}. The model of a NS accreting a Roche-lobe filling (RLF) WD companion with orbital period in the range of $\sim (10 - 80)$ minutes can well explain the observed properties of ultracompact X-ray binary stars (UCXBs), whilst the model of a NS plus a low-mass RLF He star may
also be one of the formation channels of UCXBs (e.g., \cite{2021arXiv210601369W}). Recently, UCXBs are also proposed to be dual-line GW sources for both LISA-like and LIGO-like detectors (e.g., \cite{Tauris:2018kzq,2021MNRAS.503.5495S}). Measurement of the GW signals generated by the UCXBs and NS-WD binaries will allow us to infer their orbital parameters (masses, orbital separations and eccentricities), which can provide important information and clues to understand their chemical composition, magnetic activities, tidal interaction, as well as the mass transfer process, common envelope evolution, spatial distribution and so on (e.g., \cite{Ushomirsky:2000ax,Haskell:2015psa,Tauris:2018kzq,Chen:2020wan,Yu:2021hwl}).

We studied the effects of mass transfer and GW radiation
on the orbital evolution of NS-RLF WD binaries, and the detectability of these binaries by space GW detectors (e.g., LISA; Taiji; TianQin) in a recent work \cite{Yu:2021hwl}. The GW frequency generated by a circular orbit binary, with WD mass $\sim0.05-1.4 M_{\odot}$, is in the range of $\sim0.0023-0.72$\,Hz when the Roche lobe overflow is just onset, weakly depending on the NS mass. We find that high-mass NS-WD binaries may undergo direct coalescence after unstable mass transfer if the mass of the WD component $m_{\rm wd}\gtrsim1.25~M_{\odot}$. If the WD mass $m_{\rm wd}<1.25~M_{\odot}$, NS-WD binaries may avoid direct coalescence because mass transfer after contact can lead to a reversal of the orbital evolution. The orbital evolution of the well-known UCXB source 4U\,1820--30 can be well explained in details by using the NS-RLF-WD binary model with component masses of $(1.0+0.065) M_{\odot}$. Assuming this model, the expected signal-to-noise-ratio (SNR) of 4U\,1820--30 is $\sim11.0/10.4/2.2$ by a $4$-year observations of LISA/Taiji/TianQin with  the designed sensitivities. For a $(1.4+0.5) M_{\odot}$ NS-WD binary close to contact, the expected SNR is $\sim27/40/28$ for a $7$-days observation by LISA/Taiji/TianQin. For NS-WD binaries with masses of $(1.4+\gtrsim1.1) M_{\odot}$, the significant change of GW frequencies and amplitudes may be measurable, and thus it is possible to determine the binary evolution stage. At distances up to the edge of the Galaxy ($\sim100$\,kpc), high-mass NS-WD (e.g., $\sim (1.4+0.5) M_{\odot}$) binaries will be still have SNR$\gtrsim$1. This suggests that the direct coalescence events of NS-WD binaries in the Galaxy, if any, may be detected by LISA/Taiji/TianQin in the future.

\begin{figure}
\centering
\includegraphics[width=0.8\textwidth]{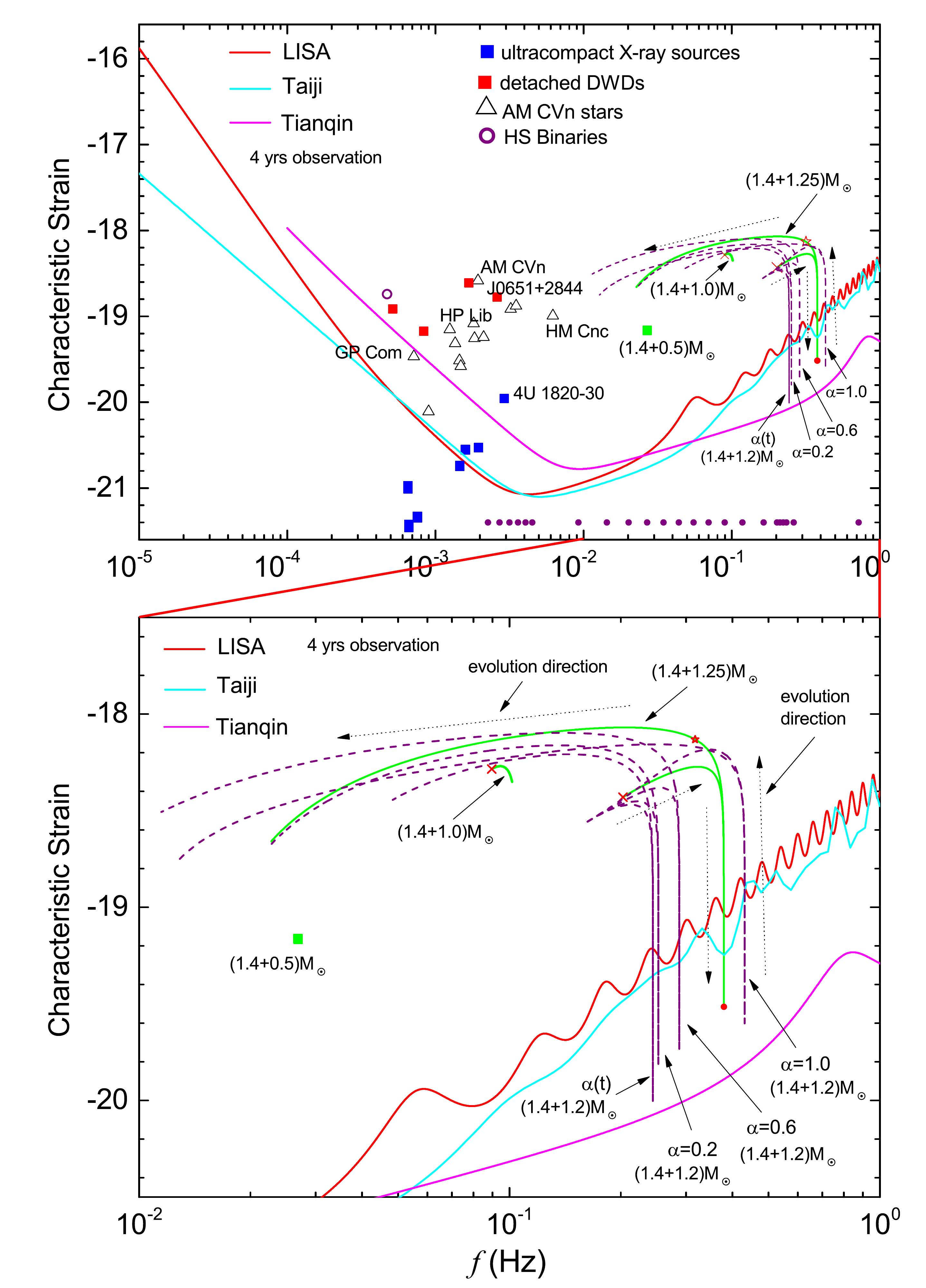}\\
\caption{Top panel: GW characteristic strains of example NS-WD binaries (assuming a $4$-year observation period). For comparison, the GW strains of the observed detached double WDs (red squares), AM CVn stars (open triangles), and hot subdwarf (HS) binaries (purple circles), are also plotted. Green squares and lines show the evolution track of NS-RLF-WD binaries with masses of $(1.4 + 0.5/1.0/1.25) M_\odot$, respectively.
The red crosses on the two green lines denote the onset of the mass transfer. Purple dashed lines show the evolution of $(1.4+1.2) M_\odot$ binary with $\alpha$ being the accretion parameter (time-dependent in our model). Red solid circle and open star mark the points where the GW frequency reaches its maximum value and the mass loss rate reaches the peak value, respectively. The sensitivity curves of LISA, Taiji, and TianQin are shown by the red, cyan, and magenta lines, respectively.
The row of purple points at the bottom denote the contact
frequencies of NS-RLF-WD binaries with different WD masses. Enlarged bottom panel: the region $f = 0.01 - 1$ (Hz) and GW strain $= -20.5 -  -17.5$ in the top panel is enlarged to show the evolution of massive NS-RLF-WD binaries more clearly. Copied from Ref. \cite{Yu:2021hwl} with permission. }
\label{fignswd}
\end{figure}

\begin{figure}
\centering
\includegraphics[width=\textwidth]{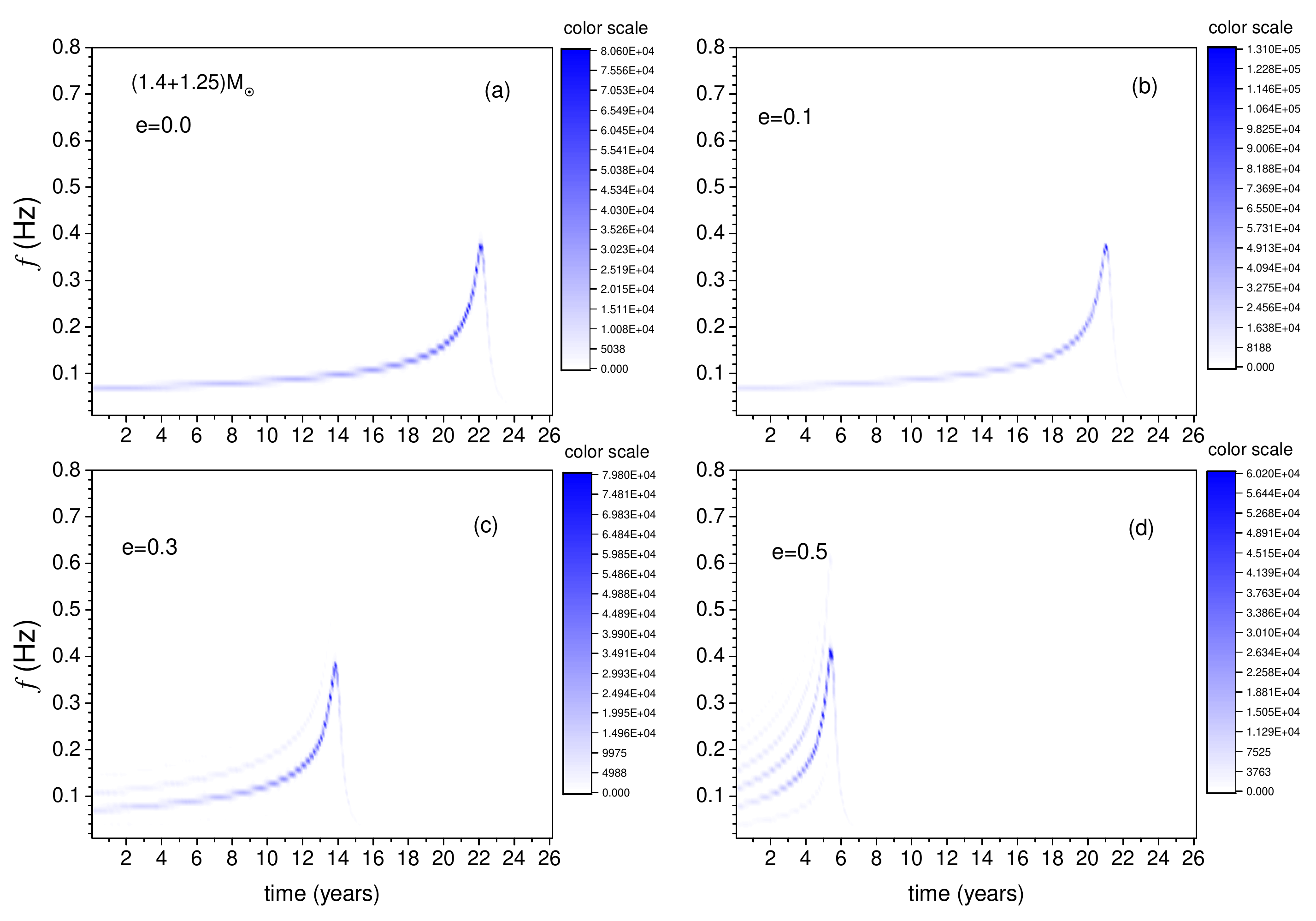}\\
\caption{The arbitrary power spectrum distribution of $(1.4+1.25){\rm M}_{\odot}$ NS-WD with eccentricity $e=0.0$, 0.1, 0.3 and 0.5 (panel (a)-(d)) as a function of GW frequency and evolution time. We here take the initial orbital period approximately 30s.}
\label{figpsd}
\end{figure}

In order to compare the GW signals from different type of sources in the Galaxy, we have also calculated the GW characteristic strain of some other UCXB sources with known distances and orbital parameters, detached double white dwarfs, AM CVn stars, and hotsubdwarf binaries, and show them Fig.~\ref{fignswd} \cite{Yu:2021hwl} with the sensitivity curves of the proposed space GW detectors LISA, Taiji, and TianQin, as well. We have also investigated the influences of eccentricities of NS-WD binaries on their GW signals. Our results indicate that the NS-WD binaries with high eccentricities evolve faster than those with low ones, and with increase of the eccentricity, the GW harmonics gradually emerge. As an example, Fig.\,\ref{figpsd} shows the evolution of the power spectrum distributions (PSD) of $(1.4+1.25){\rm M}_{\odot}$ NS-WD binaries, with eccentricities of $e=0.0$ (panel a), $0.1$ (panel b), $0.3$ (panel c), and $0.5$ (panel d), respectively, as a function of GW frequency and time, which would be compared with space-based GW observations to constrain the parameters in our models. We here take the initial orbital period approximately $30$s.

\subsection{Constraints on the neutron star maximum mass from GW170817}\label{subsec:NSmass}

Depending on the size relationship between the merger remnant (gravitational) mass and the maximum mass of a single NS, a BNS merger can give a variety of post-merger products \cite{Rezzolla:2010fd, Bartos:2012vd, Lasky:2013yaa, ravi14, Gao:2015xle}: (1) a black hole (BH) by immediate collapse after the merger; (2) a differential rotation supported hypermassive NS (HMNS), which would survive $10\sim100$ ms before collapsing; (3) a rigid rotation supported supermassive NS (SMNS), which would collapse after the NS spins down; and (4) a stable NS (SNS). Before the detection of GW170817, it has been well discussed that binary neutron star (BNS) merger events could be used to constrain the maximum mass of a non-spinning neutron star, $M_{\rm TOV}$ (see Ref. \cite{Gao:2015xle} and references therein). With the observations of GW170817, we applied some EOS-independent universal relations for rapidly spinning NS to discuss the constraints one may pose on $M_{\rm TOV}$. Here we briefly introduce our research methods and results (see Ref. \cite{Ai:2019rre} for details).

\begin{figure}
\centering
\includegraphics[width=0.9\textwidth]{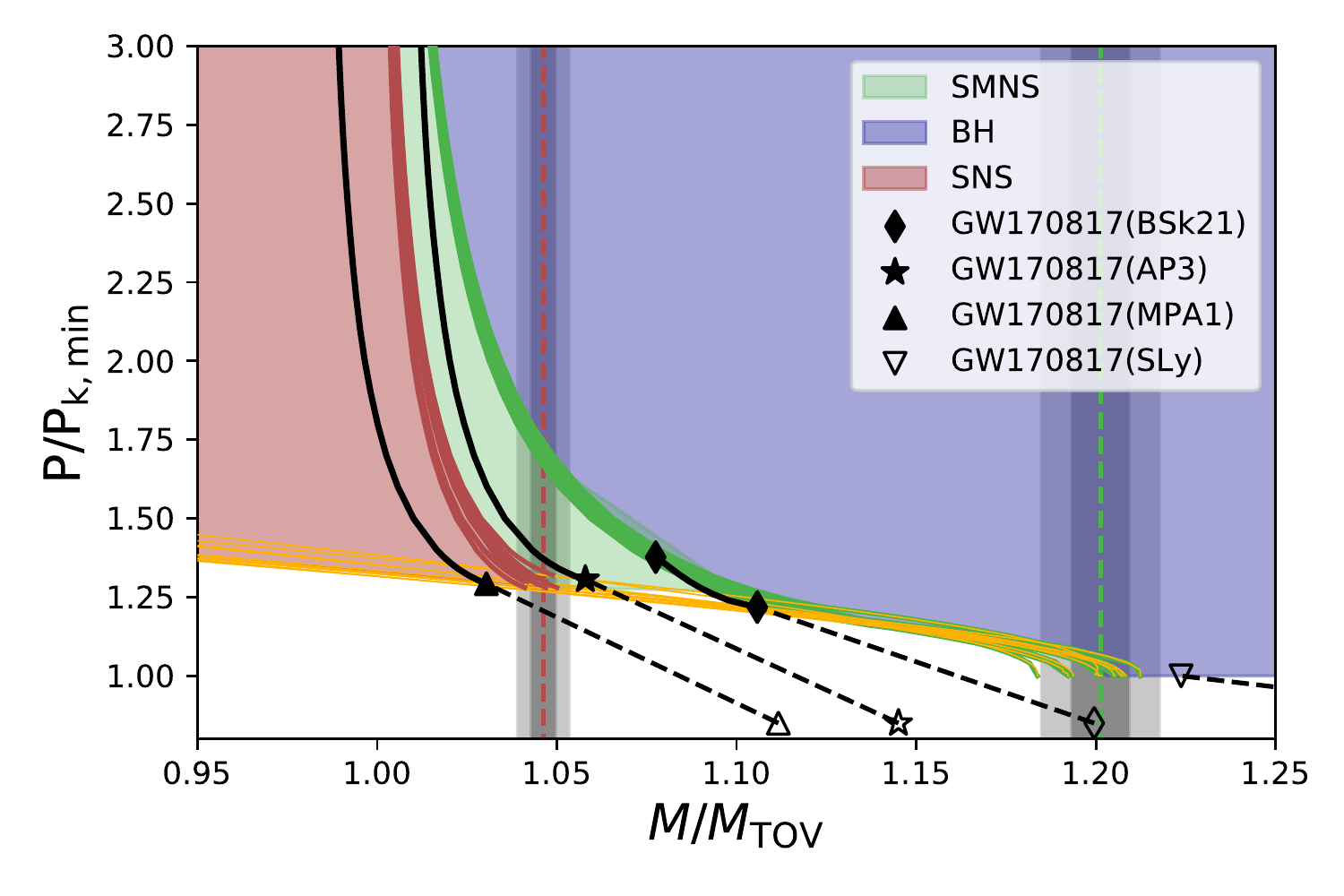}\\
\includegraphics[width=\textwidth]{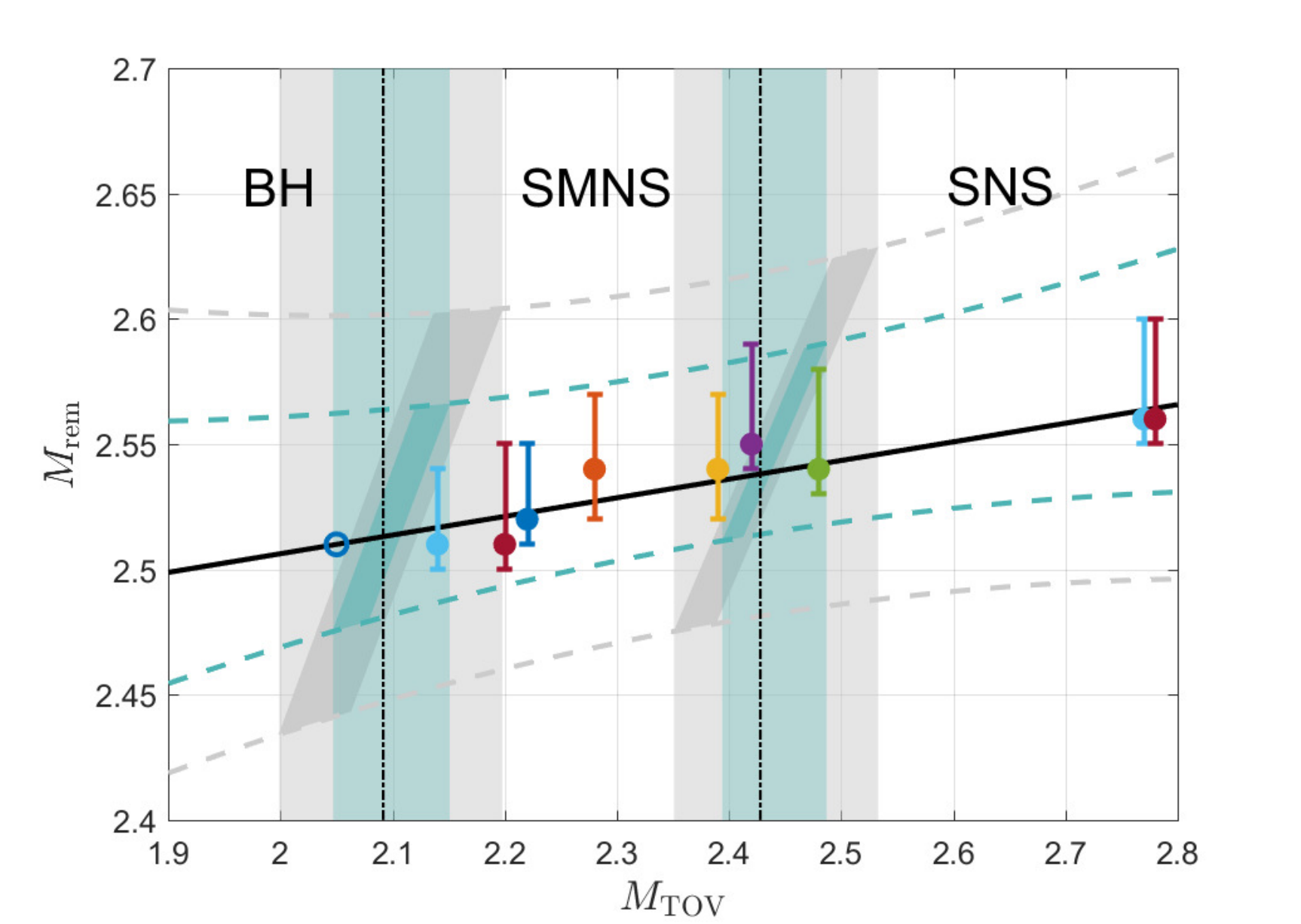}\\
\caption{Constraints on $M_{\rm TOV}$ with GW170817 observations in case of different merger products. The top panel shows the mass-dependent normalized Keplerian period ${\cal P}_k$ lines (orange), the constant $M_{\rm TOV,b}$ lines (red) and the boundary lines for SMNS collapsing into BH (green). The bottom panel shows the separation $M_{\rm TOV}$ values for different merger products (dot-dashed vertical lines) and the best fitting relation between $M_{\rm TOV}$ and $M_{\rm rem}$ (black solid line). Copied from Ref. \cite{Ai:2019rre} with permission. }
\label{fig:universal}
\end{figure}

Given an EOS, one can derive its corresponding $M_{\rm TOV}$ by solving the Tolman-Oppenheimer-Volkoff (TOV) equations \cite{Oppenheimer:1939ne}. Rotation, either rigid or differential, could enhance the gravitational mass by a certain factor $\chi = (M - M_{\rm TOV})/M_{\rm TOV}$. One can define the maximum mass of a rotating NS as
\begin{eqnarray}
M_{\rm max} & \equiv & (1+\chi_{\rm max}) M_{\rm TOV}, \label{eq:Mmax} \\
M_{\rm max,d} & \equiv & (1+\chi_{\rm d,max}) M_{\rm TOV},
\end{eqnarray}
where $\chi_{\rm max}$ and $\chi_{\rm max,d}$ are used to denote the maximum enhancement factors for uniform and differential rotations, respectively. Due to the change of rotation properties, the gravitational mass of the merger remnant will change with time. Here we define $M_{\rm rem}^0$ as the remnant mass right after the merger, $M_{\rm rem}^{\rm k}$ as the remnant mass when the NS starts to spin with a Keplerian rotation ($P=P_{\rm k}$) and $M_{\rm rem}^\infty$ as the remnant mass with no rotation ($P=\infty$). When $M_{\rm rem}^0 > M_{\rm max,d}$, the post-merger product would be a BH by immediate collapse. Otherwise if $M_{\rm max} < M_{\rm rem}^{\rm k} < M_{\rm rem}^0 \leq M_{\rm max,d}$, the post-merger product would be a HMNS and collapse into a BH after losing angular momentum as well as mass. If, however, $M_{\rm rem}^{\rm k} \leq M_{\rm max}$, the merger remnant would first go through a HMNS phase and then form a uniformly rotating NS. In this case, when $M_{\rm rem}^\infty > M_{\rm TOV}$, the merger remnant would eventually collapse into a BH after the NS spins down; when $M_{\rm rem}^\infty \leq M_{\rm TOV}$, the remnant would never collapse. The total baryonic mass of the system should be conserved during the merger. We denote the baryonic mass of the remnant as $M_b$ and the maximum baryonic mass for non-rotational NSs predicted by an EOS as $M_{\rm b,TOV}$.  In the uniformly rotating phase, we define $\chi_{\rm col}$ and $\chi_{\rm TOV}$ as the enhancement factors when SMNS collapses to BH and when $M_{\rm b} = M_{\rm b,TOV}$. Using \textit{RNS} code, we find that, for selected EoSs with $2.08M_{\odot}<M_{\rm TOV}<2.78M_{\odot}$ (\textit{SLy \cite{Douchin:2001sv}, WFF1 \cite{Wiringa:1988tp}, WFF2 \cite{Wiringa:1988tp}, AP4 \cite{Akmal:1997ft}, BSk21 \cite{goriely10}, AP3 \cite{Akmal:1997ft}, DD2 \cite{Typel:2009sy}, MPA1 \cite{Muther:1987xaa}, Ms1 \cite{Mueller:1996pm}, Ms1b \cite{Mueller:1996pm}}), $\chi_{\rm col}$ and $\chi_{\rm TOV}$ as functions of $P$, are almost independent on the EOS adopted, which could be expressed as, 
\begin{equation}
\log \chi_{\rm col} = (-2.740\pm0.045) \log {\cal P} + {\log (0.201 \pm 0.005)},\label{eq:chi_ucol}
\end{equation}

\begin{align}
\log \chi_{\rm TOV}= &(1.804\pm0.268)(\log{\cal P})^2 + (-3.661\pm0.190) {\log{\cal P}} \nonumber\\
&+ {\log(0.101\pm0.007)}, ~P>P_{\rm k,TOV},
\end{align}
where ${\cal P} = P/P_{\rm k, min}$ and $P_{\rm k,min}$ is the minimum Kepler period for a uniformly rotating NS. When $P = P_k$, we have $\chi_{\rm col}^{\rm k} = \chi_{\rm max} = 0.201 \pm 0.017 (0.008)$  and $\chi_{\rm TOV}^{\rm k}=0.046 \pm 0.008 (\pm 0.004)$  with the errors indicating 2$\sigma$ (1$\sigma$) confidence level, respectively. The enhancement factor of a SMNS at the time when differential rotation damps ($\chi^{\rm k}$) must satisfy $\chi_{\rm TOV}^{\rm k}<\chi^{\rm k}< \chi_{\rm max}$, as shown in Figure \ref{fig:universal} (see also figures 1-2 in Ref. \cite{Ai:2019rre}).

For GW170817, the total gravitational mass of the system at infinite binary separation is estimated as $2.74^{+0.04}_{-0.01}M_{\odot}$, and the mass ratio is constrained to the range of $(0.7-1)$ \cite{TheLIGOScientific:2017qsa}. Assuming non-spinning progenitors with equal mass, convert the gravitational mass for each NS ($1.37M_{\odot}$) to baryonic mass and add them together, then subtract the mass being ejected ($M_{\rm ejc} = 0.06 \pm 0.01$) during merger. Finally, convert the baryonic mass of the remnant back to the gravitational mass. The relationship between $M_{\rm rem}^{\rm k}$ and $M_{\rm TOV}$ are shown in Figure~\ref{fig:universal}, which reads (see also Ref. \cite{Gao:2019vby})
\begin{eqnarray}
M_{\rm rem}^{k}=(2.354 \pm 0.074)+(0.076\pm 0.032)M_{\rm TOV}, \label{eq:Mg}
\end{eqnarray}
with the errors indicating 2$\sigma$ confidence level. Comparing $M_{\rm rem}^{\rm k}$, as well $M_{\rm b}$, with the critical masses, one can determine what kind of remnant was produced. In reverse, one can put constrains on EOSs. With the universal relations shown above, we reach to the results: If GW170817 produces a short-lived HMNS, one has $M_{\rm TOV}<2.09^{+0.11}_{-0.09}(^{+0.06}_{-0.04})M_{\odot}$; If GW170817 produces a long-lived SMNS, the constraint should be $2.09^{+0.11}_{-0.09}(^{+0.06}_{-0.04})M_{\odot} \leq M_{\rm TOV}<2.43^{+0.10}_{-0.08}(^{+0.06}_{-0.04})M_{\odot}$; If GW170817 produces a stable NS, the constraints should be $M_{\rm TOV} \geq 2.43^{+0.10}_{-0.08}(^{+0.06}_{-0.04})M_{\odot}$. The quoted uncertainties are at the 2$\sigma$ (1$\sigma$) confidence level.

\subsection{Identifying supermassive binary black holes}\label{subsec:SMBBHs}

Identifying SMBBHs with separations less than a few pc is crucial for a broad range of topics in contemporary astrophysics, including the growth and evolution of SMBBHs, galaxy mergers, nano-Hertz GWs, etc. A large number of possible SMBBH candidates have been discovered through periodic variability from time-domain surveys (e.g., \cite{Graham:2015tba, Charisi:2016fqw, Liu:2016msr}) and serendipitous exploration over historical data compilations (e.g., \cite{Valtonen:2008tx, Bon:2016jtk, Li:2016hcm, Li:2017eqf}), or through other approaches with predicted signatures of SMBBHs (e.g., \cite{Liu:2014mga, Yan:2015mya}; see also \cite{WangJM2020b} for a review). However, there is currently no exclusively confirmed SMBBH because of the challenges in establishing effective criteria before the detection of nano-Hertz GW from SMBBHs via PTAs. In this sense, there is no doubt that developing new approaches for identifying SMBBH candidates is of great importance.

\begin{figure}
\centering
\includegraphics[width=\textwidth]{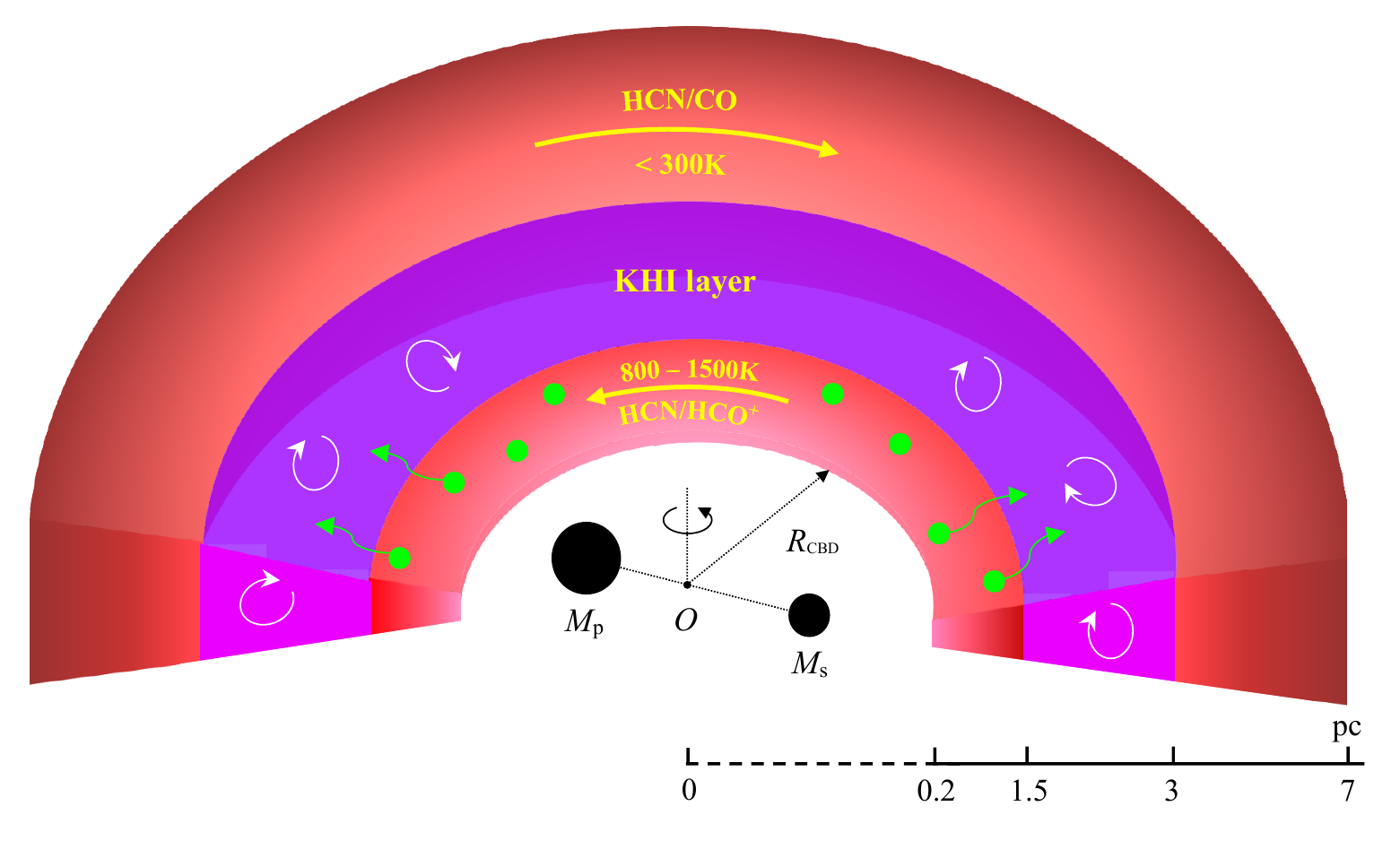}\\
\caption{Cartoon of a CB-SMBH maintaining one circumbinary disc (CBD) composed of the prograde ($R\lesssim1.5$\,pc) and the retrograde ($3{\rm pc}\lesssim R \lesssim 7{\rm pc}$ parts in NGC 1068. Copied from Ref. \cite{WangJM2020} with permission. }
\label{fig_1068}
\end{figure}

In a recent work \cite{WangJM2020}, we illustrated that a SMBBH provides necessary tidal torques to prevent the counter-rotating disk between $0.2$ and $7$\,pc detected in the nucleus of the nearby galaxy NGC 1068 (see Figure~\ref{fig_1068}) from the Helvin-Helmholtz instability, which can result in a catastrophe at the interface between the reversely rotating parts. In other words, counter-rotating disks can avoid the Helvin-Helmholtz instability in the presence of SMBBHs that supply angular momentum to the disks. Meanwhile, such angular momentum transfer can also efficiently remove the SMBBH's orbital angular momentum and expedite the orbit hardening, providing an alternative paradigm for solving the ``final pc problem''. For NGC 1068, we estimated the total  mass of the SMBBH to be $1.3\times10^7M_\odot$, mass ratio to be  $\gtrsim0.3$ and orbit separation to be $\sim 0.1$\,pc (see \cite{WangJM2020} for the detail). With these fiducial estimates, the SMBBH is radiating GWs at frequency $f\approx0.1$\,nHz with an intrinsic strain of $h_{\rm s} = 5\times10^{-19}$. The GW backgrounds emitted from NGC 1068-like SMBBHs are likely to be detected through the forthcoming PTA observations of the Square Kilometer Array (\cite{Barack:2018yly}). The putative SMBBH in NGC 1068 has an angular size of $2$\,mas, which could be resolved by the Event Horizon Telescopes if both the black holes radiate radio emissions. Such an identification basis as in NGC 1068, that is, counter-rotating disks, can serve as a new strategy to search for SMBBH candidates in nearby galaxies. A thorough high-resolution survey with Atacama Large Millimeter/Submillimeter Array (ALMA) over the nuclei of nearby galaxies is therefore highly worthwhile to find out those NGC 1068-like counter-rotating disks and identify the SMBBH candidates residing therein.

\subsection{GW background from stellar compact binaries}\label{subsec:SCB}

There are numerous sBBHs and BNSs inspiralling and merging across the cosmic time as predicted by theoretical models \cite{Belczynski:2001uc} and supported by ground-based GW observations \cite{Abbott:2016blz, TheLIGOScientific:2017qsa, Abbott:2020gyp}. LIGO and Virgo have put strong observational constraints on the local merger rate densities of sBBHs and BNSs after the observing run O3a, i.e., $19.1^{+15.9}_{-9.1}\,{\rm Gpc^{-1} yr^{-1}}$ for sBBHs and $320^{+490}_{-240}\,{\rm Gpc^{-1} yr^{-1}}$ for BNSs \cite{Abbott:2020gyp}. Currently, there are several different models for the origin of the sBBHs, the leading ones are: 1) sBBHs are the final products of the evolution of massive binary stars (hereafter denoted as the EMBS channel) \cite{Belczynski:2005mr}, 2) sBBHs are originated from the dynamical interactions in dense stellar systems, including globular cluster and galactic nuclei (hereafter denoted as the dynamical channel) \cite{Rodriguez:2016kxx}. The dynamical channel becomes even more important with the recent discovery of the most massive sBBHs GW190521 \cite{Romero-Shaw:2020thy, Fragione:2020han}, and it is anticipated that the contribution from the dynamical channel to the formation of sBBHs is significant or even dominant. The dynamical channel may produce sBBHs with extremely high eccentricities (close to 1), while the EMBS channel produces sBBHs with small eccentricities. At the low frequency range ($10^{-4}-10^{-2}$\,Hz), many inspiralling sBBHs may have significant eccentricities, if originated from the dynamical channel, because their orbits cannot be immediately circularized at this frequency range \cite{Zhao:2020iew}. BNSs are mostly, if not all, formed via the evolution of binary massive stars and it may have small eccentricities in the frequency range $>10^{-4}$\,Hz.

\begin{figure}
\centering
\includegraphics[width=0.8\textwidth]{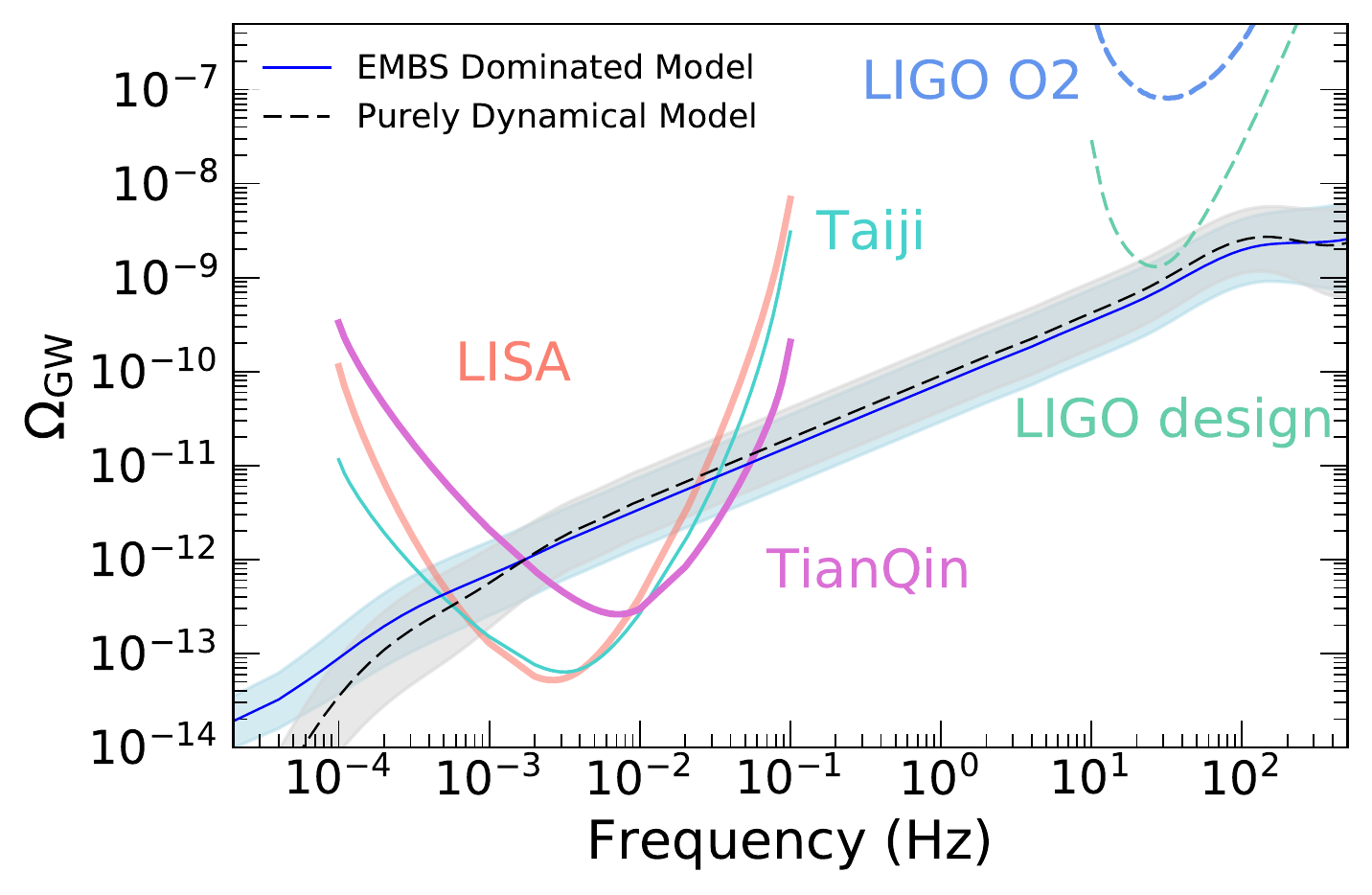}\\
\caption{The energy density spectrum of stochastic SGWB from sBBHs and BNSs. Blue solid and black dashed lines represent the total $\Omega_{\rm GW}$ resulting from the models with sBBHs formed mainly from the EMBS channel and the purely dynamical channel, respectively. Cyan solid, red solid, magenta solid, blue dashed, and green dashed curves show the sensitivity curves of Taiji, LISA, TianQin, LIGO O2, and LIGO design, respectively. Copied from Ref. \cite{Zhao:2020iew} with permission. }
\label{fig:fgwb}
\end{figure}

GWs radiated from cosmic sBBHs and BNSs form a SGWB at frequencies spanning $10^{-4}-1000$\,Hz). This SGWB is the main target for both the ground-based GW detectors (LIGO/VIGO/KAGRA/ET/CE) with working frequency at $10-1000$\,Hz and the space GW detectors (LISA, Taiji, and TianQin) with working frequency at $10^{-4}-1$\,Hz. GWs from eccentric binaries may result in detectable signatures in the SGWB.
We have considered various models for the sBBH formation and the BNS formation to predict the SGWB and estimate the SNR of this SGWB that may be detected by space GW detectors (LISA/Taiji/TianQin) and ground-based GW detector LIGO. Here for simplicity, we show the results from two  models for sBBHs: 1) the EMBS-origin-dominant model: we assume $75\%$ of cosmic sBBHs are formed from the EMBS channel and the rest $25\%$ are originated from the dynamical channel; 2) the dynamical-origin-only model: all cosmic sBBHs are from the dynamical channel.
Figure~\ref{fig:fgwb} shows the resulting SGWB energy density spectrum due to sBBHs and BNSs from EMBS-origin-dominant model and dynamical-origin-only model, respectively \cite[c.f.][]{Zhao:2020iew}. The total SGWB energy density at $25$\,Hz is predicted to be $\Omega_{\rm GW} = 6.57^{+7.65}_{-3.95}\times 10^{-10}$ and ${8.17^{+8.98}_{-4.71} \times10^{-10}}$ for the EMBS-origin-dominant and dynamical-origin-only model, respectively. Note that the values of $\Omega_{\rm GW}$ obtained here are substantially smaller than those estimated in \cite{TheLIGOScientific:2016wyq, Abbott:2017xzg, Zhao:2020iew} because the local sBBH and BNS merger rate densities are re-scaled to the LIGO/VIRGO O3a constraints \cite{Abbott:2020gyp} but not the LIGO/VIRGO previous constraints.  As seen from figure~\ref{fig:fgwb}, the SGWB spectrum deviates from a single power law with the canonical slope of $2/3$ at frequency  $\lesssim 10^{-3}$\,Hz because of the contribution from highly eccentric sBBHs originated from the dynamical channel. The deviation is quite significant in the dynamical-origin-only model and the SGWB may be better described by a broken power law. We suggest that the future detection of SGWB and its shape at low frequency may put a strong constraint on the origin of sBBHs and distinguish different models for sBBHs.

The SGWB signal from inspiralling sBBHs and BNSs can be detected by LISA, Taiji, and TianQin, but may be difficult for LIGO to detect it with the designed sensitivity. Assuming an observation time period of $4$ years, we predicted that the SGWB is expected to be detected by LISA, Taiji, and TianQin with SNRs of $92^{+109}_{-56}$, $86^{+102}_{-52}$, and $7.29^{+8.61}_{-4.42}$, respectively, for the EMBS dominated model. For the SGWB from sBBHs alone, the expected SNRs for LISA, Taiji, and TianQin are $46^{+39}_{-22}$, $44^{+36}_{-21}$, and $3.83^{+3.19}_{-1.83}$, respectively. For the model with all sBBHs originated from the dynamical channel, the estimates of SNR of total SGWB detected by LISA, Taiji, and TianQin are $104^{+119}_{-62}$, $99^{+114}_{-58}$, and $8.81^{+9.86}_{-5.14}$, respectively. The SGWB from sBBHs detected by LISA, Taiji, and TianQin are $59^{+49}_{-28}$, $58^{+48}_{-28}$, and $8.81^{+9.86}_{-5.14}$, respectively. We also estimated that the SNR of the SGWB monitored by LIGO with the designed sensitivity over an observation period of $4$\,years would be $1.59^{+1.84}_{-0.95}$ and $2.00^{+2.17}_{-1.14}$ for the EMBS-origin-dominant model and the dynamical-origin-only model, respectively. This apparently suggests that the detection of the SGWB for LIGO needs to wait for many more years, different from previous expectation \cite{TheLIGOScientific:2016wyq, Abbott:2017xzg}.

\subsection{PTA probing solar system dynamics and constructing planet ephemeris}\label{subsec:PTA1}

Pulsars, particularly the millisecond pulsars (MSPs), are of high rotational stability, which is comparable to that of time-keeping atomic clock. The stability had been revealed and confirmed using the pulsar timing method, where the difference (called timing residuals) between the measured time of arrival of pulsed signals and a low-order polynomial predication can reach sub-$100$\,ns level.  By monitoring multiple MSPs, one can study the correlation between timing residuals of pulsar pairs. The correlated common signal provides opportunities to probe fundamental physics, which includes detecting nano-Hertz GWs \cite{HD83}, investigates the stability of reference terrestrial time standards \cite{Hobbs:2019ktp}, and studies the Solar System dynamics \cite{Champion:2010zz}. These applications make use of so-called PTAs, which are an ensemble of pulsars, typically millisecond pulsars, in different sky positions \cite{FB90}.

One of the noise sources preventing from successful GW detection using PTA is the uncertainties in modeling the Solar system dynamics. It introduces correlated signals in PTA data, and can mimic, to a certain level, the properties of GW. The error in the Solar System ephemeris (SSE) will lead to dipolar correlations in the residuals of pulsar timing data for widely separated pulsars. We have performed serial work in order to solve the SSE problem. As a first step, we tackled the problem with a perturbative approach \cite{Guo:2018rpw}. The SSE error induces a dipole correlation. We utilized the signals and constructed a Bayesian data-analysis framework to detect the unknown mass in the Solar System and to measure the orbital parameters. In this way, we can probe if there is any unknown point mass in our Solar system \cite{Caballero:2018lvc}.  We expect that the future PTA data can limit the unknown massive objects in the Solar System to be lighter than $10^{-11}$ to $10^{-12} M_{\odot}$, or measure the mass of Jovian system to fractional precision of $10^{-8}-10^{-9}$.  Using the data from international PTA, we also measured the mass of planets and heavy asteroids using the perturbative method. Then we started to look into the nonlinear aspect of Solar system dynamics. To better understand the effects on pulsar timing caused by the uncertainties of SSE, we implemented the fully dynamical model of the Solar system \cite{Guo:2019sbp}, based on the SSE of Guangyu Li's group.  Under the same initial condition, we demonstrated that the planetary positions and velocities are compatible with DE435 at centimeter and $10^{-4}$\,cm/s level over a $20$-year timespan.  We noted that the dominant effects on the inner and outer planets are different between the perturbative and full-dynamic models.  For the outer planets, the timing residuals are dominated by the SSE shift, the two models produce similar results. However, for inner planets, the variations in the orbit of the Earth are more prominent, which makes the leading-order approaches insufficient and leads to larger effects on pulsar timing.  Furthermore, the power spectrum of planet ephemeris induced signal is much more complex than simple harmonics assumed before. Armed with the full dynamic model of Solar system dynamics, we studied \cite{Guo:2019sbp} 1) how to mitigate the Solar system noise in pulsar timing data processing, and 2) how to constrain the Solar system model itself by using PTA data. We are now able to systematically account for the effects of errors in the orbital elements, which making our search for GWs, errors in SSE parameters and planetary mass constraints, more robust. Because most of planet ephemeris are closed-source, our work in Solar system model becomes one of the major support for our independent PTA data analysis in the future.

\subsection{Detecting GWs using PTAs in the SKA era}\label{subsec:PTA2}

PTA is the most promising experiment to open the very low frequency window ($\sim 1-100$ nHz) of GW astronomy. 
There are three major regional PTA consortia that have been in operation for more than a decade: NANOGrav, the European Pulsar Timing Array (EPTA), and the Parkes Pulsar Timing Array (PPTA), which are currently monitoring 47 \cite{Alam:2020laa}, 42 \cite{Desvignes:2016yex} and 26 \cite{Kerr:2020qdo} millisecond pulsars, respectively. The International Pulsar Timing Array (IPTA), as the umbrella of the three PTAs, contains 65 pulsars in its most recent data release (DR2) \cite{Perera:2019sca}. Meanwhile, the Chinese PTA (CPTA) \cite{2016ASPC..502...19L} and Indian PTA (InPTA) \cite{Joshi:2018ogr} are both in rapid development and will join in the IPTA's effort in detecting GWs.

In the past several years, PTAs have already put astrophysical meaningful constraints on the stochastic GW background \cite{Lentati:2015qwp, Shannon:2015ect, Arzoumanian:2018saf}, continuous wave signals from resolvable SMBBHs \cite{Zhu:2014rta, Babak:2015lua, Aggarwal:2018mgp}, and bursts with memory \cite{Wang:2014zls, Aggarwal:2019ypr}. Recently, based on the $12.5$-yr data collected between 2004 and 2017, the NANOGrav has found strong evidence of a common-spectrum stochastic process with a median characteristic GW strain amplitude of $1.92\times 10^{-15}$ at $f=$1/yr for the fiducial $f^{-2/3}$ spectrum. However, no statistically significant evidence has been found for the inter-pulsar quadruple spatial correlation (Hellings-Downs curve \cite{Hellings:1983fr}) of the timing residuals induced by the GW from an unresolved SMBBH population \cite{Arzoumanian:2020vkk}. This result has been confirmed by the other PTAs.

While the current PTAs are on the verge of making a first detection of the stochastic GW background in the coming years, this nascent GW astronomy in the very low frequency will experience a leap with the next generation large-scale radio telescopes, namely FAST \cite{Hobbs:2014tqa} and SKA \cite{Janssen:2014dka}, which will increase the number of the pulsar timed with a precision of $\lesssim 100$~ns to $O(10^3)$ \cite{Smits:2008cf} and allow us to observe GWs generated by individual SMBBHs, early cosmic phase transition, cosmic string decay and primary black holes \cite{Burke-Spolaor:2018bvk}. These anticipated discoveries will have a profound impact on our understanding of the evolution of the early universe, large-scale structure and galaxy formation,  and fundamental theory of gravitation.

With the growing timing precision and the number of pulsars, the data analysis challenges in PTA also become more difficult. For example, the volume of the search space is enlarged exponentially with increasing number of pulsars, $N_{\rm p}$, as there are $N_{\rm p}$ unknown pulsar phase parameters, brought by the so-called pulsar terms in the GW-induced timing residual signal.  Using fully coherent methods that can handle these phase parameters semi-analytically \cite{Wang:2015nqa, Wang:2017jxw}, we have assessed the performance of the SKA-era PTA with $10^3$ pulsars timed to $100$\,ns level \cite{Wang:2016tfy}. Our work shows that, for the SKA-era PTA, the sky-averaged upper limit on GW strain amplitude will be
\begin{equation}
\label{eq:rho2hG}
h= 5.2 \times 10^{-16} \times \left(\frac{f_{\text{gw}}}{2\times10^{-8}~\text{Hz}}\right)  \,,
\end{equation}
if the network SNR $\rho = 30$ is adopted as a detection threshold. At the frequency $f_{\text{gw}}=2\times 10^{-8}$~Hz, $h=5.2 \times 10^{-16}$ which is about 2 orders of magnitude improvement over the existing limits. Given a redshifted chirp mass of $4\times 10^{9}~M_{\odot}$ ($4\times 10^{10}~M_{\odot}$), the SMBBH will be visible out to the cosmological redshift $z\approx 1$ (28). With this distance reach, some of the SMBBH candidates that are found by the electromagnetic observations, such as PG 1302-102 and PSO J334+01, can be either confirmed or ruled out. In addition, our work unveiled the relationship between the parameter estimation uncertainty and the condition number of the response matrix in the signal model. The latter indicates the ill-posedness that is inherent in coherent GW data analysis as is already known for ground-based detector networks \cite{Klimenko:2005xv, Mohanty:2006ha, Rakhmanov:2006qm}. For example, the estimated right ascension and declination of the sources at locations with increasing condition numbers tend to have larger variance and bias for a given $\rho~ (\neq 0)$. For the noise-only case ($\rho = 0$), the estimated locations of sources are attracted towards the Galactic North and South poles where the condition numbers approach unity.


For the resolvable sources, we have demonstrated that the high frequency reach of PTAs is not limited by the Nyquist frequency of single pulsar observations \cite{Wang:2020hfh}. Actually, one can take advantage of asynchronicity, a feature inherent in PTA, to reconstruct the high frequency component of the GW signals that is preserved in the data due to aliasing in the observation sequences of an array of pulsars. Using asynchronous observations which we call \textit{staggered sampling}, one can effectively extend the  GW search frequency range by a factor of up to $N_{\rm p}$ without increasing the total allocated time for pulsar timing observations and the average observation cadence per pulsar. Given the typical average observation cadence of 1/(2 weeks), the staggered sampling can increase the Nyquist frequency from $4 \times 10^{-7}$~Hz to $2 \times 10^{-5}$~Hz for the existing PTAs with about $50$ pulsars and $4 \times 10^{-4}$~Hz for the SKA-era PTA with $10^3$ pulsars \cite{Wang:2020hfh}. This will bridge the $\mu$Hz band between the conventional PTAs and the space-borne interferometric detectors, such as LISA \cite{Audley:2017drz}, TianQin \cite{Luo:2015ght} and Taiji \cite{Gong:2014mca}. Taking the SKA-era PTA as an example, we have shown the significant astrophysical implications in the light of this frequency increase: (1) the GW strain upper limit in [10, 400]$\mu$Hz will be improved by around 3 orders of magnitude over the current high-cadence experiments \cite{Dolch:2015zba}; (2) PTA will not only sensitive to the GWs from SMBBHs in the early inspiral phase, but also the more dynamic merger and ringdown phases, which can be used to test GR with high precision. For example, the no-hair theorem can be tested to $\approx 2\%$ level, compared to the $\approx 10\%$ level archived by LIGO \cite{Isi:2019aib}; and (3) measuring the Hubble constant by using only GW observations (no need for an electromagnetic counterpart). This is realized by measuring the additional timing residuals rooted in the curvature of the GW wavefront and inferring the co-moving distance of the source $D_c$. $D_c = D_L/(1+z)$, where $D_L$ is the luminosity distance obtained from GW signal simultaneously. From the Fresnel criterion, it follows that this effect will manifest itself more clearly at higher frequencies \cite{DOrazio:2020tne}.

\subsection{Forecasting SKA-PTA detection of individual SMBBHs}\label{subsec:PTA3}

\begin{figure}
\centering
\includegraphics[width=\textwidth]{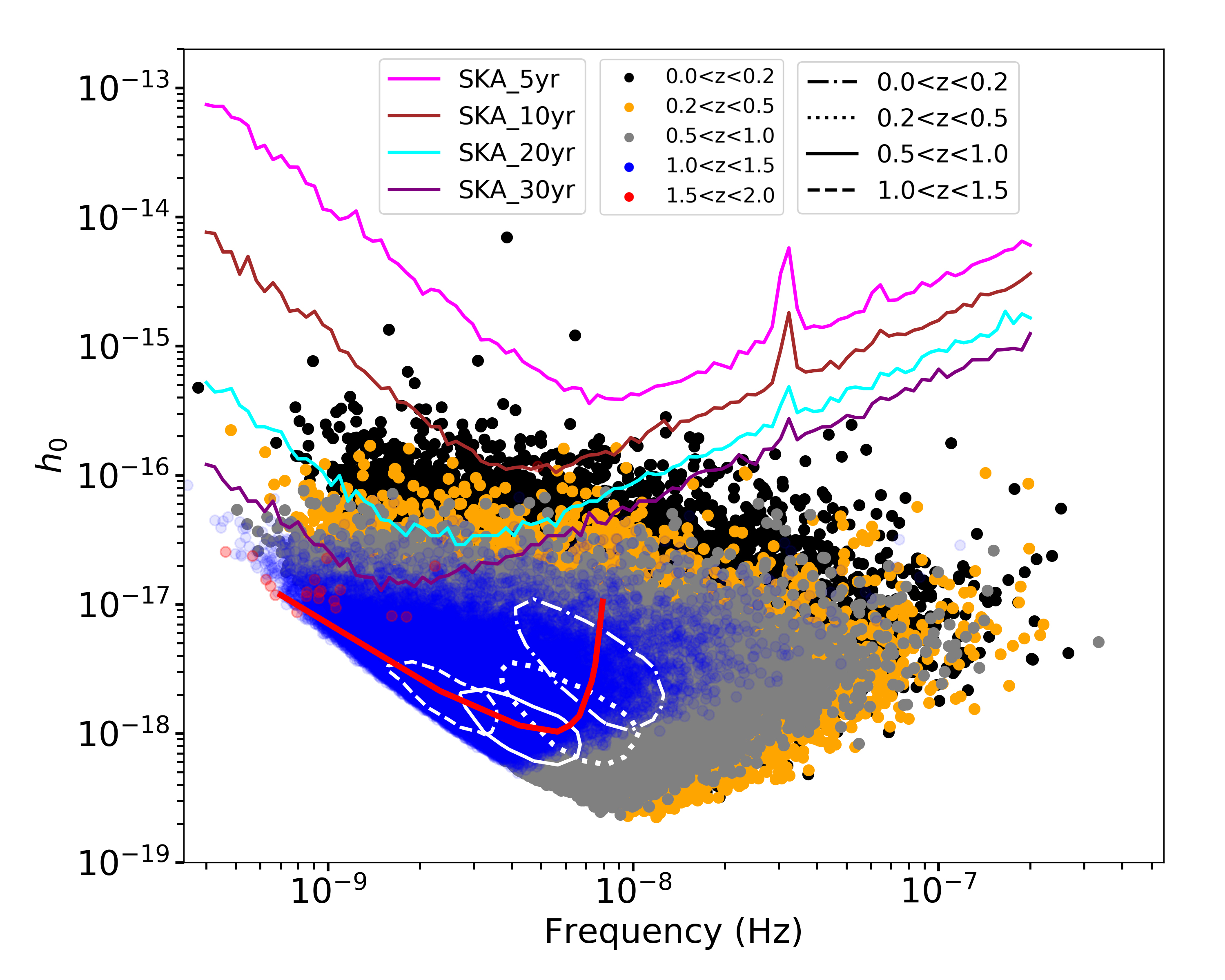}\\
\caption{Colored curves show the detection threshold for SKA-PTA at different year mark from the initiation of its operation . Black, orange, gray, blue, red dots represent SMBBH population hosted by $10^6$ galaxies from $0.0<z<0.2$, $0.2<z<0.5$, $0.5<z<1.0$, $1.0<z<1.5$, $1.5<z<2.0$ respectively. Dash-dotted, dotted, solid, dashed number density contours represent 50\% of the peak value for $0.0<z<0.2$, $0.2<z<0.5$, $0.5<z<1.0$, $1.0<z<1.5$ respectively. The red curve crossing the contour centers shows the evolution trend of SMBBH population from low to high redshifts: the GW frequencies increase; the GW strains decrease first and then increase. Copied from Ref. \cite{Feng:2020nyw} with permission. }
\label{fig:ska}
\end{figure}

As mentioned before, PTA is not only aiming at detecting the stochastic GW background from numerous cosmic SMBBHs but also individual SMBBHs in nearby and faraway Universe. We have quantified the potential of detecting GWs radiated from individual SMBBHs by the SKA-PTA \cite{Feng:2020nyw}. Our calculations demonstrate, for the first time, that even a small number (about 20) of high-quality MSPs monitored by SKA will deliver valuable information about the redshift evolution of SMBBHs. Based on infrared galaxy samples and statistical assumptions of the SMBBH population, we carried out a semi-analytical numerical simulation to estimate the number of detectable SMBBHs with SKA-PTA. Different from the hundreds of pulsars commonly assumed to be necessary in previous work, the new calculations demonstrate that a SKA-PTA consisting of merely $\sim$20 pulsars is capable of  detecting single GW sources within $5$\,years and approaching a  $\sim 100$ SMBBHs/yr detection rate within $10$\,years. A 30-year SKA-PTA operation will detect about $60$ individual SMBBHs with $z<0.05$ and more than $10^4$ within $z<1$ (Fig.~\ref{fig:ska}). With such a substantial number of expected detections, SKA-PTA
will open a new window into the SMBBHs, their host galaxies, and the evolution of the SMBBH population through redshifts.

More than $10^4$ SMBBHs detected in the Square Kilometre Array era (in an optimistic case) are a treasure trove for nano-Hertz multi-messenger astronomy. We are preparing for such an amazing party:
\begin{itemize}
\item Use available IPTA data to get a good understanding of PTA data and carry out PTA sciences. We investigated the long-term timing observations of the pulsar PSR~J1909$-$3744 from the Parkes radio telescope and managed to derive the most stringent  constraints to date on the chirp masses of a set of  SMBBH candidates \cite{Feng:2019vnf}. The previous limits are 2 to 7 times larger than our results. The estimated detection thresholds are still much larger than expected chirp masses, with that of 3C 66B being the closest at about 3 times more massive. Our analysis also demonstrated that, unlike the stochastic GW background,  the detection of single GW sources can be impervious to uncertainties in the solar system ephemeris and can benefit significantly from a priori knowledge of the SMBBH orbits.

\item As part of the CPTA efforts, we investigated methods to improve timing precision. With FAST, We found multiple jitter modes in PSR~J1022+1001\cite{2021ApJ...908..105F}.  This marks the initial step of understanding single pulse behaviours of PTA MSPs with FAST and the FAST's potential to better reveal GW \cite{Hobbs:2014tqa, li19}.
\item We developed machine learning procedures for classification and detection of GW signals from specific SMBBHs in simulated PTA data sets. Our convolutional neural network achieves high accuracy when the combined SNR is $>1.33$ \cite{Li:2020kgf}. Due to the lack of confirmed SMBBH sources and their weak estimated strain, the mainstream studies on PTAs have been focusing on the stochastic background GW, which is essentially a noise term.  Even when these PTAs are detected, they will provide only limited, congregated astrophysical information. Preparation for LIGO type individual source-detection pipeline is thus necessary for GW astrophysics using PTAs.
\end{itemize}

\section{Numerical relativity and gravitational waveform template}\label{sec:NR}

Compact binary coalescences are the most important and promising GW sources for both the ground-based and space-based GW detectors. In the past years, LIGO and Virgo have completed three observation runs. And more than 50 GW events have been announced. All of these events and other GW event candidates are compact binary coalescences.

Due to the weakness of the GW strain, the detected data by the GW detectors are typically weak signals hidden in strong noises. Consequently, a special data analysis technique, matched filtering, is required to dig out the weak GW signal. In order to let the matched filtering scheme work, accurate and complete waveform templates are indispensable.

On the other hand, the GW sources are extremely general relativistic on the GW generation side. The typical characters include extremely strong gravitational force and highly dynamical evolution. Because of these issues, numerical relativity is almost the unique method to treat GW source modeling problems. Since numerical relativity admits no approximation to Einstein equation, numerical relativity is very reliable to model GW sources.

The task of numerical relativity is solving Einstein equation with a numerical method. At first glance, one need only code the Einstein equation and put it in a supercomputer. Afterwards, the numerical relativists just sit and wait for the results. The real life for numerical relativists is much harder \cite{cao2009}. The first issue numerical relativists need to face is how to make the numerical solution process stable. Otherwise, the code will break down soon and nothing can be obtained except a mount of `NaN' (not a number). From the 1960s to 2005, numerical relativists worked hard to solve this stability problem. After 2005, stability problem of numerical relativity is solved \cite{Pretorius:2005gq,Campanelli:2005dd,Baker:2005vv} in the sense that properly implemented numerical techniques can make specific binary merger calculation stable \cite{Cao:2008wn,ZhouJian2016,baumgarte2010}. On the other hand, there is no concrete mathematical theorem to guarantee a sufficient condition for the stability of numerical relativity. Numerical relativists need to extend existing experiences to new problems and make the calculation stable.

Along with the ground-based detection development for GW, several GW templates have been constructed including post-Newtonian approximation ones, numerical relativity surrogate ones, effective one body (EOB) numerical relativity series, and IMRPhenomenon series \cite{cai2016}. The state of art for the gravitational waveform template of compact binary coalescence is as following. The binary's total mass can be regarded as a unit to make all of the involved quantities in the system dimensionless. Consequently, the waveform models are available for any total mass. The mass ratio is an essential parameter. If the mass difference between the two objects is too large, the computational requirement is huge. Reliable numerical relativity results are available till mass ratio 1 to 20. Recently many efforts have been paid to binary black hole simulations with a mass ratio around 1 to 100 \cite{Lousto:2010ut,Lousto:2020tnb,Fernando:2018mov}. The mass ratio problem provides a computational efficiency challenge to numerical relativity. Before our project, complete inspiral-merger-ringdown waveform models only work for circular binaries. We developed complete waveform models for eccentric binaries. Based on the prediction of GR, GW admits memory. Before our project, there are some intuitive waveform models inspired by post-Newtonian approximation for GW memory. We proposed a new method to calculate GW memory accurately for the full inspiral-merger-ringdown process. In the following, we will give an introduction about our progress on the efficiency problem of numerical relativity, waveform model for eccentric binaries, and the GW memory model.

\subsection{Finite element Numerical relativity}\label{subsec:FiniteElement}

Numerical methods solving partial differential equations include finite difference method, spectral method, and finite element method. Numerical relativity codes based on finite difference method or spectral method have been widely used to simulate binary black holes for waveform model construction. There is no finite element code for numerical relativity having been used for binary black hole merger yet.

The existing numerical relativity codes based on finite difference method and the ones based on spectral method have their advantages and limitations respectively. The adaptive mesh refinement (AMR) is necessary to treat the multi-scale problem met in the binary black hole problem for finite difference code. AMR technique is powerful, but the strong parallel scaling ability is highly limited by the grid numbers on each mesh level. Current numerical relativity commonly uses about $100\times100\times100$ grid boxes. Consequently finite difference code can not use too many cores to simulate binary black hole systems \cite{LOFFLER201679}. The parallel scaling ability of  pseudo-spectral code is much worse due to the global data change among the spectral domain. The advantage of pseudo-spectral code is the high convergence property. The pseudo-spectral code needs much less cores than finite difference code for binary black hole system simulations \cite{Scheel:2008rj}. Unfortunately when the mass ratio increases, it is quite hard to tune the pseudo-spectral code to make it work.

The finite element has a local data property as finite difference when discriminating the space. So finite element method has comparably high parallel scaling advantage as finite difference method. The high order polynomial function basis and/or spectral function basis in each element are similar to the  spectral method (spectral element method). Consequently, the finite element method could combine the high parallel scaling property as the finite difference method and the high convergence property of the pseudo-spectral method. So it is possible to use the finite element method to treat the unsmooth region with small element and to treat the smooth region with large element but high order basis or spectral basis. The finite difference method has to transfer data between different mesh levels. So the size of single mesh limits the strong parallelization scalability for the finite difference method. All elements in the finite element method are treated as the same level. Consequently the finite element method may admit higher strong parallelization scalability than both the spectral method and the finite difference method.

Although the finite element method admits above attractive properties, it is still unclear how to construct a finite element code for binary black hole merger. Especially it is not clear yet how to stablize the numerical calculation based on the finite element method. The key details include that the weak form of Einstein equation is difficult to design, the gauge condition and the boundary condition are highly nontrivial to construct.

When we use the finite element method to solve Einstein equations numerically, we decompose the Einstein equations into the evolution part and constraint part and taking the whole task as a Cauchy problem. As the first step, we apply the finite element method to solve the constraint part for the initial data of the Cauchy problem \cite{Cao:2015via}.

Based on 3-metric $\gamma_{ij}$ and external curvature $K_{ij}$, the constraint equations can be written as following
\begin{eqnarray}
&&{\cal H}\equiv R-K_{ij}K^{ij}+K^2-16\pi \rho=0,\\
&&{\cal M}_i\equiv D_jK^j{}_i-D_iK-8\pi s_i=0,
\end{eqnarray}
where the $R$ is the scalar curvature with respect to the spatial metric $\gamma_{ij}$. $D$ is the covariant derivative operator consistent with the spatial metric. $\rho$ and $s_i$ are the mass density and momentum density of matter. We refer our reader to \cite{Cao:2008wn} for a detailed description of the quantities involved in the above constraint equations. Borrowing the experience in the previous numerical relativity schemes, we consider the puncture scheme proposed in \cite{Brandt:1997tf} with conformally flat assumption. In this scheme, the momentum constraints are solved analytically. After that, the 3-metric and the extrinsic curvature can be determined by the spins and velocities of the two black holes. In addition, a conformal factor $\psi$ is involved. And the conformal factor $\psi$ is determined by the Hamiltonian constraint equation
\begin{equation}
-(\partial^2_x+\partial^2_y+\partial^2_z)\psi=\frac{1}{8}\hat{K}^{ij}\hat{K}_{ij}\psi^{-7}+2\pi\rho\psi^{-3}.
\end{equation}
The positions of the black holes are singular and they are called puncture points. We can transform to a regular variable $u$. The authors in \cite{Brandt:1997tf} have proved that the solution $u$ is only $C^4$ around the puncture points. This unsmoothed behavior can be well treated by the weak form of our finite element method. Binary black hole systems are theoretically described by a vacuum, so $\rho=s_i=0$.

Using regular variable $u$ the Hamiltonian equation can be written as an non-linear Poisson equation
\begin{eqnarray}
-\nabla^2u=f(u)\text{ in } \Omega.\label{nlpoisson}
\end{eqnarray}
Based on the finite element scheme we approximate the domain $\mathbb{R}^3$ with some finite domain $\Omega$. At the boundary of $\Omega$ we consider both Dirichlet boundary condition and Robin boundary condition.

For the non-linear Poisson equation (\ref{nlpoisson}) and the Robin boundary condition, we use integration by part to construct the weak form. With the basis function of the finite element $\phi_i$, we expand the unknown function as $u=u^i\phi_i$. Accordingly, we discretize the above weak form equation and use the Newton iteration method to solve the set of non-linear equations for $u^i$. Noting that all the involved matrices are symmetric, we use the preconditioned conjugate gradient to solve the linear equations. The diagonal elements of the matrix in question are used as the preconditioner. Aided with the mentioned preconditioner we can get the converged solution soon. The numerical scheme for the Dirichlet boundary condition is similar. We refer our reader to the reference \cite{Cao:2015via} for details.

After the initial data is ready, we come to the evolution equations \cite{Cao:2018vhw}. There are well developed finite element methods for Hamilton-Jacobi-like equations \cite{yan2011local}
\begin{align}
u_t+H(u_x)=S(u)
\end{align}
where $u$ denotes general unknown functions. $H$ and $S$ are some specific functions depending on $u_x$ and $u$ respectively.

Denote the discretized domain as $I_j=(x_{j-\tfrac{1}{2}},x_{j+\tfrac{1}{2}}),j=1,...,N$. $x_j=\tfrac{1}{2}(x_{j-\tfrac{1}{2}}+x_{j+\tfrac{1}{2}})$ corresponds to the center of the cell $I_j$ and $\Delta x_j=x_{j+\tfrac{1}{2}}-x_{j-\tfrac{1}{2}}$ as the size of the cell. The function space corresponding to the numerical solution is defined as a piecewise polynomial space, There is no continuity requirement at the interfaces $x_{j\pm\tfrac{1}{2}}$. This non-continuity property means a discontinuous Garlerkin method.

The Legendre polynomials are used to decompose functions in the approximation space. Our local discontinuous Garlerkin finite element method includes two steps. As the first step, we calculate the derivative $u_x$ by solving the equation $\psi=u_x$. We have two numerical solutions $p_1$ and $p_2$ to the equation of $\psi$. Then the Lax-Friedrichs numerical Hamiltonian can be constructed 
\begin{align}
\hat{H}(p_1,p_2)=H(\frac{p_1+p_2}{2})-\frac{1}{2}(p_1-p_2).\label{eq:xi}
\end{align}

As the second step, we calculate $u_t$ through
\begin{align}
\int_{I_j}u_tvdx+\int_{I_j}\hat{H}(p_1,p_2)vdx=\int_{I_j}S(u)vdx.
\end{align}
We use this $u_t$ to update $u$ with respect to time based on the fourth-order Runge-Kutta method.

A limiter is used during the evolution to alleviate the high frequency numerical error.

Following the idea we proposed in \cite{Hilditch:2012fp}, a buffer cell is added for boundary condition treatment.

For a spherically symmetric spacetime, we can write the Einstein equation as \cite{Cao:2018vhw,Hilditch:2015aba}
\begin{align}
\partial_tu^\mu+A^{\mu}{}_\nu\partial_r u^\nu&=S^\mu,\label{problem_eq}
\end{align}
where the unknown variables $u^{\mu}$ and the related $A^{\mu}{}_\nu(u^\sigma)$ and $S^\mu(u^\sigma)$ are given in \cite{Cao:2018vhw}. The Eq.~(\ref{problem_eq}) is Hamilton-Jacobi-like. We apply directly the above finite element numerical scheme to the Eq.~(\ref{problem_eq}). We realized stable evolution of a Schwarzschild black hole based on Kerr-Schild coordinate, isotropic coordinate and Painleve-Gullstrand-like (PG) coordinate.

For a general spacetime without symmetry, we note that the generalized harmonic formalism of Einstein equations is Hamilton-Jacobi-like \cite{Ji:2019rue}. Accordingly, we apply the finite element method to the generalized harmonic formalism of Einstein equations based on the Parallel Hierarchical Grid (PHG) library \cite{zhang2009parallel,zhang2009set,Cao:2015via}. Since the continuous Galerkin module is more well developed than the discontinuous Galerkin module in PHG library, we used continuous Galerkin instead of discontinuous Galerkin in \cite{Cai:2018cgd}. Similar to the above filter, we adopt the filter \cite{FISCHER2001265,PASQUETTI2002646}
\begin{align}
F_{\alpha}=\alpha F_{N-1}+(1-\alpha){\rm I}, \label{eq:filter}
\end{align}
where $F_{N-1}$ is the interpolation operator from the space of the polynomials of maximum degree $N$ to the space of the polynomials of maximum degree $N-1$, ${\rm I}$ is the identity operator, and $\alpha\in (0,1]$ is the relaxation parameter which allows us to filter only a fraction of the highest mode. Since this filter is based on interpolations in physical space, the filtered solution will still be continuous at the boundaries between each two elements. In another word, this filter is consistent with the continuous Galerkin implemented here. With these numerical skills, we can evolve single Schwarzschild black hole stably.

\subsection{Gravitational waveform models for eccentric compact binaries}\label{subsec:ECB}

More than 50 compact binary coalescence (CBC) events have been detected by LIGO and VIRGO. One possible channel for the formation of merging binaries is isolated evolution in the field. Another possible channel is through dynamical interaction in dense stellar environments such as globular clusters or galactic nuclei. Currently, it is not clear the detected binaries are formed through which channel. The binaries formed in the field can radiate away their eccentricity. The dynamically formed binaries may still have significant residual eccentricity when their GWs enter the LIGO-Virgo band. People may infer the formation channel of the binary through the eccentricity detection. For space-based detectors including LISA \cite{Bender1998LISALI,Armano:2016bkm,Audley:2017drz}, Taiji \cite{Hu:2017mde,Guo:2018npi,Taiji-1} and TianQin \cite{Luo:2015ght,Luo:2020bls,Mei:2020lrl}, the orbit of the concerned binary black hole systems may be highly eccentric  \cite{hils1995gradual,Shibata:1994xk}. Consequently more and more attention is paid to the eccentricity detection recently \cite{Mikoczi:2012qy,Huerta:2013qb,Loutrel:2014vja,Coughlin:2014swa,Sun:2015bva,Ma:2017bux,Tanay:2019knc,Loutrel:2019kky,Loutrel:2016cdw,Loutrel:2017fgu,Loutrel:2018ssg,Moore:2018kvz,Loutrel:2018ydu,Gondan:2018khr,Gondan:2017wzd,Hoang:2017fvh,Gondan:2017hbp,Moore:2019xkm,Moore:2019vjj,Chiaramello:2020ehz,Nitz:2019spj,Lenon:2020oza,Ramos-Buades:2019uvh,Ramos-Buades:2020eju}.

In order to estimate the eccentricity, an accurate waveform model for the eccentric binary system is needed. Most waveform models for eccentric binary are based on post-Newtonian approximation and are consequently valid only for the inspiral part. Currently,  there are three waveform models that can cover the whole inspiral-merger-ringdown process for BBH. One is the Eccentric, Nonspinning, Inspiral, Gaussian-process Merger Approximated waveform model (ENIGMA) and the other two are based on EOB framework including the Effective-One-Body Numerical-Relativity waveform model for Spin-aligned binary black holes along Eccentric orbit (SEOBNRE) and the extended TEOBiResumS\_SM model. Among all of these waveform models for eccentric binary, only SEOBNRE and the extended TEOBiResumS\_SM model can treat spinning black holes. Such kind of complete waveform model is important to treat parameter degeneracy, especially between the black hole spin and orbit eccentricity.

When one tries to generalize the waveform template of CBC used by ground-based detectors to the waveform template used by space-based detectors, two key issues are involved. One is about the parameter completeness and the other is accuracy. Regarding the parameter completeness, it is the mass ratio of the binary and the eccentricity of the orbit that are involved. For ground-based detectors, only stellar massive black holes are involved. So current waveform template, which is valid from mass ratio 1 to 1 till 1 to 20, is enough. For space-based detectors, since all supermassive black holes, intermediate massive black holes and stellar massive black holes are involved, the mass ratio should cover from 1 to 1 till 1 to 10 million. Regarding the eccentricity, the binary black hole systems may become near-circular when they enter the LIGO frequency band due to the gravitational wave \cite{Peters:1964zz}. But when mass ratio increases, such circularization effect becomes less effective and consequently the eccentricity will be significant when the binary enter the detection band.

One may take the binary system with large mass ratio as a perturbation of the big black hole. Consequently the GW problem can be decomposed into a trajectory problem and a related waveform problem. Han \cite{Han:2014ana} used the Teukolsky equation to treat the waveform problem and used the conserved EOB dynamics with numerical energy flux to treat the trajectory \cite{Han:2011qz}. The Teukolsky equation is solved numerically \cite{Han:2011qz,Han:2014ana}. Teukolsky equation can also be solved through some analytical method \cite{Sasaki:2003xr} or post-Newtonian approximation \cite{Forseth:2015oua}. The authors of \cite{Drasco:2005kz} used the geodesic equation to treat the trajectory and used the Teukolsky equation to treat the waveform problem. Geodesic equation indicates that the eccentricity may increase \cite{Apostolatos:1993nu,Kennefick:1998ab}. In contrast post-Newtonian approximation found the eccentricity always decay \cite{Peters:1964zz}. Interesting transient resonance phenomena was reported in \cite{Flanagan:2010cd,Berry:2016bit}. When a binary  passes through a transient resonance, the radial frequency and polar frequency become commensurate, and the orbital parameters will show a jump behavior. In contrast, the post-Newtonian approximation method has not found the eccentricity increasing and the transient resonance behavior. Possibly this is because the available post-Newtonian result is not accurate enough. But there is also another possibility that the perturbation method breaks down. Ideally, numerical relativity simulation can answer this question. But current numerical relativity techniques can not simulate such large mass ratio systems \cite{Lousto:2010ut,Lousto:2020tnb} (but see \cite{Lewis:2016lgx}). Alternatively, the  effective-one-body-numerical-relativity (EOBNR) method may also answer this question. On the side of almost equal mass cases, EOBNR framework has been calibrated against numerical relativity; on the side of extreme mass ratio cases, EOB framework can in principle be used to describe the dynamics and the gravitational waveform \cite{Yunes:2009ef}. We try to use EOBNR framework to aid numerical relativity to solve large mass ratio problem. Based on the above consideration we plan to use EOBNR framework to fill the parameter gap of the current waveform template to let it satisfy the requirement of space-based detectors. During the past years, we have constructed such a waveform model called SEOBNRE \cite{Cao:2017ndf,Liu:2019jpg} where the last letter E represents eccentricity.

The EOB technique is a well known technique to treat the two-body problem in the central force situation of classical mechanics, especially for Newtonian gravity theory \cite{goldstein2002classical}. Buonanno and Damour proposed the seminal idea of EOB approach for a two-body problem in GR \cite{Buonanno:1998gg}. EOB approach has adopted many inputs from post-Newtonian approximation, but it is different to the post-Newtonian approximation. Post-Newtonian approximation diverges before the plunge stage. In contrast, the EOB approach works well till the merger. The EOB approach has also adopted the result of perturbation method \cite{Yunes:2009ef} and the results of numerical relativity. Such a combination results in effective-one-body numerical relativity (EOBNR) model \cite{Buonanno:2007pf}. The EOBNR model family includes version 1 \cite{Taracchini:2012ig}, version 2 \cite{Taracchini:2013rva}, version 3 \cite{Abbott:2016izl,Babak:2016tgq,Abbott:2017vtc} and version 4 \cite{Bohe:2016gbl}.

The EOB approach includes three building blocks: (1)  a Hamiltonian describing the conservative part of the dynamics of two compact bodies which is represented by; (2) the radiation-reaction force describing the dissipation force; and (3) the asymptotic gravitational waveform. The first part has nothing to do with the involved orbit. That is to say the first part is valid no matter the orbit is circular or eccentric. Since the radiation-reaction force can be related to energy and angular momentum carried away by GW, the second part is closely related to the third part. So the key issue of SEOBNRE model is the construction of the waveform for eccentric orbit.

The EOB approach describes the conservative dynamics of the two-body problem in GR as a geodesic motion (more precisely Mathisson-Papapetrou-Dixon equation \cite{Barausse:2009aa}) on the top of an effective spacetime of the reduced one body. The Finsler-type term may possibly be involved besides the geodesic motion \cite{Damour:2017zjx,He:2021cix}. The reduced one body spacetime is a deformed Kerr black hole \cite{Barausse:2009xi}.

The Hamiltonian of the geodesic motion can be written as \cite{Barausse:2011ys,Taracchini:2012ig,He:2021cix}
\begin{align}
H&=M\sqrt{1+2\eta(\frac{H_{eff}}{M\eta}-1)},\label{SEOBNREHtotal}\\
H_{eff}&=H_{NS}+H_S+H_{SC}.\label{SEOBNREHterms}
\end{align}

The equation of motion corresponding to the conservative part can be written as
\begin{align}
\dot{\vec{r}}&=\frac{\partial H}{\partial\vec{\tilde{p}}},\label{dynr}\\
\dot{\vec{\tilde{p}}}&=-\frac{\partial H}{\partial\vec{r}}.\label{dynp}
\end{align}

Since the part 2 is related to the part 3, we describe our waveform formula first, and introduce the back reaction force afterwards. The gravitational waveform can be decomposed as spin-weighted $-2$ spherical harmonic modes. This kind of decomposition has been widely used in numerical relativity \cite{Bai:2011za}. Only the dominate modes $(l,m)=(2,\pm2)$ are available in \cite{Cao:2017ndf,Liu:2019jpg}. Later \cite{Liu:2021pkr} we rebuild the waveform modes for $(l,m)=(2,\pm2),(2,\pm1),(3,\pm3),(4,\pm4)$. Only the positive $m$ modes are needed and the negative $m$ modes can be got through relation $h_{l m}=(-1)^l \bar{h}_{l,-m}$ \cite{Pan:2011gk,Cao:2017ndf}. Here the over bar means the complex conjugate.

When constructing the eccentric waveform, we divide the waveform into quasi-circular part and eccentric part. Following \cite{Huerta:2016rwp} we treat the eccentric part as a perturbation. We borrow the the quasi-circular part from SEOBNR models. In \cite{Cao:2017ndf,Liu:2019jpg}, SEOBNRv1 was used. In \cite{Liu:2021pkr} SEOBNRv4 was used. The waveform is divided into two segments including quasi-normal modes and inspiral-plunge waveform. The inspiral-plunge waveform is written as \cite{Pan:2011gk}
\begin{align}
h_{l m}^{(C)}&=h_{l m}^{(N,\epsilon)}\hat{S}_{eff}^{(\epsilon)}T_{l m}e^{i\delta_{l m}}(\rho_{l m})^l N_{l m},\label{CEOB}\\
h_{l m}^{(N,\epsilon)}&=\frac{M\eta}{R}n_{l m}^{(\epsilon)}c_{l+\epsilon}V_\Phi^l Y^{l-\epsilon,-m}(\frac{\pi}{2},\Phi),
\end{align}
where $R$ is the luminosity distance of the source; $\Phi$ denotes the orbital phase; $Y^{l m}(\Theta,\Phi)$ are the usual spherical harmonics. In the non-quasi-circular correction term $N_{l m}$ \cite{Cao:2017ndf,Liu:2019jpg} depends on the parameters $a_{1}^{h_{l m}}$, $a_{2}^{h_{l m}}$, $a_{3}^{h_{l m}}$, $b_{1}^{h_{l m}}$, $b_{2}^{h_{l m}}$ and $a_{3S}^{h_{l m}}$, $a_{4}^{h_{l m}}$, $a_{5}^{h_{l m}}$, $b_3^{h_{l m}}, b_4^{h_{l m}}$. These parameters depend on $a$ and $\eta$. Following SEOBNRv1, we construct data tables for $a_{i}^{h_{l m}}$, $a_{3S}^{h_{l m}}$, $b_{1}^{h_{l m}}$ and $b_{2}^{h_{l m}}$. Then we get the wanted values for the $a$ and $\eta$ through interpolation. Then we solve the conditions (21)-(25) of \cite{Taracchini:2012ig} for $b_3^{h_{l m}}$ and $b_4^{h_{l m}}$. In \cite{Liu:2021pkr} a completely different scheme was taken to treat the non-quasi-circular correction term $N_{l m}$ which follows closely the SEOBNRv4 model.

The post-Newtonian (PN) result is valid till second PN order for the eccentric part \cite{Will:1996zj}. We refer our reader to \cite{Cao:2017ndf} for detailed eccentric part waveforms. The eccentric correction means
\begin{align}
h_{22}^{\rm (PNE)}=h_{22}-h_{22}|_{\dot{r}=0}.\label{h22willelip}
\end{align}
Then our inspiral-plunge waveform reads
\begin{align}
h_{22}^{insp-plun}=h_{22}^{(C)}+h_{22}^{\rm (PNE)}, \label{Ewave}
\end{align}
where $h_{22}^{(C)}$ is given in Eq.~(\ref{CEOB}).

In \cite{Cao:2017ndf,Liu:2019jpg} we took the above (\ref{h22willelip}) directly as the eccentric waveform part. In \cite{Liu:2021pkr} we borrow the factorization and resummation skills developed in SEOBNR quasi-circular waveform models to treat (\ref{h22willelip}). If we denote the resulted eccentric part waveform $h_{22}^{\rm (FPNE)}$, the waveform used in \cite{Liu:2021pkr} can be written as
\begin{align}
h_{22}^{insp-plun}=h_{22}^{(C)}+h_{22}^{\rm (FPNE)}.
\end{align}

The idea to construct the other modes than $(2,2)$ is similar. We refer our reader to \cite{Liu:2021pkr} for details. Based on these waveforms, we have the energy flux $\frac{dE}{dt}$ of GW \cite{Pan:2010hz}
\begin{align}
-\frac{dE}{dt}=\frac{1}{16\pi}\sum_l\sum_{m=-l}^{l}|\dot{h}_{l m}|^2.
\end{align}
We assume that $h_{l m}$ behaves as a harmonic oscillation. Consequently $\dot{h}_{l m}\approx m\Omega h_{l m}$ and $\Omega$ if the orbital frequency. Then
\begin{align}
-\frac{dE}{dt}&=\frac{1}{16\pi}\sum_l\sum_{m=-l}^{l}(m\Omega)^2|h_{l m}|^2\\
&=\frac{1}{8\pi}\sum_l\sum_{m=1}^{l}(m\Omega)^2|h_{l m}|^2.\label{waveformflux}
\end{align}

Then, the radiation-reaction force $\vec{\mathcal{F}}$ can be written as \cite{Taracchini:2012ig}
\begin{align}
\vec{\mathcal{F}}&=\frac{1}{M\eta\omega_\Phi|\vec{r}\times\vec{\tilde{p}}|}\frac{dE}{dt}\vec{\tilde{p}},\\
\omega_\Phi&=\frac{|\vec{r}\times\dot{\vec{r}}|}{r^2}.
\end{align}
$E$ here means the energy of the binary system. $E$ decreases due to the gravitational wave, $\frac{dE}{dt}<0$. The negative sign corresponds to dissipation. For quasi-circular cases without precession, since $|\vec{r}\times\vec{\tilde{p}}|\approx \tilde{p}_\phi$ our above relation reduces to
\begin{align}
\vec{\mathcal{F}}&=\frac{1}{M\eta\omega_\Phi}\frac{dE}{dt}\frac{\vec{\tilde{p}}}{\tilde{p}_\phi},
\end{align}
which is used by SEOBNR series waveform models.

Together with the force $\vec{\mathcal{F}}$, the whole EOB dynamics becomes
\begin{align}
\dot{\vec{r}}&=\frac{\partial H}{\partial\vec{\tilde{p}}},\label{fdynr}\\
\dot{\vec{\tilde{p}}}&=-\frac{\partial H}{\partial\vec{r}}+\vec{\mathcal{F}}.\label{fdynp}
\end{align}
So the above dynamical equations coupled to waveform are self-contained. Along the evolution of the dynamics, the waveform in time can be constructed. This is the basic strategy of our SEOBNRE waveform models. Through comparing to numerical relativity simulations, we find that our SEOBNRE waveform model works well for highly spinning ($\chi=0.99$), highly eccentric ($e_0$ at $Mf_0=0.02$ can reach 0.7) and large mass ratio ($q\leq10$) BBH. In the near future, we will use more numerical relativity simulations to test even larger mass ratio. Especially we expect our SEOBNRE model can work for the whole mass ratio range \cite{Yunes:2009ef}.

\subsection{GW memory models for compact binaries}\label{subsec:GWmemory}

The same as other waveform models, our SEOBNRE described in the last section does not include $m=0$ modes. Such modes are closely related to the GW memory. GW memory is an interesting prediction of GR. The detection of the GW memory can be used to test GR and to recover the property of the GW source. A quantitative model is needed for such detection and parameters reconstruction.

In \cite{Cao:2016maz} we borrowed the idea of SEOBNR waveform models for the non-memory mode to divide the $(2,0)$ waveform mode into three segments including inspiral, merger, and ringdown. The inspiral part uses the PN approximation. The ringdown part uses the quasi-normal modes. Due to the memory effect, the quasi-normal modes oscillate around a final memory value instead of 0. The merger part is the most difficult part to model. Not like the non-memory modes, the inspiral waveform can join the ringdown waveform perfectly. For $(2,0)$ mode, there is a jump between the inspiral waveform and the final ringdown waveform. This is why we introduce a stand-alone merger segment in \cite{Cao:2016maz}.

The post-Newtonian approximation for GW memory reads \cite{Favata:2008yd,Blanchet:1992br}
\begin{eqnarray}
&&h^{(\text{mem})}=\sum_{l=2}^\infty\sum_{m=-l}^lh^{lm}_{(\text{mem})}{}_{-2}Y_{lm}(\theta,\phi),\\
&&h^{lm}_{(\text{mem})}=R\sqrt{\frac{(l-2)!}{(l+2)!}}\sum_{l'=2}^\infty\sum_{l''=2}^\infty\sum_{m'=-l'}^{l'}
\sum_{m''=-l''}^{l''}(-1)^{m+m''}\nonumber\\
&&\times\int_{-\infty}^t<\dot{h}_{l'm'}\dot{\bar{h}}_{l''m''}>dt'\int{}_{2}Y_{l'm'}{}_{-2}\overline{Y}_{l''m''}\overline{Y}_{lm}d\Omega.\nonumber\\
\label{eq:Kip}
\end{eqnarray}
The angle-bracket $<...>$ in the above equation means the average over several wavelengths. Here $\theta$ and $\phi$ are the angular coordinates of the detector. $R$ is the luminosity distance of the GW source. $Y_{lm}$ and ${}_{-2}Y_{lm}$ are the usual spherical harmonic function and the spin weighted spherical harmonic function with spin-weight $-2$, respectively. $h_{lm}$ are the spin weighted spherical harmonic modes of the non-memory part of the GW. The dot means the time derivative. The overbar means the complex conjugate. Note that $h_{lm}$ is proportional to $\frac{1}{R}$. Therefore $h^{lm}_{(\text{mem})}$ behaves as $\frac{1}{R}$ also.

The non-memory part of PN waveform can be written as [the Eq.~(B1) of \cite{Favata:2008yd}]
\begin{eqnarray}
h_{l m}&=&8\sqrt{\frac{\pi}{5}}\frac{\eta M x}{R}e^{-im\psi}\hat{h}_{lm}.\label{eqeobwave}
\end{eqnarray}
Here $x=(M\omega)^{2/3}$ is the PN parameter with $\omega$ the orbital frequency. The phase variable $\psi$ is related to orbital phase $\varphi$ through [the Eq.~(B2) of \cite{Favata:2008yd}]
\begin{eqnarray}
\psi&=&\varphi-3x^{3/2}[1-\frac{\eta}{2}x]\ln(\frac{x}{x_0}),\\
\ln x_0&=&\frac{11}{18}-\frac{2}{3}\gamma_E-\frac{4}{3}\ln 2,
\end{eqnarray}
where $\gamma_E$ is Euler's constant with approximated value $0.577216$. The time $t$ and the PN parameter $x$ can be related through
\begin{eqnarray}
&&\frac{dx}{dt}=\frac{-{\mathcal F}}{dE/dx},\label{eq:dxdt}
\end{eqnarray}
where ${\mathcal F}$ is the GW luminosity and $E$ the orbital energy. Then we can reduce Eq.~(\ref{eq:Kip}) to
\begin{align}
h^{lm}_{(\text{mem})}=&R\sqrt{\frac{(l-2)!}{(l+2)!}}\sum_{l'=2}^\infty\sum_{l''=2}^\infty\sum_{m'=-l'}^{l'}
\sum_{m''=-l''}^{l''}(-1)^{m+m''}\nonumber\\
&\int_{0}^x<\frac{dh_{l'm'}}{dx'}\frac{d\bar{h}_{l''m''}}{dx'}>\frac{dx'}{dt}dx'
\int{}_{2}Y_{l'm'}{}_{-2}\overline{Y}_{l''m''}\overline{Y}_{lm}d\Omega,
\end{align}
where we have used $x'$ to denote the dummy integration variable.

Combining all above results we achieve
\begin{align}\label{PNmodel}
\frac{R}{M}h^{(\text{mem})}_{20}=&\frac{4}{7}\sqrt{\frac{5\pi}{6}}\eta
x\left\{1+x(-\frac{4075}{4032}+\frac{67}{48}\eta)\right.\nonumber\\
&\left.+x^{3/2}(\frac{-2813+756\eta}{2400}\chi_S+\frac{3187}{2400}\delta\chi_A)\right\}.
\end{align}
This is the inspiral segment of the memory waveform model we constructed in \cite{Cao:2016maz}.

Now we come to the quasi-normal segment of the memory waveform. Respect to time we identify this part as the one after the peak of the (2,2) mode. Different to the $m\neq0$ quasi-normal modes model, the QNM oscillates around an nonzero value due to the ``memory" effect. This nonzero ``memory" value $h^{\infty}_{20}$ (corresponding to the notation $h^{tot}_{20}$ of \cite{Pollney:2010hs,Liu:2021zys}) has been investigated in \cite{Pollney:2010hs} for equal mass non-precession binary black hole mergers
\begin{align}
\frac{R}{M}h^{\infty}_{20}=&0.0996+0.0562\chi_S+0.0340\chi_S^2\nonumber\\
&+0.0296\chi_S^3+0.0206\chi_S^4.\label{eq:A2}
\end{align}
One caution is in order for the above equation. The numerical relativity results in \cite{Pollney:2010hs} only include equal mass BBH. So the above equation is only valid for equal mass BBH cases. We have extended the above results to general spin-aligned binary black hole mergers in \cite{Liu:2021zys}
\begin{align}
\frac{D}{M}h^{\infty}_{20}=&[0.0969+0.0562\chi_{\rm up}+0.0340\chi_{\rm up}^2\nonumber\\
&+0.0296\chi_{\rm up}^3+0.0206\chi_{\rm up}^4](4\eta)^{1.65}.\label{meq6}
\end{align}
\begin{align}
\chi_{\rm up}\equiv\chi_{\rm eff}+\frac{3}{8}\sqrt{1-4\eta}\chi_{\rm A}.\label{meq7}
\end{align}

The above result only considers the asymptotic memory value. Respect to time, the center value of the QNM oscillation increases gradually. In order to find out this gradual increasing behavior, we first construct a simple QNM model for $(2,0)$ as following
\begin{align}
\frac{R}{M}h_{20,QNM}=\frac{R}{M}h^{\infty}_{20}-\rho
e^{-\frac{t-T_R}{\tau_{200}}}\cos[\omega_{200}(t-T_R)],\label{QNMmodel}
\end{align}
where $T_R$ describes the time when the ringdown waveform begins, $\rho$ describes the amplitude of the QNM. $\sigma_{200}=\frac{1}{\tau_{200}}+i\omega_{200}$ corresponds to the complex frequency of the quasi-normal mode $l=2,m=0,n=0$. The final mass and the final spin of the binary black hole merger can be determined by the initial parameters of the two black holes.

We have fine tuned the parameters $h^{\infty}_{20}$, $\rho$ and $T_R$ in Eq.~(\ref{QNMmodel}) to best fit the numerical relativity results of \cite{Pollney:2010hs}
\begin{eqnarray}
\rho&=&0.41667\chi_S^4-0.52083\chi_S^3\nonumber\\
&&+0.19583\chi_S^2-0.036667\chi_S+0.015,\label{eq:rho}\\
T_R&=&528.65\chi_S^4-751.04\chi_S^3\nonumber\\
&&+326.35\chi_S^2-42.958\chi_S+2.5.\label{eq:TR}
\end{eqnarray}
It is interesting to ask whether these fitted parameters work for the numerical relativity results of \cite{Mitman:2020pbt,Mitman:2020bjf}. That's our future work.

Although we have constructed a simple model for the merger part of memory waveform, it's not good enough we think. So we do not describe that part in detail here. The interesting readers can refer \cite{Cao:2016maz} for the detail.

In \cite{Liu:2021zys} we designed a new method to accurately calculate the memory. Our method does not need slow motion and weak field approximations of GW sources. Our method can accurately calculate memory based on non-memory waveform. Many studies including \cite{Favata:2008yd,Favata:2009ii,Favata:2008ti,Favata:2010zu,Favata:2011qi,Talbot:2018sgr} applied the above PN results (\ref{eq:Kip}) to binary black hole merger to get the gravitational waveform of memory. And later this GW memory waveform was used to analyze LIGO data \cite{Lasky:2016knh,Boersma:2020gxx,Hubner:2019sly,Ebersold:2020zah}. In \cite{Liu:2021zys} we use our accurate calculation method to confirm that this naive approximated waveform for memory is quite accurate till merger actually.

Using Bondi-Sachs (BS) coordinate $(u,r,\theta,\phi)$, Bondi-Metzner-Sachs (BMS) theory describes gravitational radiation with the concept of null infinity. Here $u$ corresponds to the time of observer very far away from the GW source. In BMS theory, the GW source is looked as an isolated spacetime. The gravitational waveform depends only on $(u,\theta,\phi)$. And more the waveform is proportional to $\frac{1}{D}$ where $D$ is the luminosity distance. For simplicity we use notation `$t$' for the Bondi time $t=u$. Aided with some mathematical skills and after a tedious calculation, we get for $l\geq2$
\begin{align}
&h_{lm}\bigg{|}_{t_1}^{t_2}=-\sqrt{\frac{(l-2)!}{(l+2)!}}\left[\left.\frac{4}{D}\int\Psi_2^{\circ}[{}^0Y_{lm}]\sin\theta d\theta d\phi\right|_{t_1}^{t_2}\right.-\nonumber\\
&\,\,\,\,D\sum_{l'=2}^{\infty}\sum_{l''=2}^{\infty}\sum_{m'=-l'}^{l'}\sum_{m''=-l''}^{l''}
\Gamma_{l'l''lm'-m''-m}\times\nonumber\\
&\,\,\,\,\,\left(\int_{t_1}^{t_2}\dot{h}_{l'm'}\dot{\bar{h}}_{l''m''}dt-\dot{h}_{l'm'}(t_2)\bar{h}_{l''m''}(t_2)+\left.\dot{h}_{l'm'}(t_1)\bar{h}_{l''m''}(t_1)\right)\right],\label{meq4}
\end{align}
This set of coupled equations can be looked as unknowns $h_{l0}$ respect to $h_{lm},m\neq0$. The GW memory is dominated by modes $h_{l0}$ for non-precession binary black holes. That is why $h_{l0}$ are called GW memory modes and $h_{lm},\,m\neq0$ are called non-memory modes.

Because $\dot{h}_{l0}\approx0$ \cite{Favata:2008yd}, we get
\begin{align}
&h_{l0}\bigg{|}_{t_1}^{t_2}=-\sqrt{\frac{(l-2)!}{(l+2)!}}\Re\left[\left.\frac{4}{D}\int \Psi_2^{\circ}[{}^0Y_{l0}]\sin\theta d\theta d\phi\right|_{t_1}^{t_2}\right.-\nonumber\\
&D\sum_{l'=2}^{\infty}\sum_{l''=2}^{\infty}\sum_{\mbox{\tiny$\begin{array}{c}
m'=-l',\\
m'\neq0\end{array}$}}^{l'}\sum_{\mbox{\tiny$\begin{array}{c}
m''=-l'',\\
m''\neq0\end{array}$}}^{l''}
\Gamma_{l'l''lm'-m''0}\times\nonumber\\
&\,\,\,\,\,\left(\int_{t_1}^{t_2}\dot{h}_{l'm'}\dot{\bar{h}}_{l''m''}dt-\dot{h}_{l'm'}(t_2)\bar{h}_{l''m''}(t_2)+\right.\nonumber\\
&\,\,\,\,\,\,\,\,\,\,\,\,\,\,\,\,\,\,\,\,\,\,\,\,\,\,\,\left.\left.\dot{h}_{l'm'}(t_1)\bar{h}_{l''m''}(t_1)\right)\right].\label{meq2}
\end{align}

In the mass center frame of the whole system, we have $\Psi_2^\circ(-\infty,\theta,\phi)=M_0$. Here $M_0$ corresponds to the Bondi mass and also ADM mass \cite{Ashtekar:2019viz}. Later, the Bondi mass $M$ decreases because the GW carries out some energy $E_{\rm GW}$, $M=M_0-E_{\rm GW}$. The spacetime will at last becomes a Kerr black hole with mass $\tilde{M}$. But the mass center frame of the Kerr black hole is different from the mass center frame at beginning due to the kick velocity. These two asymptotic inertial frames are related by a boost transformation. So we have $\tilde{M}=M/\gamma$, where $\gamma$ is the Lorentz factor.

Since the kick velocity is small $v\ll c$, the contribution $\left.\dot{h}_{l'm'}\bar{h}_{l''m''}\right|_{t_1}^{t_2}$ is small, our Eq.~(\ref{meq2}) recovers the PN approximation result (\ref{eq:Kip}). Detail investigations done in \cite{Liu:2021zys} the corrections introduced by kick velocity $\vec{v}$ are negligible for the whole inspiral-merger-ringdown process. Such corrections contribute between 0.01\% and 1\% for all kinds of spin-aligned binary black holes.

But we need to note that the accurately calculated waveform for memory (\ref{meq2}) can not capture the behavior of quasi-normal ringing. So it is interesting to combine the above results (\ref{QNMmodel}) and (\ref{meq2}) together to construct a complete waveform for memory. After that add these $m=0$ modes into our SEOBNRE waveform model to get a most general waveform model for binary black hole systems. When such tasks are done, the waveform of binary black holes with arbitrary mass ratio, arbitrary spin, and arbitrary orbit shape can be described by our SEOBNRE waveform model. Then naturally it can work for GW data analysis for space detectors.

\section{Conclusions}\label{sec:conclusion}

In this brief review, we have summarized the important progress of GW studies over the past five years but with a special focus on some of our own work within a key project supported by the National Natural Science Foundation of China, due to the limited time and space, compared to the massive literature in this field, for which we should apologize sincerely here if we have missed any such important papers. The main results we achieved are summarized below:

(1) We have carried out a 3D numerical simulation for the phase transition involving a gauge field to account for the primordial magnetic field. We have found a new GW generation mechanism from preheating with a cuspy potential. We have studied the induced GWs from non-Gaussian curvature perturbations. We have found large anisotropies in the stochastic GW background from cosmic domain walls. We have proposed LISA-Taiji network for a joint detection and fast and accurate localization of GW events, which also doubles the number of EM counterparts compared to a single Taiji mission so as to provide much tighter constraints on cosmological parameters. 

(2) We have obtained strong constraints on the neutron star maximum mass by using the GW and electromagnetic observations of GW170817 with consideration of different EOSs. We have also helped to organize the Chinese-PTA collaboration to search for nano-Hertz GW radiation and investigated the detection of individual SMBBHs by current and future PTAs.

(3) We have developed finite element numerical relativity algorithm and a waveform model SEOBNRE for binary black hole coalescence along an eccentric orbit.

As the flourishing progress has been made from the scientific side, the engineering side has also achieved significant progress. In 2015  the LISA pathfinder was launched with great success. In 2019, two Chinese space-borne GW detector projects, Taiji and TianQin, also launched  Taiji-1 and TianQin-1 separately, going the first step to GW detection in space. By the time of the 2030s, it is expected that the space-based GW detectors will run in space. Certainly, updated ground-based GW detectors, PTA measurements and even B-mode polarization measurements of CMB will give us unexpected surprises in a near future. By the time, one is  able to resolve those important theoretical issues in GW cosmology and GW astrophysics from the possible detections of massive binary coalescence events, EMRIs events, SMBBHs event, GWs from preheating and first-order phase transition, induced GWs background, and more GW events with EM counterparts.

\acknowledgments
This review is a status progress report supported by the National Natural Science Foundation of China Grants No.11690021, No.11690022, No.11690023, No.11690024.

\bibliographystyle{JHEP}
\bibliography{ref}
\end{document}